\documentclass[abstract=true,headings=standardclasses]{scrartcl}

\usepackage[utf8]{inputenc}
\usepackage[T1]{fontenc}
\usepackage{moreverb}
\usepackage[intlimits]{amsmath}
\usepackage{amssymb}
\usepackage{booktabs}
\usepackage{wrapfig}
\usepackage{graphicx}
\usepackage{siunitx}
\usepackage{algorithm}
\usepackage{algpseudocode}
\usepackage[acronyms]{glossaries}
\usepackage{hyperref}
\usepackage{subfig}
\usepackage{bbm}
\usepackage{xcolor}
\usepackage{xspace}

\newacronym{ABC}{\protect ABC}{absorbing boundary condition}
\newacronym{AWE}{\protect AWE}{asymptotic waveform evaluation}
\newacronym{BC}{\protect BC}{boundary condition}
\newacronym{CCCS}{\protect CCCS}{current controlled current source}
\newacronym{CCVS}{\protect CCVS}{current controlled voltage source}
\newacronym{CFL}{\protect CFL}{\textsc{Courant}\xspace-\textsc{Friedrichs}\xspace-\textsc{Levy}\xspace}
\newacronym{EQS}{\protect EQS}{electroquasistatic}
\newacronym{EM}{\protect EM}{electromagnetic}
\newacronym{EMs}{\protect EM}{electromagnetics}
\newacronym{ET}{\protect ET}{electrothermal}
\newacronym{FDTD}{\protect FDTD}{finite difference time domain}
\newacronym{FEM}{\protect FEM}{finite element method}
\newacronym{FIT}{\protect FIT}{finite integration technique}
\newacronym{KCL}{\protect KCL}{\textsc{Kirchhoff}\xspace's current law}
\newacronym{KVL}{\protect KVL}{\textsc{Kirchhoff}\xspace's voltage law}
\newacronym{MGEs}{\protect MGEs}{\textsc{Maxwell}\xspace grid equations}
\newacronym{MNA}{\protect MNA}{modified nodal analysis}
\newacronym{MOR}{\protect MOR}{model order reduction}
\newacronym{MPVL}{\protect MPVL}{matrix \textsc{Pad\'{e}}\xspace via a \textsc{Lanczos}\xspace-type process}
\newacronym{PEC}{\protect PEC}{perfect electric conducting}
\newacronym{PEEC}{\protect PEEC}{partial element equivalent circuit}
\newacronym{PMC}{\protect PMC}{perfect magnetic conducting}
\newacronym{PRIMA}{\protect PRIMA}{passive reduced-order interconnect macromodeling algorithm}
\newacronym{TE}{\protect TE}{transverse electric}
\newacronym{TM}{\protect TM}{transverse magnetic}
\newacronym{VCCS}{\protect VCCS}{voltage controlled current source}
\newacronym{VCVS}{\protect VCVS}{voltage controlled voltage source}

\catcode`@=11
\newcommand{\frownfill}{\ensuremath{\scriptscriptstyle\m@th\mathord\frown}}
\newcommand{\bow}[1]{\ensuremath{\vbox{\m@th\ialign{##\crcr
      \hfil\frownfill\hfil\crcr\noalign{\kern-0.2\p@\nointerlineskip}
      $\hfil\displaystyle{#1}\hfil$\crcr}}}}
\newcommand{\bbow}[1]{\ensuremath{\vbox{\m@th\ialign{##\crcr
     \hfil\frownfill\hfil\crcr\noalign{\kern-0.7\p@\nointerlineskip}
     \hfil\frownfill\hfil\crcr\noalign{\kern-0.3\p@\nointerlineskip}
      $\hfil\displaystyle{#1}\hfil$\crcr}}}}
\newcommand{\bbbow}[1]{\ensuremath{\vbox{\m@th\ialign{##\crcr
     \hfil\frownfill\hfil\crcr\noalign{\kern-0.7\p@\nointerlineskip}
     \hfil\frownfill\hfil\crcr\noalign{\kern-0.7\p@\nointerlineskip}
     \hfil\frownfill\hfil\crcr\noalign{\kern-0.3\p@\nointerlineskip}
      $\hfil\displaystyle{#1}\hfil$\crcr}}}}
\newcommand{\widefrownfill}{\ensuremath{\m@th\mathord\frown}}
\newcommand{\widebow}[1]{\ensuremath{\vbox{\m@th\ialign{##\crcr
      \hfil\widefrownfill\hfil\crcr\noalign{\kern-0.9\p@\nointerlineskip}
      $\hfil\displaystyle{#1}\hfil$\crcr}}}}
\newcommand{\widebbow}[1]{\ensuremath{\vbox{\m@th\ialign{##\crcr
     \hfil\widefrownfill\hfil\crcr\noalign{\kern-1.8\p@\nointerlineskip}
     \hfil\widefrownfill\hfil\crcr\noalign{\kern-0.9\p@\nointerlineskip}
      $\hfil\displaystyle{#1}\hfil$\crcr}}}}
\newcommand{\widebbbow}[1]{\ensuremath{\vbox{\m@th\ialign{##\crcr
     \hfil\widefrownfill\hfil\crcr\noalign{\kern-1.8\p@\nointerlineskip}
     \hfil\widefrownfill\hfil\crcr\noalign{\kern-1.8\p@\nointerlineskip}
     \hfil\widefrownfill\hfil\crcr\noalign{\kern-0.9\p@\nointerlineskip}
      $\hfil\displaystyle{#1}\hfil$\crcr}}}}

\usepackage{authblk}

\setkomafont{title}{\normalfont}
\title{\LARGE Automated Netlist Generation for 3D Electrothermal and Electromagnetic Field Problems}
\author[1,2]{Thorben~Casper}
\author[1]{David~Duque}
\author[1,2]{Sebastian~Sch\"ops}
\author[1,2]{Herbert~De~Gersem}
\affil[1]{Institut f\"ur Theorie Elektromagnetischer Felder, Technische Universit\"at Darmstadt, Schlo{\ss}gartenstr. 8, 64289 Darmstadt Germany}
\affil[2]{Graduate School of Computational Engineering, Technische Universit\"at Darmstadt, Dolivostr. 15, 64293 Darmstadt, Germany}
\date{\vspace{-5ex}}

\newenvironment{keywords}{\begin{trivlist}\item[]{\bfseries Keywords:}}{\end{trivlist}}

\begin{document}

\maketitle

\begin{abstract}
We present a method for the automatic generation of netlists describing general three-dimensional electrothermal and electromagnetic field problems. Using a pair of structured orthogonal grids as spatial discretisation, a one-to-one correspondence between grid objects and circuit elements is obtained by employing the finite integration technique. The resulting circuit can then be solved with any standard available circuit simulator, alleviating the need for the implementation of a custom time integrator. Additionally, the approach straightforwardly allows for field-circuit coupling simulations by appropriately stamping the circuit description of lumped devices. As the computational domain in wave propagation problems must be finite, stamps representing absorbing boundary conditions are developed as well. Representative numerical examples are used to validate the approach. The results obtained by circuit simulation on the generated netlists are compared with appropriate reference solutions.


\end{abstract}

\section{Introduction}
\label{sec:introduction}

In order to analyse complex \gls*{EM} problems, two main routes can be identified.
First, a numerical approach for solving the full set of \textsc{Maxwell}\xspace's field equations offers the advantage of capturing all relevant effects but can be very demanding in terms of computational resources.
One of the first methods for \gls*{EM} analysis was the \gls*{FDTD} scheme proposed by Yee~\cite{Yee_1966aa} in the 1960s.
This method became popular and is still a standard approach for high-frequency \gls*{EM} simulations~\cite{Taflove_1998aa}.
About ten years later, in the 1970s, Weiland~\cite{Weiland_1977aa} introduced the \gls*{FIT} as an extension to the \gls*{FDTD} scheme using integral unknowns and allowing for non-Cartesian and unstructured grids~\cite{Rienen_1985aa}.
Additionally, later developments of the method led to the usage of integral unknowns to obtain an exact implementation of \textsc{Maxwell}\xspace's equations~\cite{Weiland_1996aa}.
The method proved to be more efficient in terms of memory requirements and computing time.
While the \gls*{FEM} was mainly used in structural mechanics for many years, the introduction of edge elements made it applicable for \gls*{EM} problems as well~\cite{Bossavit_1988ab}.
Secondly, one may use compact models to obtain an efficient representation of a complex system.
For example, electrical engineers employ circuits to model and describe the behaviour of complex devices.
Nevertheless, the generation of such compact models can be a tedious task requiring empirical know-how to apply the appropriate approximations.
To generate such circuit models, different techniques are available.
For instance, mathematical analysis and physical insight allows to construct circuits representing the problem at hand as is done by Choi et al~\cite{Choi_2000aa}.
A circuit's topology and the required component values can also be obtained from experimental results as is done by Moumouni and Baker~\cite{Moumouni_2015aa} and other groups.
Another approach, e.g. followed by Codecasa et al~\cite{Codecasa_2016aa}, Eller~\cite{Eller_2017ae} and Wittig et al~\cite{Wittig_2002aa}, is to apply \gls*{MOR} techniques directly to the field formulation of the problem from which a circuit description can be found more easily~\cite{Codecasa_2016aa}.
However, the resulting elements may have non-physical values.
Now, if one is able to represent an \gls*{EM} problem by means of an electric circuit, one can use any circuit simulator to obtain the solution.
The most popular representatives are SPICE\xspace programs that were introduced in the 1970s~\cite{Nagel_1973aa} and are still used as a synonym for circuit solvers.
Later, extensions to deal with \gls*{ET} simulations using circuits were developed within the SPICE\xspace framework~\cite{Vogelsong_1989aa,Hefner_1993aa}.
The mathematical tool employed by most SPICE\xspace-like programs is still the \gls*{MNA} presented by Ho et al~\cite{Ho_1975aa} in the 1970s.

First approaches to combine numerical field simulation with circuit elements were proposed in the 1990s.
These were based on the \gls*{FDTD} scheme~\cite{Sui_1992aa,Tsuei_1993aa,Piket-May_1994aa,Thomas_1994aa} because of the topological similarities between circuits and the finite difference scheme.
Later, the insertion of lumped elements into \gls*{FEM} schemes in time or frequency domain was developed by Guillouard et al~\cite{Guillouard_1996aa,Guillouard_1999aa}.
These approaches became known as field-circuit coupling and have evolved into an important research topic~\cite{Kettunen_2001aa,Benderskaya_2004aa,Schops_2013aa,Schops_2011ac}.
One possible field-circuit coupling method is the direct insertion of lumped circuit elements into the field model by applying these to edges of the discretisation grid~\cite{Witting_1997aa}.

To extract compact circuits representing \gls*{EM} problems in a generic way, numerous approaches can be found in the literature.
In the \gls*{PEEC} method presented by Ruehli~\cite{Ruehli_1974aa,Ruehli_2015aa}, equivalent circuits are derived from integral equations, allowing for a combined \gls*{EM}-circuit solution both in frequency and time domain.
However, \gls*{PEEC} requires empirical approximations in addition to the applied discretisation.
For quasistatic approximations, automated circuit generation based on the boundary element method was presented by Milsom~\cite{Milsom_1999aa}.
The method presented therein yields circuits whose size depends on the electrical dimensions of the problem.
Many methods for application-specific circuit extraction based on device or system responses are also available~\cite{Antonini_2003aa,Russer_1994aa,Choi_2002aa}.
A methodology for the generation of equivalent circuits based on the semi-discrete \textsc{Maxwell}\xspace's field equations was also proposed by Ramachadran et al~\cite{Ramachandran_2008aa}.
Therein, \textsc{Yee}\xspace's discretisation scheme is employed to cast \textsc{Maxwell}\xspace's curl equations as the concatenation of interacting fundamental circuits in which voltages and currents model the sought electric and magnetic fields.
In this manner, circuit stamps are required for both primal and dual edges.
Their interaction is organised by \glspl*{VCVS} and \glspl*{CCCS}, respectively.

In the design of electronic devices, enhancing their functionalities is always of high interest.
The according volume shrinking may give rise to high power densities that can lead to thermal issues.
As an example, the introduction of stacked 3D chips intensifies the heat issue since the heat can be trapped between the stacked layers.
Therefore, \gls*{ET} modelling is of great importance for device engineers.
To handle the \gls*{ET} coupling in a circuit simulation framework, two general approaches are mainly available:
the relaxation method which consists of the iterative coupling of an electric circuit with an external thermal-only field simulator~\cite{Chvala_2014aa,Wunsche_1997aa,Van-Petegem_1994aa}, and the monolithic approach which consists of the direct coupling of the electric circuit with a thermal circuit~\cite{Simpson_2014ab}.
The latter allows to run the simulation directly on the full \gls*{ET} circuit without any software package and thereby avoids the weak coupling between solvers.
To extract \gls*{ET} circuits from a given 3D problem, various methods were proposed.
Some of them were based on existing \gls*{EM} simulation methods and have been extended in functionality to also cover the \gls*{ET} case, as done by Lombardi et al~\cite{Lombardi_2017aa} for the \gls*{PEEC}.
Generating compact models from the calculated or measured response function is another popular approach and has been followed by Evans et al and Bernardoni et al~\cite{Evans_2013aa,Bernardoni_2018aa}.
For thermal problems, methods that derive an equivalent circuit directly from the mesh can also be found~\cite{Culpo_2010aa,Wilkerson_2004aa}, but none of them accounts for the \gls*{ET} coupling.
Karagol and Bikdash~\cite{Karagol_2010aa} presented an approximate representation obtained from a graph-partitioning algorithm of an \gls*{FEM} mesh resulting in a medium-sized \gls*{ET} circuit.
A lumped-element representation of every \gls*{FEM} element has also been proposed for \gls*{ET} simulations~\cite{Hsu_1996aa}.
For an exact representation of a semi-discretised 3D \gls*{ET} field problem, Casper et al~\cite{Casper_2016ab} developed an automatic netlist generation method based on the \gls*{FIT}.

\begin{figure}[t]
	\centering
    \subfloat[\label{fig:lumpedElementsOnCube}]{\includegraphics{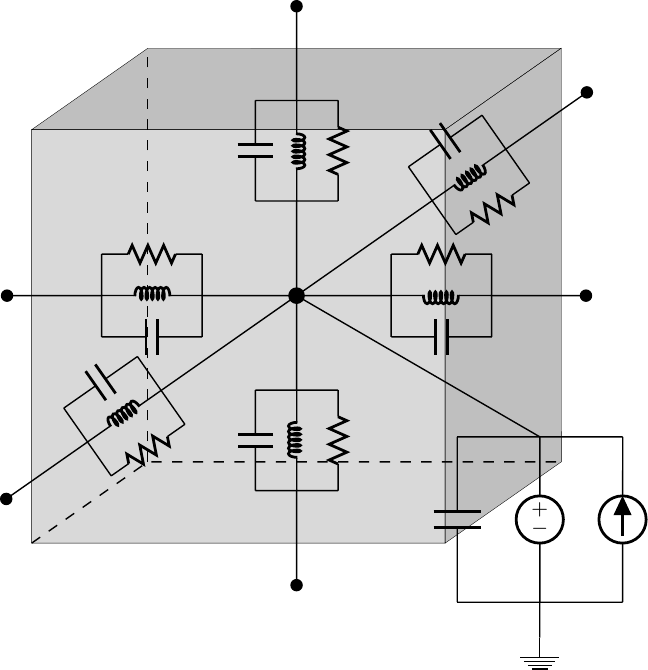}}
    \hspace{2em}
    \subfloat[\label{fig:methodology}]{\includegraphics{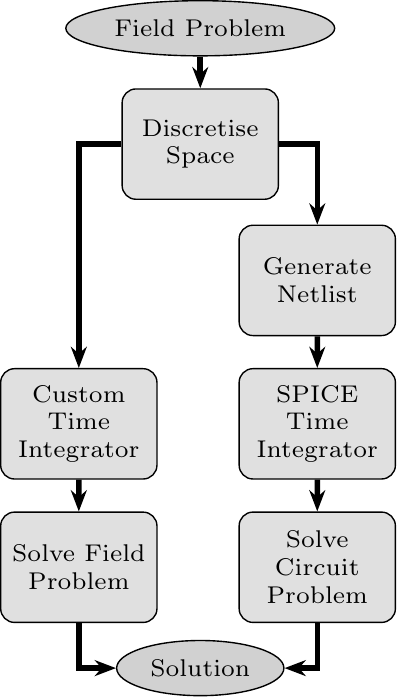}}
    \caption{(a) One grid cell is represented by a circuit node. Connections to neighbouring cells are carried out via lumped elements. (b) The left branch of the diagram shows the standard solution approach by using a field solver. The right branch illustrates the approach described herein to generate a netlist that is then fed to a circuit simulator.}
\end{figure}

In this paper, we present a method to automatically generate netlists representing general 3D \gls*{ET} and \gls*{EM} coupled field problems.
In our approach, neighbouring cells in the primal grid interact via the parallel connection of circuit elements as illustrated in Figure~\ref{fig:lumpedElementsOnCube}.
Thus, the size of the resulting circuit depends on the geometrical size of the problem and the fineness of the discretisation grid.
This allows to use any available circuit simulator.
Hence, the need for a dedicated field solver with a custom time integrator is alleviated.
To accomplish this, we employ the \gls*{FIT} for discretising the relevant continuous field equations. In contrast to \textsc{Yee}\xspace's finite difference scheme employed by Ramachandran et al~\cite{Ramachandran_2008aa}, \gls*{FIT} is a structure-preserving discretisation strategy which does not require further approximations in dealing with the field and material quantities.
Moreover, the concept of integral quantities used in the \gls*{FIT} translates naturally into the framework of circuit descriptions.
In this manner, we obtain an exact grid representation of the field equations that we map transparently into circuit stamps.
In these stamps, which are associated only with primal edges, lumped elements are directly taken from the entries of the material matrices.
These entries are properly integrated constitutive parameters.
In Figure~\ref{fig:methodology}, we summarise our approach.
In addition to previous work on \gls*{ET} problems~\cite{Casper_2016ab}, we also present a more elaborated description and implementation of boundary conditions and excitations.
We also point out that the methodology presented herein allows for straightforward field-circuit coupling and is especially useful for an accurate representation of small devices in a larger circuit.
In order to simulate wave propagation problems on a finite computational domain by means of circuit simulation, we present \glspl*{ABC}~\cite{Engquist_1977aa,Mur_1981ab}.

The outline of this paper is as follows.
In Section~\ref{sec:FIT}, we provide the required basics of the \gls*{FIT}.
The fundamentals of the \gls*{MNA} are summarised in Section~\ref{sec:MNA}.
Then, the main part of the paper starts with the circuit representation of \gls*{ET} field problems in Section~\ref{sec:circuitsET}.
How to extract circuit stamps for \gls*{EM} field problems is presented in Section~\ref{sec:circuitsEM}.
Finally, we show numerical examples in Section~\ref{sec:numerics} and conclude the paper in Section~\ref{sec:conclusion}.

\section{Continuous Thermal and Electromagnetic Formulations}
\label{subsec:EMproblem}

Let us consider a domain $\mathcal{D}$ with boundary $\partial\mathcal{D}$ and characterised by the constitutive parameters $\lbrace\varepsilon,\nu,\sigma\rbrace$, where $\varepsilon$ is the electric permittivity, $\nu$ is the magnetic reluctivity and $\sigma$ is the electric conductivity.
For every facet $\ensuremath{A}$, every volume $\ensuremath{V}$ and with impressed electric sources $\ensuremath{\mathbf{J}_{\mathrm{i}}}$, the \gls*{EM} field $\lbrace\ensuremath{\mathbf{E}},\ensuremath{\mathbf{H}}\rbrace$ in $\mathcal{D}$ is given by \textsc{Maxwell}\xspace's equations
\begin{subequations}
    \begin{align}
        -\oint\limits_{\partial\ensuremath{A}}\ensuremath{\mathbf{E}}\cdot\mathrm{d}\ensuremath{\mathbf{\ensuremath{L}}}&=
        \frac{\mathrm{d}}{\mathrm{d}t}\int_{\ensuremath{A}}\ensuremath{\mathbf{B}} \cdot\mathrm{d}\ensuremath{\mathbf{S}},\\
                \oint\limits_{\partial\ensuremath{A}}\ensuremath{\mathbf{H}}\cdot\mathrm{d}\ensuremath{\mathbf{\ensuremath{L}}}&=
        \frac{\mathrm{d}}{\mathrm{d}t}\int_{\ensuremath{A}}\ensuremath{\mathbf{D}}+\ensuremath{\mathbf{J}_{\mathrm{c}}}+\ensuremath{\mathbf{J}_{\mathrm{i}}}\cdot\mathrm{d}\ensuremath{\mathbf{S}},\\ 
                \int_{\partial\ensuremath{V}}\ensuremath{\mathbf{D}} \cdot\mathrm{d}\ensuremath{\mathbf{S}}&=\int_{\ensuremath{V}}\varrho\,\mathrm{d} V,\\
                \int_{\partial\ensuremath{V}}\ensuremath{\mathbf{B}} \cdot\mathrm{d}\ensuremath{\mathbf{S}}&=0.
    \end{align} 
    \label{eq:MaxwellContInt}\end{subequations}
We call henceforth~\eqref{eq:MaxwellContInt} the E-H formulation for conciseness.
Above, $\ensuremath{\mathbf{D}}$ and $\ensuremath{\mathbf{B}}$ are the electric and magnetic flux density, respectively, $\ensuremath{\mathbf{J}_{\mathrm{c}}}$ is the electric conduction current and $\varrho$ is the electric charge density.
To guarantee the uniqueness of the solution, \textsc{Maxwell}\xspace's equations are supplemented with the constitutive relations, viz.
\begin{alignat*}{4}
      \ensuremath{\mathbf{D}}  &= \varepsilon\ensuremath{\mathbf{E}},\qquad
    & \ensuremath{\mathbf{H}}  &= \nu\ensuremath{\mathbf{B}},\qquad
    & \ensuremath{\mathbf{J}_{\mathrm{c}}} &= \sigma\ensuremath{\mathbf{E}},
\end{alignat*}
with suitable initial and \glspl*{BC} on $\partial\mathcal{D}$. 

We also deem it convenient to obtain \textsc{Maxwell}\xspace's equations involving the auxiliary magnetic vector potential $\ensuremath{\mathbf{A}}$.
To this end, we recall that $\ensuremath{\mathbf{E}}=-\nabla\varphi-\partial\ensuremath{\mathbf{A}}/\partial t$ and $\nabla \cdot(\sigma_{\text{g}}\ensuremath{\mathbf{A}})=f$, with an auxiliary scalar potential $\varphi$, a gauging material parameter $\sigma_{\text{g}}$ and an arbitrary \emph{scalar} gauging function $f$.
Substitution of these definitions in \eqref{eq:MaxwellContInt} and applying \textsc{Stokes}\xspace' theorem yields
\begin{subequations}
    \begin{align}
        -\int_{\ensuremath{A}}\nabla \times\ensuremath{\mathbf{E}}\cdot\mathrm{d}\ensuremath{\mathbf{S}}&=\frac{\mathrm{d}}{\mathrm{d}t}\int_{\ensuremath{A}}\nabla \times\ensuremath{\mathbf{A}}\cdot\mathrm{d}\ensuremath{\mathbf{S}},\\
                \int_{\ensuremath{A}}\nabla \times\left(\nu\nabla \times\ensuremath{\mathbf{A}}\right)\cdot\mathrm{d}\ensuremath{\mathbf{S}}&=
        \frac{\mathrm{d}}{\mathrm{d}t}\int_{\ensuremath{A}}\ensuremath{\mathbf{D}}+\ensuremath{\mathbf{J}_{\mathrm{c}}}+\ensuremath{\mathbf{J}_{\mathrm{i}}}\cdot\mathrm{d}\ensuremath{\mathbf{S}},\\
                \int_{\partial\ensuremath{V}}\ensuremath{\mathbf{D}}\cdot\mathrm{d}\ensuremath{\mathbf{S}}&=\int_{\ensuremath{V}}\varrho\,\mathrm{d} V,\\
        \int_{\partial\ensuremath{V}}\ensuremath{\mathbf{B}}\cdot\mathrm{d}\ensuremath{\mathbf{S}}&=0,
    \end{align}
    \label{eq:FaradayAmpere}\end{subequations}
which we refer to as the E-A formulation.

Whenever conducting materials are involved, electric currents result in \textsc{Joule}\xspace losses $Q_{\text{J}}=\sigma\left(\nabla\varphi\right)^{2}$ that enter as source term into the heat equation which is given by
\begin{equation}
    \int_{\ensuremath{V}}\left(\rho c\dot{T}-Q_{\text{J}}\right)\,\mathrm{d} V
    =\int_{\partial\ensuremath{V}}\left(\lambda\nabla T-\ensuremath{\mathbf{q}_{\mathrm{i}}}\right)\cdot\mathrm{d}\ensuremath{\mathbf{S}},
    \label{eq:thermalContInt}
\end{equation}
where $\rho c$ is the volumetric heat capacity, $T$ is the temperature, $\lambda$ is the thermal conductivity and $\ensuremath{\mathbf{q}_{\mathrm{i}}}$ represent any impressed thermal flux density. 
In general, when thermal effects are considered, all constitutive parameters are also a function of the temperature.
To perform circuit extraction, we shall consider the above \gls*{ET} and \gls*{EM} formulations separately as described in Sections~\ref{sec:circuitsET} and \ref{sec:circuitsEM}, respectively.

\section{Discretising the Thermal and Electromagnetic Formulations}
\label{sec:FIT}

\begin{figure}
	\centering
	\includegraphics{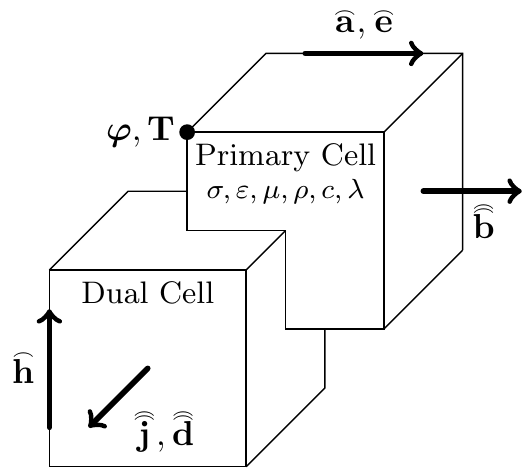}
	\caption{Staggered pair of primal and dual cell for a regular hexahedral grid with the allocation of electric, magnetic and thermal quantities.}
	\label{fig:gridStaggering}
\end{figure}

We discretise the domain $\mathcal{D}$ into a pair of orthogonal grids given by the primal grid $G$ and its dual $\protect\widetilde{G}$.
The grid $G$ consists of primal points $\ensuremath{P}_{i},\,i=1,\dots,N_{\text{P}}$, primal edges (lines) $\ensuremath{L}_{n},\,n=1,\ldots,N_{\text{E}}$, primal facets (areas) $\ensuremath{A}_{n},\,n=1,\ldots,N_{\text{F}}$, and primal volumes $\ensuremath{V}_{i},\,i=1,\ldots,N_{\text{V}}$.
Similarly, the grid $\protect\widetilde{G}$ consists of dual points $\ensuremath{\protect\widetilde{P}}_{i},\,i=1,\dots,\protect\widetilde{N}_{\text{P}}$, dual edges $\ensuremath{\protect\widetilde{L}}_{n},\,n=1,\ldots,\protect\widetilde{N}_{\text{E}}$, dual facets $\ensuremath{\protect\widetilde{A}}_{n},\,n=1,\ldots,\protect\widetilde{N}_{\text{F}}$, and dual volumes $\ensuremath{\protect\widetilde{V}}_{i},\,i=1,\ldots,\protect\widetilde{N}_{\text{V}}$.
The grids $G$ and $\protect\widetilde{G}$ are dual to each other in the sense that a primal edge $\ensuremath{L}_{n}$ intersects a dual facet $\ensuremath{\protect\widetilde{A}}_{n}$ and a primal point $\ensuremath{P}_{i}$ is located inside a dual volume $\ensuremath{\protect\widetilde{V}}_{i}$ and vice versa.
For a regular hexahedral grid, the grid staggering is depicted in Figure~\ref{fig:gridStaggering}.
Due to this duality, the number of primal and dual grid objects fulfils
\begin{equation*}
	N_{\text{E}}=\protect\widetilde{N}_{\text{F}},\quad
	N_{\text{F}}=\protect\widetilde{N}_{\text{E}},\quad
	N_{\text{V}}=\protect\widetilde{N}_{\text{P}},\quad
	N_{\text{P}}=\protect\widetilde{N}_{\text{V}}.
\end{equation*}
As mentioned above, we write $\ensuremath{L}_{n}$ for the $n$-th edge of the primal grid and we emphasise the duality of edges and facets by using the same index.
Thus, $\ensuremath{\protect\widetilde{A}}_{n}$ is the dual facet corresponding to the primal edge $\ensuremath{L}_{n}$ and $\ensuremath{\protect\widetilde{L}}_{n}$ is the dual edge corresponding to the primal facet $\ensuremath{A}_{n}$. 
Let us further introduce a short (index) notation for geometric objects.
If $\ensuremath{L}_{n}$ or $\ensuremath{A}_{n}$ are used as an index, we simply write $n$ instead.
Whether $n$ refers to an edge or a facet should become clear from the context.
For the dual objects, we use $\protect\tilde{n}$ instead of $\ensuremath{\protect\widetilde{L}}_{n}$ or $\ensuremath{\protect\widetilde{A}}_{n}$.
The notation for points and volumes and their duals is done accordingly.
In Table~\ref{tab:gridNotation}, we summarise this notation.

\begin{table}
    \centering
    \begin{tabular}{ccc}\toprule
    Description           & Normal            & Index\\\bottomrule\toprule
    $i$-th primal point  & $\ensuremath{P}_{i}$   & $i$\\
    $i$-th dual point     & $\ensuremath{\protect\widetilde{P}}_{i}$  & $\protect\tilde{i}$\\
    $n$-th primal edge   & $\ensuremath{L}_{n}$    & $n$\\
    $n$-th dual edge      & $\ensuremath{\protect\widetilde{L}}_{n}$   & $\protect\tilde{n}$\\
    $n$-th primal facet  & $\ensuremath{A}_{n}$   & $n$\\
    $n$-th dual facet     & $\ensuremath{\protect\widetilde{A}}_{n}$  & $\protect\tilde{n}$\\
    $i$-th primal volume & $\ensuremath{V}_{i}$  & $i$\\
    $i$-th dual volume    & $\ensuremath{\protect\widetilde{V}}_{i}$ & $\protect\tilde{i}$\\\bottomrule
\end{tabular}
     \caption{Normal and index notation format for different entities of the grid.}
    \label{tab:gridNotation}
\end{table}

The grid counterparts of the field quantities are allocated to points, edges, facets or volumes and collected in column vectors.
Typical examples from \gls*{EMs} are the discrete electric potentials \ensuremath{\boldsymbol{\mathrm{\varphi}}}, fields \ensuremath{\protect\bow{\mathrm{\mathbf{e}}}}\xspace, currents \ensuremath{\protect\bbow{\mathrm{\mathbf{j}}}}\xspace and charges $\bbbow{\ensuremath{\mathrm{\mathbf{q}}}\xspace}$ that are allocated to primal points, edges, facets and volumes, respectively.
The number of bows indicates the dimension of the corresponding geometric object.
Nevertheless, we typically write \ensuremath{\mathrm{\mathbf{q}}}\xspace instead of $\bbbow{\ensuremath{\mathrm{\mathbf{q}}}\xspace}$ for conciseness. 
Defining all other grid quantities accordingly, their allocation used in this paper is shown in Figure~\ref{fig:gridStaggering}.
Furthermore, if we want to indicate a grid quantity, e.g. an electric field, to be allocated to an edge $\ensuremath{L}_{n}$ that is part of the boundary of a facet $\ensuremath{A}_{k}$ ($\ensuremath{L}_{n}\cap\partial\ensuremath{A}_{k}=\ensuremath{L}_{n}$), we use the indexed notation $\ensuremath{\protect\bow{e}}_{k;n}$.
To relate grid quantities on either the primal or dual mesh, topological matrices, corresponding to the continuous topological operators, have to be defined.
The incidence between grid points and edges is given by the discrete gradient matrix $\ensuremath{\mathbf{G}}\xspace$.
For an oriented edge $\ensuremath{L}_{n}$, $G_{ni}=-1$ if $\ensuremath{P}_{i}$ is the starting point of the edge, $G_{ni}=1$ if $\ensuremath{P}_{i}$ is the ending point of the edge and $G_{ni}=0$ if $\ensuremath{P}_{i}$ is neither starting nor ending point of the edge ($\ensuremath{P}_{i}\cap\partial\ensuremath{L}_{n}=\emptyset$).
To express the incidence between grid edges and facets, we use the discrete curl matrix $\ensuremath{\mathbf{C}}\xspace$.
Given a primal facet $\ensuremath{A}_{k}$ and its oriented boundary $\partial\ensuremath{A}_{k}$, $\ensuremath{C}_{kn}=-1$ if edge $\ensuremath{L}_{n}$ is oriented in opposite direction than $\partial\ensuremath{A}_{k}$, $\ensuremath{C}_{kn}=1$ if edge $\ensuremath{L}_{n}$ is oriented in the same way as $\partial\ensuremath{A}_{k}$ and $\ensuremath{C}_{kn}=0$ if $\ensuremath{L}_{n}$ does not touch $\ensuremath{A}_{k}$ ($\ensuremath{L}_{n}\cap\partial\ensuremath{A}_{k}=\emptyset$).
Finally, the discrete divergence matrix $\ensuremath{\mathbf{S}}\xspace$ denotes the incidence between grid facets and volumes.
For a primal volume $\ensuremath{V}_{i}$ and its oriented boundary $\partial\ensuremath{V}_{k}$, the entries of $\ensuremath{\mathbf{S}}\xspace$ are defined in analogy to those of $\ensuremath{\mathbf{G}}\xspace$ and $\ensuremath{\mathbf{C}}\xspace$.
Additionally, we have the dual gradient, curl and divergence matrices \ensuremath{\protect\widetilde{\mathbf{G}}}\xspace, \ensuremath{\protect\widetilde{\mathbf{C}}}\xspace and \ensuremath{\protect\widetilde{\mathbf{S}}}\xspace, respectively, defined accordingly.
Useful relations between the topological matrices on the primal and dual grids are given by $\ensuremath{\mathbf{G}}\xspace=-\ensuremath{\protect\widetilde{\mathbf{S}}^\top}$, $\ensuremath{\protect\widetilde{\mathbf{G}}}\xspace=-\ensuremath{\mathbf{S}}\xspace^{\top}$ and $\ensuremath{\mathbf{C}}\xspace=\ensuremath{\protect\widetilde{\mathbf{C}}}\xspace^{\top}$~\cite{Clemens_2001aa}.

To relate quantities on the primal grid to quantities on the dual grid and vice versa, material relations are employed.
For the problem formulated in Section~\ref{subsec:EMproblem}, the following three different kind of relations can be identified:
\begin{itemize}
    \item Quantities allocated to primal facets must be related to quantities allocated to dual edges.
    \item Quantities allocated to primal edges must be related to quantities allocated to dual facets.
    \item Quantities allocated to primal points must be related to quantities allocated to dual volumes.
\end{itemize}
As representatives for the above listed constitutive relations, we formulate three discrete material laws as
\begin{equation*}
    \ensuremath{\protect\bow{\mathrm{\mathbf{h}}}}\xspace=\ensuremath{\mathbf{M}_{\nu}}\xspace\ensuremath{\protect\bbow{\mathrm{\mathbf{b}}}}\xspace,\quad\ensuremath{\protect\bbow{\mathrm{\mathbf{d}}}}\xspace=\ensuremath{\mathbf{M}_{\varepsilon}}\xspace\ensuremath{\protect\bow{\mathrm{\mathbf{e}}}}\xspace,\quad \ensuremath{ \mathbf{Q} }=\ensuremath{\mathbf{M}_{\rho c}}\xspace\dot{\ensuremath{\mathbf{T}}},\\
\end{equation*}
where $\ensuremath{\mathbf{M}_{\nu}}\xspace$ is the magnetic reluctance matrix mapping the discrete magnetic flux $\ensuremath{\protect\bbow{\mathrm{\mathbf{b}}}}\xspace$ allocated to primal facets to the discrete magnetic field $\ensuremath{\protect\bow{\mathrm{\mathbf{h}}}}\xspace$ allocated to dual edges, $\ensuremath{\mathbf{M}_{\varepsilon}}\xspace$ is the electric capacitance matrix mapping the discrete electric field $\ensuremath{\protect\bow{\mathrm{\mathbf{e}}}}\xspace$ allocated to primal edges to the discrete electric flux density $\ensuremath{\protect\bbow{\mathrm{\mathbf{d}}}}\xspace$ allocated to dual facets, and
$\ensuremath{\mathbf{M}_{\rho c}}\xspace$ is the thermal capacitance matrix mapping the time derivative of the grid temperature $\dot{\ensuremath{\mathbf{T}}}$ allocated to primal points to the discrete heat power $\ensuremath{ \mathbf{Q} }$ allocated to dual volumes.
For other constitutive parameters, the material matrices are defined following these three cases.

Having established the grid constructs $G$ and $\protect\widetilde{G}$ and the corresponding topological and material matrices, the E-H formulation~\eqref{eq:MaxwellContInt} of \textsc{Maxwell}\xspace's equations upon such a grid pair can be written as~\cite{Weiland_1996aa}
\begin{subequations}
    \begin{align}
        -\ensuremath{\mathbf{C}}\xspace\ensuremath{\protect\bow{\mathrm{\mathbf{e}}}}\xspace    &=\ensuremath{\mathbf{M}_{\nu}}\xspace^{-1}\frac{\mathrm{d}\ensuremath{\protect\bow{\mathrm{\mathbf{h}}}}\xspace}{\mathrm{d}t},\label{eq:MGEmagChargeFaraday}\\
        \ensuremath{\protect\widetilde{\mathbf{C}}}\xspace\ensuremath{\protect\bow{\mathrm{\mathbf{h}}}}\xspace    &=\ensuremath{\mathbf{M}_{\varepsilon}}\xspace\frac{\mathrm{d}\ensuremath{\protect\bow{\mathrm{\mathbf{e}}}}\xspace}{\mathrm{d}t}+\ensuremath{\mathbf{M}_{\sigma}}\xspace\ensuremath{\protect\bow{\mathrm{\mathbf{e}}}}\xspace+\ensuremath{\protect\bbow{\mathrm{\mathbf{j}}}}\xspace_{\text{i}},\label{eq:MGEmagChargeAmpere}\\ 
        \ensuremath{\protect\widetilde{\mathbf{S}}}\xspace \ensuremath{\protect\bbow{\mathrm{\mathbf{d}}}}\xspace    &=\ensuremath{\mathrm{\mathbf{q}}}\xspace,\\
        \qquad\ensuremath{\mathbf{S}}\xspace\ensuremath{\protect\bbow{\mathrm{\mathbf{b}}}}\xspace&=\ensuremath{\mathbf{0}},
    \end{align} 
    \label{eq:MGEmagCharge}\end{subequations}
where \ensuremath{\mathbf{M}_{\sigma}}\xspace is the electric conductance matrix, \ensuremath{\protect\bbow{\mathrm{\mathbf{b}}}}\xspace is the discrete magnetic flux density, $\ensuremath{\protect\bbow{\mathrm{\mathbf{j}}}}\xspace_{\text{i}}$ is the discrete impressed electric current density and \ensuremath{\mathrm{\mathbf{q}}}\xspace is the discrete electric charge.
Similarly, with the discrete vector potential \ensuremath{\protect\bow{\mathrm{\mathbf{a}}}}\xspace, the discrete E-A formulation is given by
\begin{subequations}
    \begin{align}
        \ensuremath{\mathbf{C}}\xspace\ensuremath{\protect\bow{\mathrm{\mathbf{e}}}}\xspace&=-\ensuremath{\mathbf{C}}\xspace\frac{\mathrm{d}\ensuremath{\protect\bow{\mathrm{\mathbf{a}}}}\xspace}{\mathrm{d}t},\label{eq:MGEfaraday}\\
        \ensuremath{\protect\widetilde{\mathbf{C}}}\xspace\ensuremath{\mathbf{M}_{\nu}}\xspace\ensuremath{\mathbf{C}}\xspace\ensuremath{\protect\bow{\mathrm{\mathbf{a}}}}\xspace&=\ensuremath{\mathbf{M}_{\varepsilon}}\xspace\frac{\mathrm{d}\ensuremath{\protect\bow{\mathrm{\mathbf{e}}}}\xspace}{\mathrm{d}t}+\ensuremath{\mathbf{M}_{\sigma}}\xspace\ensuremath{\protect\bow{\mathrm{\mathbf{e}}}}\xspace+\ensuremath{\protect\bbow{\mathrm{\mathbf{j}}}}\xspace_{\text{i}},\label{eq:MGEampere}
    \end{align}
    \label{eq:MGE}\end{subequations}
together with the discrete form of the gauging
\begin{equation}
    \ensuremath{\protect\widetilde{\mathbf{S}}}\xspace\mathbf{M}_{\text{G}}\ensuremath{\protect\bow{\mathrm{\mathbf{a}}}}\xspace=\ensuremath{ \mathbf{F} },
    \label{eq:gaugeAfit}
\end{equation}
where $\ensuremath{ \mathbf{F} }$ is the discrete counterpart of the gauging function $f$ and $\mathbf{M}_{\text{G}}$ is a gauging matrix.
Since $\mathbf{M}_{\text{G}}$ maps from quantities on primal edges to dual facets, it shares properties with the material matrices (e.g. $\ensuremath{\mathbf{M}_{\varepsilon}}\xspace$) and thus can be interpreted as a material matrix with a material value equal to $\sigma_{\text{g}}$.

Casting also the heat equation \eqref{eq:thermalContInt} into a spatially discrete form, we obtain
\begin{equation*}
    \ensuremath{\mathbf{M}_{\rho c}}\xspace\dot{\ensuremath{\mathbf{T}}}+\ensuremath{\protect\widetilde{\mathbf{S}}}\xspace\ensuremath{\mathbf{M}_{\lambda}}\xspace\ensuremath{\protect\widetilde{\mathbf{S}}}\xspace^{\top}\ensuremath{\mathbf{T}}=\mathbf{Q}_{\text{J}}-\ensuremath{\protect\widetilde{\mathbf{S}}}\xspace\ensuremath{\protect\bbow{\mathrm{\mathbf{q}}}}\xspace_{\text{i}},
\end{equation*}
where $\ensuremath{\mathbf{M}_{\lambda}}\xspace$ is the thermal conductance matrix and $\mathbf{Q}_{\text{J}}$ is the discrete vector of the \textsc{Joule}\xspace losses.
For details on the computation of $\mathbf{Q}_{\text{J}}$, we refer the reader to the work by Casper et al~\cite{Casper_2016ab}.
The impressed thermal fluxes are given by their discrete representative $\ensuremath{\protect\bbow{\mathrm{\mathbf{q}}}}\xspace_{\text{i}}$.

\subsection{Finite Integration Technique and its Relation to Other Discretisation Schemes}

For the circuit extraction from 3D field models as presented in this paper, we require diagonal, symmetric and positive definite material matrices.
To fulfil this requirement, we choose to use the \gls*{FIT} and assume the material to coincide with the primal grid cells, see Figure~\ref{fig:gridStaggering}.
While the \gls*{FIT} has also been formulated for anisotropic materials~\cite{Kruger_2001aa}, diagonal material matrices are obtained only in the isotropic case or in the case when the principal axes of anisotropy coincide with the coordinate axes.
Then, the entries of the different material matrices are given by
\begin{alignat*}{2}
    M_{\varepsilon;\protect\tilde{n} n}=\varepsilon_{\protect\tilde{n} n}\frac{|\ensuremath{\protect\widetilde{A}}_{n}|}{|\ensuremath{L}_{n}|},\quad
    M_{\lambda;\protect\tilde{n} n}&=\lambda_{\protect\tilde{n} n}\frac{|\ensuremath{\protect\widetilde{A}}_{n}|}{|\ensuremath{L}_{n}|},\quad
    M_{\sigma;\protect\tilde{n} n}&&=\sigma_{\protect\tilde{n} n}\frac{|\ensuremath{\protect\widetilde{A}}_{n}|}{|\ensuremath{L}_{n}|},\\
    M_{\nu;\protect\tilde{n} n}&=\nu_{\protect\tilde{n} n}\frac{|\ensuremath{\protect\widetilde{L}}_{n}|}{|\ensuremath{A}_{n}|},\quad
    M_{\rho c;\protect\tilde{i} i}&&=\rho c_{\protect\tilde{i} i}|\ensuremath{\protect\widetilde{V}}_{i}|,
\end{alignat*}
where $\varepsilon_{\protect\tilde{n} n}$, $\sigma_{\protect\tilde{n} n}$, $\lambda_{\protect\tilde{n} n}$, $\nu_{\protect\tilde{n} n}$, and $\rho c_{\protect\tilde{i} i}$ are obtained by a suitable averaging scheme~\cite{Clemens_2001aa,Casper_2016aa} and $|\cdot|$ represents the measure (i.e. area, length or volume) of the corresponding geometrical object.

When using the \gls*{FIT} as the discretisation scheme with a canonical numbering of the grid nodes, the curl and material matrices exhibit the block structure
\begin{align}
    \ensuremath{\mathbf{C}}\xspace=
    \begin{pmatrix}
        \ensuremath{\mathbf{0}}              & -\ensuremath{\mathbf{P}}\xspace_{z}  & \ensuremath{\mathbf{P}}\xspace_{y} \\
        \ensuremath{\mathbf{P}}\xspace_{z}   & \ensuremath{\mathbf{0}}              & -\ensuremath{\mathbf{P}}\xspace_{x}\\
        -\ensuremath{\mathbf{P}}\xspace_{y}  & \ensuremath{\mathbf{P}}\xspace_{x}   & \ensuremath{\mathbf{0}} 
    \end{pmatrix},\quad
    \ensuremath{\mathbf{M}_{\varepsilon}}\xspace&=
    \begin{pmatrix}
        \ensuremath{\mathbf{M}}_{\varepsilon;x} & \ensuremath{\mathbf{0}}                 & \ensuremath{\mathbf{0}}\\   
        \ensuremath{\mathbf{0}}                 & \ensuremath{\mathbf{M}}_{\varepsilon;y} & \ensuremath{\mathbf{0}}\\
        \ensuremath{\mathbf{0}}                 & \ensuremath{\mathbf{0}}                 & \ensuremath{\mathbf{M}}_{\varepsilon;z}  
    \end{pmatrix},
    \label{eq:fitMatBlockStruct}
\end{align}
where $\ensuremath{\mathbf{P}}\xspace_{\xi}\in\{0,\pm 1\}^{N_{\text{E}}\times N_{\text{P}}}$, $\xi\in\{x,y,z\}$ are the grid differential operators for the different coordinate directions.
Here, \ensuremath{\mathbf{M}_{\varepsilon}}\xspace was used as an example whereas an equivalent block structure applies for all other material matrices.
Within the theory of the \gls*{FIT}, the discrete quantities \ensuremath{\boldsymbol{\mathrm{\varphi}}}, \ensuremath{\protect\bow{\mathrm{\mathbf{e}}}}\xspace, \ensuremath{\protect\bbow{\mathrm{\mathbf{j}}}}\xspace, \ensuremath{\mathrm{\mathbf{q}}}\xspace that were introduced in Section~\ref{sec:FIT} are defined by means of integration with respect to their corresponding geometrical object, such that
\begin{align*}
    \left(\ensuremath{\boldsymbol{\mathrm{\varphi}}}\right)_{i}:=\varphi(\ensuremath{P}_{i}),\quad
    \ensuremath{\protect\bow{e}}_{n}:=\int_{\ensuremath{L}_{n}}\ensuremath{\mathbf{E}}\cdot\mathrm{d}\ensuremath{\mathbf{\ensuremath{L}}},\quad
    \ensuremath{\protect\bbow{j}}_{\protect\tilde{n}}:=\int_{\ensuremath{\protect\widetilde{A}}_{n}}\ensuremath{\mathbf{J}}\cdot\mathrm{d}\ensuremath{\mathbf{S}},\quad
    q_{i}:=\int_{\ensuremath{V}_i}\varrho\ \mathrm{d} V,
\end{align*}
where an analogous definition applies for \ensuremath{\protect\bow{\mathrm{\mathbf{a}}}}\xspace, \ensuremath{\protect\bow{\mathrm{\mathbf{h}}}}\xspace, \ensuremath{\protect\bbow{\mathrm{\mathbf{d}}}}\xspace and \ensuremath{\protect\bbow{\mathrm{\mathbf{b}}}}\xspace.
Due to the applied integration, one speaks of grid voltages instead of discrete fields, of grid currents (fluxes) instead of discrete current (flux) densities and of grid charges instead of discrete charge densities.

As an alternative to the \gls*{FIT}, equivalent approaches such as the cell method~\cite{Alotto_2006aa} or the \gls*{FEM} can be used as long as the orthogonality and one-to-one relation between primal and dual grid objects is guaranteed.
When using \gls*{FEM}, linear basis functions together with an appropriate mass lumping for the material matrices must be used to obtain an equivalent scheme~\cite{Bondeson_2005aa}.

\section{A Primer on Circuit Theory and the Modified Nodal Analysis}
\label{sec:MNA}

In this section, we briefly review some fundamentals about circuit theory and the \gls*{MNA}~\cite{Ho_1975aa,Gunther_2005aa}.
\textsc{Kirchhoff}\xspace's current and voltage laws are derived and form the basics for circuit analysis.

Any circuit can be understood as a directed graph consisting of interconnected nodes and branches.
Let $v_{i}$ be one of the $N_{\text{n}}$ nodal potentials in a circuit and $b_{n}$ one of $N_{\text{b}}$ directed branches.
With the incidence matrix $\ensuremath{ \mathbf{A} }\in\{-1,0,1\}^{N_{\text{n}}\times N_{\text{b}}}$ linking nodes and branches, the voltage-potential relation is given by
\begin{equation*}
    \ensuremath{ \mathbf{V} } = \ensuremath{ \mathbf{A} }^{\top}\mathbf{v}.
\end{equation*}
The entries of $\ensuremath{ \mathbf{A} }$ are defined such that $a_{in}=+1$ if the branch $b_{n}$ is directed away from node $n_{i}$ and $a_{in}=-1$ if $b_{n}$ is directed towards $n_{i}$.
If $n_{i}$ is neither starting nor ending point of $b_{n}$, then $a_{in}=0$.
With this definition, the exemplary voltage $V_{n}$ on the branch $b_{n}$ directed from $n_{i}$ to $n_{j}$ is given by $V_{n}=v_{i}-v_{j}$.

\begin{figure}[t]
    \centering
    \subfloat[\label{fig:KCLonenode}]{\raisebox{3ex}{\includegraphics[width=0.23\columnwidth]{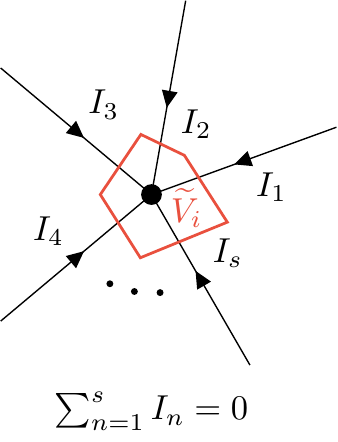}}}
    \subfloat[\label{fig:KCLmanynodes}]{\raisebox{3ex}{\includegraphics[width=0.23\columnwidth]{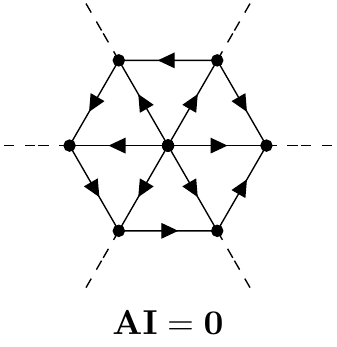}}}
    \subfloat[\label{fig:circuitExample}]{\includegraphics[width=0.53\columnwidth]{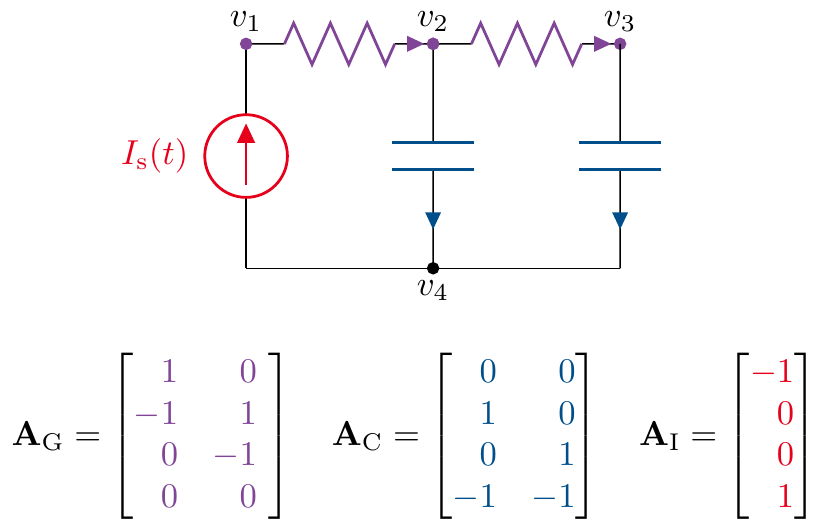}}
    \caption{KCL example for (a) one node and (b) several nodes. The net current flow into a node must be zero. (c) Circuit example and its branch matrices.}
\end{figure}

For time invariant geometries, the current continuity equation reads
\begin{equation}
    \int_{\partial V}\ensuremath{\mathbf{J}}\cdot\ \mathrm{d}\mathbf{A}+\int_V\dot{\varrho}\ \mathrm{d} V=0,
    \label{eq:continuityEquation}
\end{equation}
for an arbitrary volume $V$.
By considering a volume $\ensuremath{\protect\widetilde{V}}_{i}$ around an arbitrary circuit node $n_{i}$ and assuming that capacitive charges are located either fully inside or outside of $\ensuremath{\protect\widetilde{V}}_{i}$, the total charge and also the charge's change rate in $\ensuremath{\protect\widetilde{V}}_{i}$ is zero.
Therefore, \eqref{eq:continuityEquation} becomes
\begin{equation*}
    \int_{\partial\ensuremath{\protect\widetilde{V}}_{i}}\ensuremath{\mathbf{J}}\cdot\ \mathrm{d}\mathbf{A} = 0.
\end{equation*}
If $\partial\ensuremath{\protect\widetilde{V}}_{i}$ is composed by a finite number $s$ of conductors with cross-sectional areas $\ensuremath{\protect\widetilde{A}}_{n}$, \gls*{KCL} is obtained as
\begin{equation}
    \sum_{n=1}^{s}I_{n}=\sum_{n=1}^{s}\int_{\ensuremath{\protect\widetilde{A}}_{n}}\ensuremath{\mathbf{J}}\cdot\ \mathrm{d}\mathbf{A}=0,
    \label{eq:KCLoneNode}
\end{equation}
where $I_{n}$ is the total current through the facet $\ensuremath{\protect\widetilde{A}}_{n}$.
This relation is also depicted in Figure~\ref{fig:KCLonenode}.
To express~\eqref{eq:KCLoneNode} for all nodes in the circuit (cf. Figure~\ref{fig:KCLmanynodes}), the incidence matrix $\ensuremath{ \mathbf{A} }$ can be used such that
\begin{equation}
    \ensuremath{ \mathbf{A} }\ensuremath{ \mathbf{I} }=\ensuremath{\mathbf{0}},
    \label{eq:KCLallNodes}
\end{equation}
where $\ensuremath{ \mathbf{I} }\in\mathbb{R}^{N_{\text{b}}}$ is a vector of all currents allocated to the branches and $\ensuremath{\mathbf{0}}$ is a vector of zeros of suitable dimension.

In a circuit, the basic branch elements are conductors, capacitors, inductors as well as voltage and current sources.
As these elements are allocated to branches, sets of $N_{\text{G}}$, $N_{\text{C}}$, $N_{\text{L}}$, $N_{\text{V}}$ and $N_{\text{I}}$ branches are defined.
Hence, $\ensuremath{ \mathbf{A} }$ can be arranged into a block matrix with sub-blocks for these elements~\cite{Gunther_2005aa}, viz. 
\begin{equation*}
    \ensuremath{ \mathbf{A} } =
    \begin{bmatrix}
        \ensuremath{\ensuremath{\mathbf{A}}_\text{G}} & \ensuremath{\ensuremath{\mathbf{A}}_\text{C}} & \ensuremath{\ensuremath{\mathbf{A}}_\text{L}} & \ensuremath{\ensuremath{\mathbf{A}}_\text{V}} & \ensuremath{\ensuremath{\mathbf{A}}_\text{I}}
    \end{bmatrix}.
\end{equation*}
A circuit example consisting of a current source, two resistors and two capacitors together with the corresponding block structure of $\ensuremath{\mathbf{A}}$ is shown in Figure~\ref{fig:circuitExample}.
A similar subdivision is done for the current and voltage vectors,
\begin{equation*}
    \ensuremath{\mathbf{I}}^{\top} =
    \begin{bmatrix}
        \ensuremath{\ensuremath{\mathbf{I}}_\text{R}}^{\top} & \ensuremath{\ensuremath{\mathbf{I}}_\text{C}}^{\top} & \ensuremath{\ensuremath{\mathbf{I}}_\text{L}}^{\top} & \ensuremath{\ensuremath{\mathbf{I}}_\text{V}}^{\top} & \ensuremath{\ensuremath{\mathbf{I}}_\text{I}}^{\top}
    \end{bmatrix},\quad
    \ensuremath{\mathbf{V}}^{\top} =
    \begin{bmatrix}
        \ensuremath{\ensuremath{\mathbf{V}}_\text{G}}^{\top} & \ensuremath{\ensuremath{\mathbf{V}}_\text{C}}^{\top} & \ensuremath{\ensuremath{\mathbf{V}}_\text{L}}^{\top} & \ensuremath{\ensuremath{\mathbf{V}}_\text{V}}^{\top} & \ensuremath{\ensuremath{\mathbf{V}}_\text{I}}^{\top}
    \end{bmatrix}.
\end{equation*}
With these definitions, the voltages are given by
\begin{equation}
    \ensuremath{\ensuremath{\mathbf{V}}_\text{G}}=\ensuremath{\ensuremath{\mathbf{A}}_\text{G}}^{\top}\ensuremath{\mathbf{v}},\quad
    \ensuremath{\ensuremath{\mathbf{V}}_\text{C}}=\ensuremath{\ensuremath{\mathbf{A}}_\text{C}}^{\top}\ensuremath{\mathbf{v}},\quad
    \ensuremath{\ensuremath{\mathbf{V}}_\text{L}}=\ensuremath{\ensuremath{\mathbf{A}}_\text{L}}^{\top}\ensuremath{\mathbf{v}},\quad
    \ensuremath{\ensuremath{\mathbf{V}}_\text{V}}=\ensuremath{\ensuremath{\mathbf{A}}_\text{V}}^{\top}\ensuremath{\mathbf{v}},\quad
    \ensuremath{\ensuremath{\mathbf{V}}_\text{I}}=\ensuremath{\ensuremath{\mathbf{A}}_\text{I}}^{\top}\ensuremath{\mathbf{v}},
    \label{eq:currentVoltageBlock}
\end{equation}
and~\eqref{eq:KCLallNodes} becomes
\begin{equation}
    \ensuremath{\ensuremath{\mathbf{A}}_\text{G}}\ensuremath{\ensuremath{\mathbf{I}}_\text{R}} + \ensuremath{\ensuremath{\mathbf{A}}_\text{C}}\ensuremath{\ensuremath{\mathbf{I}}_\text{C}} + \ensuremath{\ensuremath{\mathbf{A}}_\text{L}}\ensuremath{\ensuremath{\mathbf{I}}_\text{L}} + \ensuremath{\ensuremath{\mathbf{A}}_\text{V}}\ensuremath{\ensuremath{\mathbf{I}}_\text{V}} + \ensuremath{\ensuremath{\mathbf{A}}_\text{I}}\ensuremath{\ensuremath{\mathbf{I}}_\text{I}} = \ensuremath{\mathbf{0}}.
    \label{eq:KCLblock}
\end{equation}
The relation between voltages and currents for the different branches is established by constitutive diagonal matrices that contain the element-wise material parameters.
These are the conductance, capacitance, and inductance matrices $\ensuremath{\mathbf{G}}$, $\ensuremath{\mathbf{C}}$ and $\ensuremath{\mathbf{L}}$, respectively.
Expressing the corresponding source branches by means of the source voltages $\ensuremath{\ensuremath{\mathbf{V}}_\text{s}}$ and source currents $\ensuremath{\ensuremath{\mathbf{I}}_\text{s}}$, this relation becomes
\begin{equation}
    \ensuremath{\ensuremath{\mathbf{I}}_\text{R}}=\ensuremath{\mathbf{G}}\ensuremath{\ensuremath{\mathbf{V}}_\text{G}},\quad
    \ensuremath{\ensuremath{\mathbf{I}}_\text{C}}=\ensuremath{\mathbf{C}}\ensuremath{\dot{\ensuremath{\mathbf{V}}}_\text{C}},\quad
    \ensuremath{\ensuremath{\mathbf{I}}_\text{L}}=\ensuremath{\mathbf{L}}^{-1}\int\ensuremath{\ensuremath{\mathbf{V}}_\text{L}}\ \mathrm{d} t,\quad
    \ensuremath{\ensuremath{\mathbf{V}}_\text{V}}=\ensuremath{\ensuremath{\mathbf{V}}_\text{s}},\quad
    \ensuremath{\ensuremath{\mathbf{I}}_\text{I}}=\ensuremath{\ensuremath{\mathbf{I}}_\text{s}}.
    \label{eq:branchRelations}
\end{equation}
Combining~\eqref{eq:currentVoltageBlock}, \eqref{eq:KCLblock} and~\eqref{eq:branchRelations}, we obtain the \gls*{MNA} formulation
\begin{equation}
    \ensuremath{\ensuremath{\mathbf{A}}_\text{C}}\ensuremath{\mathbf{C}}\ensuremath{\ensuremath{\mathbf{A}}_\text{C}}^{\top}\dot{\ensuremath{\mathbf{v}}}+\ensuremath{\ensuremath{\mathbf{A}}_\text{G}}\ensuremath{\mathbf{G}}\ensuremath{\ensuremath{\mathbf{A}}_\text{G}}^{\top}\ensuremath{\mathbf{v}}+\ensuremath{\ensuremath{\mathbf{A}}_\text{L}}\ensuremath{\ensuremath{\mathbf{I}}_\text{L}}=\ensuremath{\ensuremath{\mathbf{A}}_\text{I}}\ensuremath{\ensuremath{\mathbf{I}}_\text{s}}-\ensuremath{\ensuremath{\mathbf{A}}_\text{V}}\ensuremath{\ensuremath{\mathbf{I}}_\text{V}},\quad
    \ensuremath{\mathbf{L}}\dot{\ensuremath{\mathbf{I}}}_{\text{L}}=\ensuremath{\ensuremath{\mathbf{A}}_\text{L}}^{\top}\ensuremath{\mathbf{v}},\quad
    \ensuremath{\ensuremath{\mathbf{A}}_\text{V}}^{\top}\ensuremath{\mathbf{v}}=\ensuremath{\ensuremath{\mathbf{V}}_\text{s}}.
    \label{eq:MNA}
\end{equation}
Note that in contrast to the standard \gls*{MNA} theory, \eqref{eq:MNA} still requires regularisation, typically done by the introduction of a reference (\emph{ground}) node.

\section{Circuit Representation of Electrothermal Field Problems}
\label{sec:circuitsET}

In this section, we derive the circuit representation for transient \gls*{ET} field problems.
We apply the \gls*{EQS} approximation~\cite{Clemens_2004ab} to \textsc{Maxwell}\xspace's equations given by \eqref{eq:MaxwellContInt} and consider the coupling with the transient heat equation given by~\eqref{eq:thermalContInt}.
Then, the bi-directionally coupled system in differential form reads
\begin{subequations}
    \begin{align}
        -\nabla \cdot\left(\varepsilon\nabla\dot{\varphi}\right)-\nabla \cdot\left(\sigma(T)\nabla\varphi\right) &= -\nabla \cdot\ensuremath{\mathbf{J}_{\mathrm{i}}},\label{eq:EQScont}\\
        \rho c \dot{T} - \nabla \cdot\left(\lambda(T)\nabla T\right) &= Q_{\text{J}}(\varphi,T)-\nabla \cdot\ensuremath{\mathbf{q}_{\mathrm{i}}},\label{eq:thermalCont}
    \end{align}
    \label{eq:ETcont}\end{subequations}
with suitable initial and boundary conditions.
The coupling is manifested by the \textsc{Joule}\xspace heating given by $Q_{\text{J}}=\sigma(\nabla\varphi)^{2}$ in one way and by the temperature dependent conductivity $\sigma(T)$ in the opposite way.
Due to the \gls*{EQS} approximation, this formulation does not account for inductive effects but does consider resistive and capacitive effects.
For simplicity, we neglect the temperature dependency of the permittivity $\varepsilon$ and of the volumetric heat capacity $\rho c$.
Applying \gls*{FIT} upon the \gls*{ET} system of~\eqref{eq:ETcont}, the semi-discrete formulation reads
\begin{subequations}
    \begin{align}
        \ensuremath{\protect\widetilde{\mathbf{S}}}\xspace\ensuremath{\mathbf{M}_{\varepsilon}}\xspace\ensuremath{\protect\widetilde{\mathbf{S}}}\xspace^{\top}\dot{\ensuremath{\boldsymbol{\mathrm{\varphi}}}}+\ensuremath{\protect\widetilde{\mathbf{S}}}\xspace\ensuremath{\mathbf{M}_{\sigma}}\xspace(\ensuremath{\mathbf{T}})\ensuremath{\protect\widetilde{\mathbf{S}}}\xspace^{\top}\ensuremath{\boldsymbol{\mathrm{\varphi}}}&=-\ensuremath{\protect\widetilde{\mathbf{S}}}\xspace\ensuremath{\protect\bbow{\mathrm{\mathbf{j}}}}\xspace_{\text{i}},\label{eq:EQSdisc}\\
        \ensuremath{\mathbf{M}_{\rho c}}\xspace\dot{\ensuremath{\mathbf{T}}}+\ensuremath{\protect\widetilde{\mathbf{S}}}\xspace\ensuremath{\mathbf{M}_{\lambda}}\xspace(\ensuremath{\mathbf{T}})\ensuremath{\protect\widetilde{\mathbf{S}}}\xspace^{\top}\ensuremath{\mathbf{T}}&=\mathbf{Q}_{\text{J}}(\ensuremath{\boldsymbol{\mathrm{\varphi}}},\ensuremath{\mathbf{T}})-\ensuremath{\protect\widetilde{\mathbf{S}}}\xspace\ensuremath{\protect\bbow{\mathrm{\mathbf{q}}}}\xspace_{\text{i}},\label{eq:thermalDisc}
    \end{align}
    \label{eq:ETdisc}\end{subequations}
with initial and boundary conditions yet to be applied.
Note that the system \eqref{eq:EQSdisc} requires a regularisation which is typically done by choosing a reference (\emph{ground}) node $\varphi_{\text{gnd}}=0$, where $\varphi_{\text{gnd}}$ is one of the entries of $\ensuremath{\boldsymbol{\mathrm{\varphi}}}$.

To generate the netlist for formulation~\eqref{eq:ETdisc}, the electric and thermal sub-problems are considered separately.
The connection is subsequently established by the \textsc{Joule}\xspace losses and the temperature dependent electric conductivity.
Next, in Section~\ref{subsec:EQSextract} and Section~\ref{subsec:thermalExtract}, the netlist generation for the \gls*{EQS} case and the thermal case is presented, respectively.
Temperature dependent materials are discussed in Section~\ref{subsec:ETnonlinearMaterials} and finally, the implementation of initial and boundary conditions is described in Section~\ref{subsec:ETinitialBoundary}.
This allows us to formulate an algorithm for the \gls*{ET} netlist generation as presented in Section~\ref{subsec:netlistGeneration}.

\subsection{Electroquasistatic Circuit Representation}
\label{subsec:EQSextract}

Let us now concentrate on the \gls*{EQS} sub-problem given by~\eqref{eq:EQScont}.
Since inductances are neglected in the \gls*{EQS} case, \eqref{eq:MNA} simplifies to
\begin{align}
    \ensuremath{\ensuremath{\mathbf{A}}_\text{C}}\ensuremath{\mathbf{C}}\ensuremath{\ensuremath{\mathbf{A}}_\text{C}}^{\top}\dot{\ensuremath{\mathbf{v}}}+\ensuremath{\ensuremath{\mathbf{A}}_\text{G}}\ensuremath{\mathbf{G}}\ensuremath{\ensuremath{\mathbf{A}}_\text{G}}^{\top}\ensuremath{\mathbf{v}}=-\ensuremath{\ensuremath{\mathbf{A}}_\text{I}}\ensuremath{\ensuremath{\mathbf{I}}_\text{s}}-\ensuremath{\ensuremath{\mathbf{A}}_\text{V}}\ensuremath{\ensuremath{\mathbf{I}}_\text{V}},\quad
    \ensuremath{\ensuremath{\mathbf{A}}_\text{V}}^{\top}\ensuremath{\mathbf{v}}=\ensuremath{\ensuremath{\mathbf{V}}_\text{s}}.
    \label{eq:MNA4EQS}
\end{align}
Thus, by inspection of~\eqref{eq:EQSdisc} and~\eqref{eq:MNA4EQS}, we are led to the following equivalences:
\begin{itemize}
    \item The incidence matrices $\ensuremath{\ensuremath{\mathbf{A}}_\text{C}}$ and $\ensuremath{\ensuremath{\mathbf{A}}_\text{G}}$ coincide with the \gls*{FIT} divergence matrix $\ensuremath{\protect\widetilde{\mathbf{S}}}\xspace$.
    \item The capacitance matrix $\ensuremath{\mathbf{C}}$ coincides with the \gls*{FIT} capacitance matrix $\ensuremath{\mathbf{M}_{\varepsilon}}\xspace$.
    \item The incidence matrices $\ensuremath{\ensuremath{\mathbf{A}}_\text{V}}$ and $\ensuremath{\ensuremath{\mathbf{A}}_\text{I}}$ coincide with the identity matrix.
    \item The conductance matrix $\ensuremath{\mathbf{G}}$ coincides with the \gls*{FIT} conductance matrix $\ensuremath{\mathbf{M}_{\sigma}}\xspace$.
    \item The nodal voltages $\ensuremath{\mathbf{v}}$ correspond to the \gls*{FIT} degrees of freedom $\ensuremath{\boldsymbol{\mathrm{\varphi}}}$.
    \item The source currents $\ensuremath{\ensuremath{\mathbf{I}}_\text{s}}$ are given by the divergence of the impressed currents $\ensuremath{\protect\widetilde{\mathbf{S}}}\xspace\ensuremath{\protect\bbow{\mathrm{\mathbf{j}}}}\xspace_{\text{i}}$.
    \item The source voltages $\ensuremath{\ensuremath{\mathbf{V}}_\text{s}}$ correspond to the \textsc{Dirichlet}\xspace potentials $\ensuremath{\boldsymbol{\mathrm{\varphi}}}_{\text{Dir}}$, which are related to the reference node $\varphi_{\text{gnd}}$.
    \item These equivalences also prevail themselves in the physical units.
\end{itemize}
Summarised, the field-circuit relations for \gls*{EQS} read
\begin{subequations}
    \begin{align}
        \ensuremath{\ensuremath{\mathbf{A}}_\text{G}}\mathrel{\widehat{=}}\ensuremath{\ensuremath{\mathbf{A}}_\text{C}}             &\mathrel{\widehat{=}}\ensuremath{\protect\widetilde{\mathbf{S}}}\xspace,                \label{eq:EQSequivAGC}\\
        \ensuremath{\mathbf{G}}\mathrel{\widehat{=}}\ensuremath{\mathbf{M}_{\sigma}}\xspace,\quad\ensuremath{\mathbf{C}}  &\mathrel{\widehat{=}}\ensuremath{\mathbf{M}_{\varepsilon}}\xspace,                   \label{eq:EQSequivMats}\\
        \ensuremath{\ensuremath{\mathbf{A}}_\text{V}}\mathrel{\widehat{=}}\ensuremath{\ensuremath{\mathbf{A}}_\text{I}}             &\mathrel{\widehat{=}}\mathbbm{I},                    \label{eq:EQSequivAVI}\\
                                \ensuremath{\mathbf{v}} &\mathrel{\widehat{=}}\ensuremath{\boldsymbol{\mathrm{\varphi}}},                   \label{eq:EQSequivV}\\
                                \ensuremath{\ensuremath{\mathbf{I}}_\text{s}}&\mathrel{\widehat{=}}\ensuremath{\protect\widetilde{\mathbf{S}}}\xspace\ensuremath{\protect\bbow{\mathrm{\mathbf{j}}}}\xspace_{\text{i}},\label{eq:EQSequivI}\\
                                \ensuremath{\ensuremath{\mathbf{V}}_\text{s}}&\mathrel{\widehat{=}}\ensuremath{\boldsymbol{\mathrm{\varphi}}}_{\text{Dir}}                 \label{eq:EQSequivVdir},
    \end{align}
    \label{eq:EQSequivalences}\end{subequations}
where $\mathbbm{I}$ is the identity matrix of corresponding size and $\ensuremath{\boldsymbol{\mathrm{\varphi}}}_{\text{Dir}}$ represents the potentials on the \textsc{Dirichlet}\xspace boundary nodes.
To find the circuit stamp of each edge in the grid upon which~\eqref{eq:EQSdisc} holds, we employ the equivalences of~\eqref{eq:EQSequivalences} from which the circuit topology is derived.
From~\eqref{eq:EQSequivAGC}, conductors and capacitors are placed along the branches of the circuit.
According to~\eqref{eq:EQSequivMats}, the values of the conductors and capacitors are directly taken from the corresponding FIT material matrices.
\begin{figure}[t]
    \centering
	\includegraphics{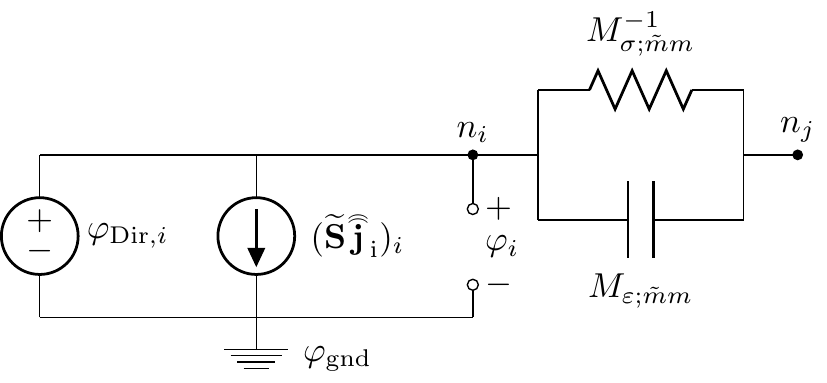}
	\caption{Equivalent electric circuit stamp for the exemplary edge $\ensuremath{L}_{m}$ between points $\ensuremath{P}_{i}$ and $\ensuremath{P}_{j}$.}
    \label{fig:EQScircuitStamp}
\end{figure}
Current and voltage sources are connected between a circuit node and ground as indicated by~\eqref{eq:EQSequivAVI}.
Furthermore, \eqref{eq:EQSequivV} shows that the circuit's nodal potentials are equal to the potentials at the grid points.
According to~\eqref{eq:EQSequivI}, the current sources in the circuit represent the divergence of the \gls*{FIT} impressed currents.
Finally, if \textsc{Dirichlet}\xspace \glspl*{BC} are imposed, voltage sources in the circuit represent \gls*{FIT} \textsc{Dirichlet}\xspace potentials as given by~\eqref{eq:EQSequivVdir}.
We further discuss \glspl*{BC} in Section~\ref{subsec:ETinitialBoundary}.
To summarise, if we consider an exemplary grid edge $\ensuremath{L}_{m}$, we obtain a representative \gls*{EQS} circuit stamp as shown in Figure~\ref{fig:EQScircuitStamp}.
Note that the temperature dependence of the materials is neglected for now and will be discussed in Section~\ref{subsec:ETnonlinearMaterials}

\subsection{Thermal Circuit Representation}
\label{subsec:thermalExtract}

In this section, we describe the circuit representation of the sub-problem described by~\eqref{eq:thermalCont}.
By comparing~\eqref{eq:thermalDisc} to~\eqref{eq:EQSdisc}, we observe a slightly different equation structure.
Thermal capacities are not subject to spatial differences and thus do not link to neighbouring nodes.
Instead, a thermal capacitance influences the change rate of the absolute temperature of a node.
Thus, thermal capacitances are placed on branches connecting the nodes to a reference node at zero temperature.
This reference node is an additional non-physical node that is introduced to obtain a consistent circuit representation.
In the literature, this approach is also referred to as the \textsc{Cauer}\xspace model representing a discretised image of the heat flow~\cite{Gerstenmaier_2007aa,Bechtold_2006ab}.
An equivalent approach is the \textsc{Foster}\xspace model, in which the capacitances are placed between the circuit nodes and the parameters are adjusted accordingly.
In the \textsc{Foster}\xspace model, the heat propagation is instantaneous and does not account for the fact that an object requires some delay before changing its temperature.

\begin{figure}
    \centering
	\subfloat[\label{fig:thermalCircuitStamp}]{\includegraphics{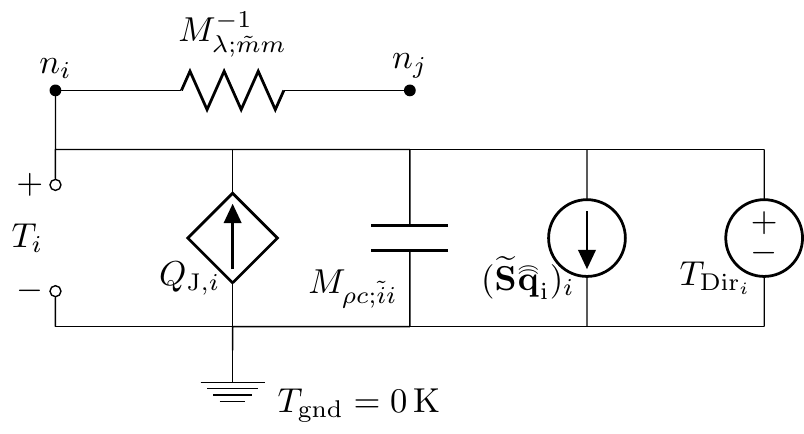}}
    \hspace{2em}
    \subfloat[\label{fig:RobinBoundary}]{\raisebox{3ex}{\includegraphics{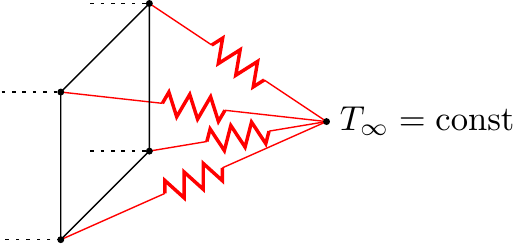}}}
	\caption{(a) Equivalent thermal circuit stamp for the exemplary edge $\ensuremath{L}_{m}$ between points $\ensuremath{P}_{i}$ and $\ensuremath{P}_{j}$. (b) Visualisation of \textsc{Robin}\xspace \glspl*{BC}}
\end{figure}

The \gls*{MNA} formulation of~\eqref{eq:MNA} must be extended by this additional reference ground node such that
\begin{equation}
    \ensuremath{\ensuremath{\widehat{\ensuremath{\mathbf{A}}}}_\text{C}}\ensuremath{\mathbf{C}}\ensuremath{\ensuremath{\widehat{\ensuremath{\mathbf{A}}}}_\text{C}}^{\top}\dot{\ensuremath{\widehat{\ensuremath{\mathbf{v}}}}}+\ensuremath{\ensuremath{\widehat{\ensuremath{\mathbf{A}}}}_\text{G}}\ensuremath{\mathbf{G}}\ensuremath{\ensuremath{\widehat{\ensuremath{\mathbf{A}}}}_\text{G}}^{\top}\ensuremath{\widehat{\ensuremath{\mathbf{v}}}}=-\ensuremath{\ensuremath{\widehat{\ensuremath{\mathbf{A}}}}_\text{I}}\ensuremath{\ensuremath{\mathbf{I}}_\text{s}}-\ensuremath{\ensuremath{\widehat{\ensuremath{\mathbf{A}}}}_\text{V}}\ensuremath{\ensuremath{\mathbf{I}}_\text{V}},\quad
    \ensuremath{\ensuremath{\widehat{\ensuremath{\mathbf{A}}}}_\text{V}}^{\top}\ensuremath{\widehat{\ensuremath{\mathbf{v}}}}=\ensuremath{\ensuremath{\mathbf{V}}_\text{s}},
    \label{eq:MNA4thermal}
\end{equation}
with
\begin{equation*}
    \ensuremath{\ensuremath{\widehat{\ensuremath{\mathbf{A}}}}_\text{C}}:=
    \begin{bmatrix}
        \mathbbm{I} & -\mathbbm{1}
    \end{bmatrix},\quad
    \ensuremath{\ensuremath{\widehat{\ensuremath{\mathbf{A}}}}_\text{G}}:=
    \begin{bmatrix}
        \ensuremath{\ensuremath{\mathbf{A}}_\text{G}} & \ensuremath{\mathbf{0}}
    \end{bmatrix},\quad
    \ensuremath{\ensuremath{\widehat{\ensuremath{\mathbf{A}}}}_\text{I}}:=
    \begin{bmatrix}
        \mathbbm{I} & -\mathbbm{1}
    \end{bmatrix},\quad
    \ensuremath{\widehat{\ensuremath{\mathbf{v}}}}^{\top}:=
    \begin{bmatrix}
        \ensuremath{\mathbf{v}}^{\top} & v_{\text{gnd}}
    \end{bmatrix},
\end{equation*}
where $\mathbbm{1}$ is a column vector of ones of appropriate size and $v_{\text{gnd}}=T_{\text{gnd}}=0$.
With these definitions, the equivalences between the \gls*{MNA} formulation of~\eqref{eq:MNA4thermal} and the thermal formulation of~\eqref{eq:thermalDisc} are readily obtained as
\begin{subequations}
    \begin{align}
        \ensuremath{\ensuremath{\mathbf{A}}_\text{C}}\mathrel{\widehat{=}}\ensuremath{\ensuremath{\mathbf{A}}_\text{V}}\mathrel{\widehat{=}}  \ensuremath{\ensuremath{\mathbf{A}}_\text{I}}&\mathrel{\widehat{=}}\mathbbm{I},                            \label{eq:thermalEquivAVIC}\\
                                \ensuremath{\ensuremath{\mathbf{A}}_\text{G}}&\mathrel{\widehat{=}}\ensuremath{\protect\widetilde{\mathbf{S}}}\xspace,                        \label{eq:thermalEquivAR}\\
        \ensuremath{\mathbf{G}}\mathrel{\widehat{=}}\ensuremath{\mathbf{M}_{\lambda}}\xspace,\quad\ensuremath{\mathbf{C}} &\mathrel{\widehat{=}}\ensuremath{\mathbf{M}_{\rho c}}\xspace,                          \label{eq:thermalEquivMats}\\
                                \ensuremath{\mathbf{v}} &\mathrel{\widehat{=}}\ensuremath{\mathbf{T}},                              \label{eq:thermalEquivT}\\
                                \ensuremath{\ensuremath{\mathbf{I}}_\text{s}}&\mathrel{\widehat{=}}-\mathbf{Q}_{\text{J}}+\ensuremath{\protect\widetilde{\mathbf{S}}}\xspace\ensuremath{\protect\bbow{\mathrm{\mathbf{q}}}}\xspace_{\text{i}}\label{eq:thermalEquivQ}\\
                                \ensuremath{\ensuremath{\mathbf{V}}_\text{s}}&\mathrel{\widehat{=}}\ensuremath{\mathbf{T}}_{\text{Dir}}                            \label{eq:thermalEquivTdir},
    \end{align}
    \label{eq:thermalEquivalences}\end{subequations}
where $\ensuremath{\mathbf{T}}_{\text{Dir}}$ represents the temperatures at \textsc{Dirichlet}\xspace boundary nodes.
To derive the circuit stamp of each edge in the grid upon which~\eqref{eq:thermalDisc} holds, we employ the equivalences in~\eqref{eq:thermalEquivalences} from which the circuit topology is derived.
From~\eqref{eq:thermalEquivAVIC}, we infer that capacitances, voltage and current sources are directly connected between a circuit node and the thermal ground.
As in the \gls*{EQS} case, conductors connect two neighbouring nodes in the grid as seen from~\eqref{eq:thermalEquivAR}.
The values of the conductors and capacitors are taken directly from the matrices $\ensuremath{\mathbf{M}_{\lambda}}\xspace$ and $\ensuremath{\mathbf{M}_{\rho c}}\xspace$ according to~\eqref{eq:thermalEquivMats}.
In this manner, the nodal potentials represent the sought temperatures as seen from~\eqref{eq:thermalEquivT}.
The current source is the sum of \textsc{Joule}\xspace losses, in fact represented by \glspl*{CCCS}, and the impressed heat flux according to~\eqref{eq:thermalEquivQ}.
Finally, if \textsc{Dirichlet}\xspace \glspl*{BC} are given, these are modelled by voltage sources in the circuit as stated in~\eqref{eq:thermalEquivTdir}.
We further comment on \glspl*{BC} in Section~\ref{subsec:ETinitialBoundary}.
To summarise, if we consider an exemplary grid edge $\ensuremath{L}_{m}$, we obtain a representative thermal circuit stamp as shown in Figure~\ref{fig:thermalCircuitStamp}.
Note that the temperature dependence of the materials is neglected until now and will be discussed in Section~\ref{subsec:ETnonlinearMaterials}.

To further highlight the equivalences between electric and thermal circuits, we would like to briefly comment on this.
Many quantities in electrical circuits can find their equivalent in thermal circuits.
For example, electric potentials are equivalent to temperatures while electric currents are equivalent to heat fluxes.
In Table~\ref{tab:electrothermalEquivalence}, some of these equivalences are summarised.

\begin{table}[t]
    \centering
    \begin{tabular}{r r}\toprule
    Electrical Circuit & Thermal Circuit\\\bottomrule\toprule
    electric potential (\si{V}) & temperature (\si{K})\\
    electric voltage (\si{V}) & temperature difference (\si{K})\\ 
    electric current (\si{A}) & thermal heat flux (\si{W})\\
    electric charge (\si{C})& thermal energy (\si{J})\\
    electric conductor (\si{S})& thermal conductor (\si{W/K})\\
    electric capacitance (\si{F})& thermal capacitance (\si{J/K})\\\bottomrule
\end{tabular} 
     \caption{Equivalences between electric and thermal quantities in circuit representations.}
    \label{tab:electrothermalEquivalence}
\end{table}

\subsection{Temperature Dependent Materials}
\label{subsec:ETnonlinearMaterials}

Most materials exhibit temperature dependent behaviour.
For the kind of considered materials, the temperature mainly influences the electric and thermal conductivity of the involved materials.
In this section, we describe an approach to account for a temperature-dependent electric conductivity when generating the corresponding SPICE\xspace netlist.
Nevertheless, the presented approach can be applied accordingly to other material's temperature dependencies as well, e.g.\ the thermal capacity.

Since the SPICE\xspace elements that represent the electric conductivities are conductors placed in circuit branches (cf. Section~\ref{subsec:EQSextract}), we first need to define the temperature of a branch $b_{m}$.
To this end, we take the average temperature $\overline{T}_{m}$ of the nodes interconnected by the branch $b_{m}$.
Assuming that the temperature dependence of the electric conductivity is known, the temperature-dependent electric conductance of branch $b_{m}$ is given by
\begin{equation}
    G_{m}\left(\overline{T}_{m}\right)=\sigma_{\protect\tilde{m} m}\left(\overline{T}_{m}\right)\frac{|\ensuremath{\protect\widetilde{A}}_{m}|}{|\ensuremath{L}_{m}|}.
    \label{eq:Gel}
\end{equation}
We implement~\eqref{eq:Gel} by means of behavioural sources in the SPICE\xspace language.

\subsection{Initial Conditions and Boundary Conditions}
\label{subsec:ETinitialBoundary}

When a coupled problem of more than one transient differential equation is considered, each sub-problem requires its own initial conditions and \glspl*{BC}.
Therefore, we impose these conditions on the \gls*{EQS} and thermal sub-problems separately.
However, certain equivalences allow to follow the same procedure for both sub-problems.
For any kind of transient problem, initial conditions are required.
As every SPICE\xspace dialect supports specifying initial conditions, these can be directly imposed by the corresponding syntax in the netlist.
For electric problems, different types of \glspl*{BC} are of interest.
In low-frequency problems as the \gls*{EQS} case, \textsc{Dirichlet}\xspace or \textsc{Neumann}\xspace conditions are typically used.
\textsc{Dirichlet}\xspace \glspl*{BC} conditions correspond to a fixed potential enforced at the boundary, while \textsc{Neumann}\xspace conditions prescribe the electric current through the boundary.
Similarly, in thermal problems, \textsc{Dirichlet}\xspace \glspl*{BC} correspond to a prescribed temperature at the boundary, while \textsc{Neumann}\xspace \glspl*{BC} prescribe thermal fluxes through the boundary.
Additionally, thermal problems commonly also involve \textsc{Robin}\xspace \glspl*{BC} that describe convective and radiative boundaries.

\textsc{Dirichlet}\xspace \glspl*{BC} are represented in the circuit by voltage sources between the ground node and the \textsc{Dirichlet}\xspace nodes.
Homogeneous \textsc{Neumann}\xspace conditions are automatically fulfilled since no edge or branch leaves the domain.
For simplicity, we do not consider inhomogeneous \textsc{Neumann}\xspace conditions in this paper.
A \textsc{Robin}\xspace \gls*{BC} can be understood as a conduction between a boundary node and an external node $n_{\infty}$ representing the fixed ambient temperature $T_{\infty}$.
Therefore, \textsc{Robin}\xspace \glspl*{BC} are represented in the circuit by conductors connected between the boundary nodes and $n_{\infty}$ as shown in Figure~\ref{fig:RobinBoundary}.
We collect the relevant \glspl*{BC} in Table~\ref{tab:bndConditions}.

\begin{table}[b]
    \centering
    \begin{tabular}{>{\centering}p{3.8cm}>{\centering\arraybackslash}p{5.2cm}}\toprule
    Boundary Condition & Implementation\\\bottomrule\toprule
    \textsc{Dirichlet}\xspace & \multicolumn{1}{c}{lumped voltage sources}\\
    hom. \textsc{Neumann}\xspace & \multicolumn{1}{c}{no edges leaving the circuit}\\
    \textsc{Robin}\xspace & additional non-physical ground\\\bottomrule
\end{tabular}
     \caption{Considered \glspl*{BC} and their implementation.}
    \label{tab:bndConditions}
\end{table}

\subsection{Electrothermal Netlist Generation}
\label{subsec:netlistGeneration}

\begin{algorithm}[h]
    \caption{\gls{ET} SPICE netlist generation.}
    \label{alg:netlistET}
    \begin{algorithmic}[1]
        \For{edge $\ensuremath{L}_{m}\in G$ between primal nodes $\ensuremath{P}_{i}$ and $\ensuremath{P}_{j}$}
            \State write BGel$m$\quad{}n$i$\quad{}n$j$\quad{}$\text{I=}V_{ij}G_{m}(\overline{T}_{m})$
            \State write Cel$m$\quad{}n$i$\quad{}n$j$\quad{}$M_{\varepsilon;\protect\tilde{m} m}$\quad{}$\text{ic}=0$
            \State write Rth$m$\quad{}n$i$T\quad{}n$j$T\quad{}$M_{\lambda;\protect\tilde{m} m}^{-1}$
        \EndFor
        \For{node $\ensuremath{P}_{i}\in G$}
            \State write Cth$i$\quad{}n$i$T\quad{}gnd\quad{}$M_{\rho c;\protect\tilde{i} i}$\quad{}$\text{ic}=T_{0}$
            \State write BLoss$i$\quad{}gnd\quad{}n$i$T\quad{}$\text{I=}Q_{\text{J},i}(t)$
            \If{$\ensuremath{P}_{i}$ is electric \textsc{Dirichlet}\xspace node}
                \State write VDirEl$i$\quad{}n$i$\quad{}gnd\quad{}$V_{\text{Dir},i}(t)$
            \EndIf
            \If{$\ensuremath{P}_{i}$ is thermal \textsc{Dirichlet}\xspace node}
                \State write VDirTh$i$\quad{}n$i$T\quad{}gnd\quad{}$T_{\text{Dir},i}(t)$
            \EndIf
            \If{an impressed current flows out of $\ensuremath{P}_{i}$}
                \State write IimpEl$i$\quad{}n$i$\quad{}gnd\quad{}$(\ensuremath{\protect\widetilde{\mathbf{S}}}\xspace\ensuremath{\protect\bbow{\mathrm{\mathbf{j}}}}\xspace_{\text{i}}(t))_{i}$ 
            \EndIf
            \If{an impressed heat flux flows out of $\ensuremath{P}_{i}$}
                \State write IimpTh$i$\quad{}n$i$T\quad{}gnd\quad{}$(\ensuremath{\protect\widetilde{\mathbf{S}}}\xspace\ensuremath{\protect\bbow{\mathrm{\mathbf{q}}}}\xspace_{\text{i}}(t))_{i}$ 
            \EndIf
        \EndFor
    \end{algorithmic}
\end{algorithm}

To finalise this section, we formulate Algorithm~\ref{alg:netlistET} to automatically generate \gls*{ET} SPICE\xspace netlists.
For every grid edge $\ensuremath{L}_{m}$ that connects grid points $\ensuremath{P}_{i}$ and $\ensuremath{P}_{j}$, where $i<j$, the thermal conductance, electric capacitance and the temperature dependent electric conductance (cf. Section~\ref{subsec:ETnonlinearMaterials}) are written to the netlist connecting nodes $n_{i}$ and $n_{j}$ of the circuit (lines 2--4).
Due to the possible non-linearity of the electric conductivity (cf.~Section~\ref{subsec:ETnonlinearMaterials}), a behavioural source is used, where $V_{ij}$ is the voltage between node $n_{i}$ and $n_{j}$.
Initial conditions (ic) for the electric part can be included using the corresponding syntax for the capacitors.
Here, we use exemplary zero initial conditions.
Additionally, for every grid point $\ensuremath{P}_{i}$, the thermal capacitance and the \glsfirst{CCCS} representing the \textsc{Joule}\xspace losses are added to the netlist connecting node $n_{i}$ and the ground node (gnd) of the circuit (lines 5 and 6)\footnote{For a straight-forward implementation, we use behavioural sources instead of \glspl*{CCCS}}.
Initial conditions (ic) for the thermal part are specified by pre-charging the thermal capacitors with the initial temperature $T_{0}$.
To specify a \gls*{CCCS} in the SPICE\xspace language, a behavioural source is used.
If $\ensuremath{P}_{i}$ is specified as an electric (thermal) \textsc{Dirichlet}\xspace node, an additional voltage source connecting node $n_{i}$ and the ground node of the circuit is inserted (lines 9--14).
Furthermore, if an impressed current (heat flux) flows out of $\ensuremath{P}_{i}$, an additional current source is added to the netlist connecting node $n_{i}$ and the ground node of the circuit (lines 15--20).

\section{Circuit Representation of Electromagnetic Field Problems} 
\label{sec:circuitsEM}

In this section, we neglect thermal effects and describe the circuit representation of general 3D \gls*{EM} field problems as given by \eqref{eq:MaxwellContInt}--\eqref{eq:FaradayAmpere}.
However, the thermo-\gls*{EM} coupling can be established analogously.
First, in Section~\ref{subsec:setsEdgesFacets}, we introduce an auxiliary set notion to collect specific edges and facets of the grid.
In Section~\ref{subsec:EMcircuitRepresentation}, the E-H formulation \eqref{eq:MGEmagCharge} and the E-A formulation \eqref{eq:MGE}--\eqref{eq:gaugeAfit} of the \gls*{MGEs} are transparently mapped into an electric circuit that fully describes the problem at hand.
Finally, in Section~\ref{subsec:theoryABC}, we extend our analysis in order to realise \glspl*{ABC} as circuit stamps.
These are typically needed to limit the computational domain while minimising unphysical reflections caused by the domain truncation.

\subsection{Auxiliary Sets of Edges and Facets}
\label{subsec:setsEdgesFacets}

\begin{figure}[h]
    \subfloat[$\mathcal{L}^{m}$\label{fig:edgeSetLm}]{\includegraphics[width=0.249\columnwidth]{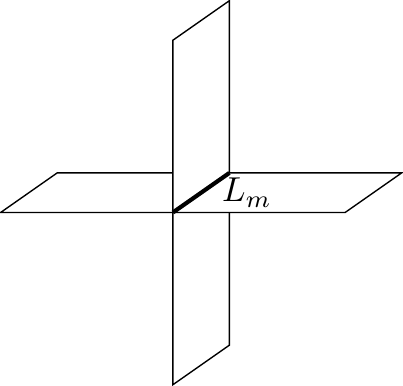}}
    \subfloat[$\mathcal{L}^{m,0}$\label{fig:edgeSetLm0}]{\includegraphics[width=0.249\columnwidth]{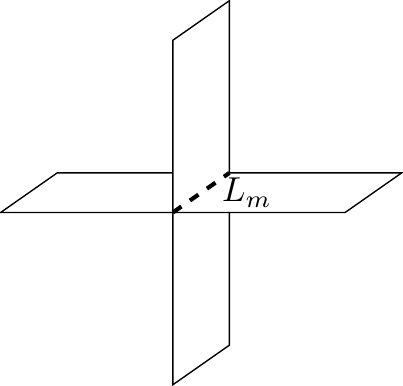}}
    \subfloat[$\mathcal{L}_{k}^{m}$\label{fig:edgeSetLm1}]{\includegraphics[width=0.249\columnwidth]{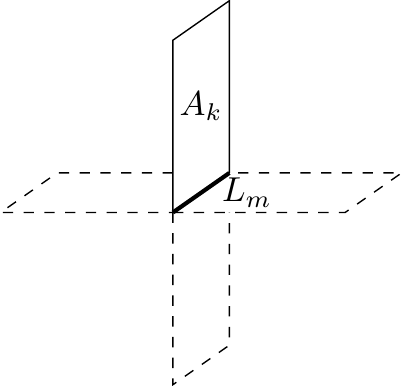}}
    \subfloat[$\mathcal{L}_{k}^{m,0}$\label{fig:edgeSetLm10}]{\includegraphics[width=0.249\columnwidth]{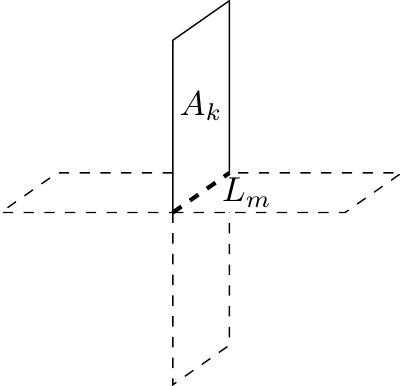}}
    \caption{Edge $\ensuremath{L}_{m}$ and the four facets collected in $\mathcal{A}^{m}=\{\ensuremath{A}_{1},\dots,\ensuremath{A}_{4}\}$ for a regular hexahedral grid. The highlighted edges illustrate the sets (a) $\mathcal{L}^{m}$, (b) $\mathcal{L}^{m,0}$, (c) $\mathcal{L}_{k}^{m}$ and (d) $\mathcal{L}_{k}^{m,0}$.}
    \label{fig:edgeSets}
\end{figure}

In the following sections, we will take sums over specific edges or facets of the grid.
Since the edges and facets in the neighbourhood of a specific edge $\ensuremath{L}_{m}$ are of interest, we introduce sets containing collections of these edges and facets and label them with the superscript~$^{m}$.
Let $\mathcal{A}^{m}$ be the set of all facets in which $\ensuremath{L}_{m}$ is embedded. 
For a regular hexahedral grid, these facets are shown in Figure~\ref{fig:edgeSets}.
All edges that are embedded in the facets contained in $\mathcal{A}^{m}$ are collected in the set $\mathcal{L}^{m}$, where this definition also implies $\ensuremath{L}_{m}\in\mathcal{L}^{m}$.
We denote by $\mathcal{L}^{m,0}$ the resulting set after extracting the very edge $\ensuremath{L}_{m}$ from $\mathcal{L}^{m}$, that is $\mathcal{L}^{m,0}:=\mathcal{L}^{m}\setminus\{\ensuremath{L}_{m}\}$.
Additionally, we denote the edges that are embedded in facet $\ensuremath{A}_{k}\in\mathcal{A}^{m}$ by the sets $\mathcal{L}_{k}^{m}$ and $\mathcal{L}_{k}^{m,0}$, respectively.
The definitions of the sets $\mathcal{L}^{m}$, $\mathcal{L}^{m,0}$, $\mathcal{L}_{k}^{m}$ and $\mathcal{L}_{k}^{m,0}$ are visualised in Figure~\ref{fig:edgeSets} for the case of a regular hexahedral grid.
For such a grid, $\mathcal{L}^{m}$, $\mathcal{L}^{m,0}$, $\mathcal{L}_{k}^{m}$ and $\mathcal{L}_{k}^{m,0}$ contain \num{13}, \num{12}, \num{4} and \num{3} edges, respectively.
In Section~\ref{subsec:circuitVectorPotForm}, an additional tree and cotree splitting is introduced.
Thereby, edges can either belong to the tree or the cotree.
This motivates the introduction of the subscripts $_{t}$ and $_{c}$ to denote tree and cotree, respectively.
The additional auxiliary sets that are used due to this splitting are denoted by $\mathcal{L}_{k;\text{t}}^{m}$, $\mathcal{L}_{k;\text{c}}^{m}$, $\mathcal{L}_{k;\text{t}}^{m,0}$ and $\mathcal{L}_{k;\text{c}}^{m,0}$.
Lastly, the dual edges that are embedded in the dual facet $\ensuremath{\protect\widetilde{A}}_{m}$ are collected in the set $\protect\widetilde{\mathcal{L}}_{\protect\tilde{m}}^{m}$.

\subsection{Circuit Representation of the \textsc{Maxwell}\xspace Grid Equations}
\label{subsec:EMcircuitRepresentation}

In this section, circuit representations of the \gls*{MGEs} given by \eqref{eq:MGEmagCharge}--\eqref{eq:gaugeAfit} are derived.
First, we start with the E-H formulation of~\eqref{eq:MGEmagCharge} to find a corresponding circuit description which is presented in Section~\ref{subsec:circuitFieldForm}.
Subsequently, in Section~\ref{subsec:circuitVectorPotForm} a circuit description based on the E-A formulation of~\eqref{eq:MGE}--\eqref{eq:gaugeAfit} is presented.
For this formulation, a \emph{tree-cotree} decomposition is used.

\subsubsection{Circuit Representation Based on the E-H Formulation}
\label{subsec:circuitFieldForm}

\begin{figure}[t]
    \centering
    \includegraphics[width=\columnwidth]{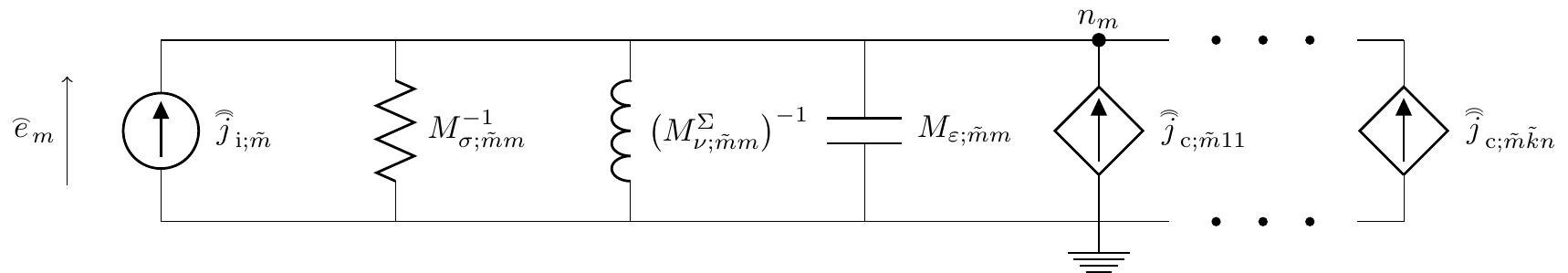}
    \caption{Circuit stamp of a primal edge $\ensuremath{L}_{m}$ with material matrices mapped into RLC lumped elements as stated by \eqref{eq:KCLfromMGEs}. \glspl*{VCCS} account for the interaction with neighbouring edges.}
    \label{fig:circuitStampEMelSource}
\end{figure}

A first electric circuit representing the \gls*{MGEs} is realised from the E-H formulation of~\eqref{eq:MGEmagCharge}.
The goal is to find an expression for the electric voltage on one primal edge such that a circuit stamp for each edge is obtained.
Let $\ensuremath{L}_{m}$ be the edge of interest with its corresponding dual facet $\ensuremath{\protect\widetilde{A}}_{m}$.
We collect all relevant quantities on the grid objects in the neighbourhood of $\ensuremath{L}_{m}$ by using the notation introduced in Section~\ref{subsec:setsEdgesFacets}.
To find the voltage $\ensuremath{\protect\bow{e}}_{m}$ on $\ensuremath{L}_{m}$, let us consider the pair of interlocked facets $\ensuremath{A}_{k}$ and $\ensuremath{\protect\widetilde{A}}_{m}$ as illustrated in Figure~\ref{fig:facetsInterlocked}.
For this part of the grid, it suffices to consider only the $k$-th row of \eqref{eq:MGEmagChargeFaraday} and the $\protect\tilde{m}$-th row of \eqref{eq:MGEmagChargeAmpere} giving
\begin{subequations}
    \begin{align}
        \sum_{n\in\mathcal{L}_{k}^{m}}\ensuremath{C}_{kn}\ensuremath{\protect\bow{e}}_{k;n}&=-M_{\nu;\protect\tilde{k} k}^{-1}\frac{\mathrm{d}\ensuremath{\protect\bow{h}}_{\protect\tilde{k}}}{\mathrm{d}t},\label{eq:FaradayOneFacet}\\
        \sum_{\protect\tilde{k}\in\protect\widetilde{\mathcal{L}}_{\protect\tilde{m}}^{m}}\ensuremath{\protect\widetilde{C}}_{\protect\tilde{m}\protect\tilde{k}}\ensuremath{\protect\bow{h}}_{\protect\tilde{m};\protect\tilde{k}}&=
        M_{\varepsilon;\protect\tilde{m} m}\frac{\mathrm{d}\ensuremath{\protect\bow{e}}_{m}}{\mathrm{d}t}+M_{\sigma;\protect\tilde{m} m}\ensuremath{\protect\bow{e}}_{m}+\ensuremath{\protect\bbow{j}}_{\text{i};\protect\tilde{m}}.\label{eq:AmpereOneFacet}
    \end{align}
\end{subequations}
From~\eqref{eq:FaradayOneFacet}, the magnetic grid voltage $\ensuremath{\protect\bow{h}}_{\protect\tilde{k}}$ allocated at edge $\ensuremath{\protect\widetilde{L}}_{k}$ (which happens to be also embedded in facet $\ensuremath{\protect\widetilde{A}}_{m}$) reads 
\begin{equation}
    \ensuremath{\protect\bow{h}}_{\protect\tilde{k}}=-M_{\nu;\protect\tilde{k} k}\sum_{n\in\mathcal{L}_{k}^{m}}\ensuremath{C}_{kn}\int\ensuremath{\protect\bow{e}}_{k;n}\ \mathrm{d} t.
    \label{eq:Faradayhfitlock}
\end{equation}
By inserting~\eqref{eq:Faradayhfitlock} in~\eqref{eq:AmpereOneFacet} for all $\ensuremath{\protect\bow{h}}_{\protect\tilde{m};\protect\tilde{k}}$, the voltage $\ensuremath{\protect\bow{e}}_{m}$ on edge $\ensuremath{L}_{m}$ is implicitly given by
\begin{equation*}
    -\sum_{\protect\tilde{k}\in\protect\widetilde{\mathcal{L}}_{\protect\tilde{m}}^{m}}\sum_{n\in\mathcal{L}_{k}^{m}}\ensuremath{\protect\widetilde{C}}_{\protect\tilde{m}\protect\tilde{k}}M_{\nu;\protect\tilde{k} k}\ensuremath{C}_{kn}\int\ensuremath{\protect\bow{e}}_{k;n}\ \mathrm{d} t=M_{\varepsilon;\protect\tilde{m} m}\frac{\mathrm{d}\ensuremath{\protect\bow{e}}_{m}}{\mathrm{d}t}+M_{\sigma;\protect\tilde{m} m}\ensuremath{\protect\bow{e}}_{m}+\ensuremath{\protect\bbow{j}}_{\text{i};\protect\tilde{m}}.
\end{equation*}
Since $\ensuremath{L}_{m}$ is contained in $\mathcal{L}_{k}^{m}$, we can extract the contribution of $\ensuremath{L}_{m}$ from the sum on the left hand side such that
\begin{multline}
        \sum_{\protect\tilde{k}\in\protect\widetilde{\mathcal{L}}_{\protect\tilde{m}}^{m}}\ensuremath{\protect\widetilde{C}}_{\protect\tilde{m}\protect\tilde{k}}M_{\nu;\protect\tilde{k} k}\ensuremath{C}_{km}\int\ensuremath{\protect\bow{e}}_{m}\,\mathrm{d} t
        +\sum_{\protect\tilde{k}\in\protect\widetilde{\mathcal{L}}_{\protect\tilde{m}}^{m}}\sum_{n\in\mathcal{L}_{k}^{m,0}}\ensuremath{\protect\widetilde{C}}_{\protect\tilde{m}\protect\tilde{k}}M_{\nu;\protect\tilde{k} k}\ensuremath{C}_{kn}\int\ensuremath{\protect\bow{e}}_{k;n}\ \mathrm{d} t\\
        +M_{\varepsilon;\protect\tilde{m} m}\frac{\mathrm{d}\ensuremath{\protect\bow{e}}_{m}}{\mathrm{d}t}+M_{\sigma;\protect\tilde{m} m}\ensuremath{\protect\bow{e}}_{m}+\ensuremath{\protect\bbow{j}}_{\text{i};\protect\tilde{m}}=0.
    \label{eq:KCLfromMGEsLong}
\end{multline}
We further define
\begin{subequations}
    \begin{align}
        M_{\nu;\protect\tilde{m} m}^{\Sigma}&:=\sum_{\protect\tilde{k}\in\protect\widetilde{\mathcal{L}}_{\protect\tilde{m}}^{m}}\ensuremath{\protect\widetilde{C}}_{\protect\tilde{m}\protect\tilde{k}}M_{\nu;\protect\tilde{k} k}\ensuremath{C}_{km}=\sum_{\protect\tilde{k}\in\protect\widetilde{\mathcal{L}}_{\protect\tilde{m}}^{m}}M_{\nu;\protect\tilde{k} k},\\
        \ensuremath{\protect\bbow{j}}_{\text{c};\protect\tilde{m}\protect\tilde{k} n}&:=\ensuremath{\protect\widetilde{C}}_{\protect\tilde{m}\protect\tilde{k}}M_{\nu;\protect\tilde{k} k}\ensuremath{C}_{kn}\int\ensuremath{\protect\bow{e}}_{k;n}\,\mathrm{d} t.
        \label{eq:abbrevCircuit}
    \end{align}
\end{subequations}
Thanks to the properties $\ensuremath{\mathbf{C}}\xspace=\ensuremath{\protect\widetilde{\mathbf{C}}}\xspace^{\top}$, $\ensuremath{\protect\widetilde{C}}_{\protect\tilde{m}\protect\tilde{k}}\in\{-1,1\}$ and $\ensuremath{C}_{km}\in\{-1,1\}$, we have that $\ensuremath{\protect\widetilde{C}}_{\protect\tilde{m}\protect\tilde{k}}=\ensuremath{C}_{km}$ and thus their product equals to unity.
However, the product $\ensuremath{\protect\widetilde{C}}_{\protect\tilde{m}\protect\tilde{k}}\ensuremath{C}_{kn}$ can be either $-1$ or $1$ as illustrated by Figure~\ref{fig:facetsInterlocked}.
Thus, we write~\eqref{eq:KCLfromMGEsLong} with the help of~\eqref{eq:abbrevCircuit} compactly as
\begin{equation}
    M_{\nu;\protect\tilde{m} m}^{\Sigma}\int\ensuremath{\protect\bow{e}}_{m}\,\mathrm{d} t
    +\sum_{\protect\tilde{k}\in\protect\widetilde{\mathcal{L}}_{\protect\tilde{m}}^{m}}\sum_{n\in\mathcal{L}_{k}^{m,0}}\ensuremath{\protect\bbow{j}}_{\text{c};\protect\tilde{m}\protect\tilde{k} n}
    +M_{\varepsilon;\protect\tilde{m} m}\frac{\mathrm{d}\ensuremath{\protect\bow{e}}_{m}}{\mathrm{d}t}+M_{\sigma;\protect\tilde{m} m}\ensuremath{\protect\bow{e}}_{m}+\ensuremath{\protect\bbow{j}}_{\text{i};\protect\tilde{m}}=0.
    \label{eq:KCLfromMGEs}
\end{equation}
Equation~\eqref{eq:KCLfromMGEs} is the \textsc{Kirchhoff}\xspace's current law (KCL) associated with the primal edge $\ensuremath{L}_{m}$ with $\ensuremath{\protect\bow{e}}_{m}$ representing the voltage drop along the edge.
In fact, according to the definitions of the field quantities in \eqref{eq:KCLfromMGEs}, it is easy to realise that
\begin{itemize}
\item $\ensuremath{\protect\bow{e}}_{m}$ has unit of voltage (\si{V}),
\item $\ensuremath{\protect\bbow{j}}_{\text{i},\protect\tilde{m}}$ has unit of current (\si{A}),
\item $M_{\varepsilon;\protect\tilde{m} m}$ is positive and has unit of capacitance (\si{F}),
\item $M_{\sigma;\protect\tilde{m} m}$ is positive and has unit of conductance (\si{S}),
\item $M_{\nu;\protect\tilde{m} m}^{\Sigma}$ is positive and has unit of reluctance (\si{H\tothe{-1}}),
\item $\ensuremath{\protect\bbow{j}}_{\text{c};\protect\tilde{m}\protect\tilde{k} n}$ has unit of current (\si{A}).
\end{itemize}
Thus, by using \glspl*{VCCS} to model $\ensuremath{\protect\bbow{j}}_{\text{c};\protect\tilde{m}\protect\tilde{k} n}$ accounting for the contributions from neighbouring edges, we can directly represent~\eqref{eq:KCLfromMGEs} with the circuit stamp depicted in Figure~\ref{fig:circuitStampEMelSource}, which preserves the voltage drop between the terminals of $\ensuremath{L}_{m}$. 
There are as many of these stamps as primal edges in $G$, and all of them interact via \glspl*{VCCS}.
The concatenation of these elementary stamps constitutes the electric circuit representing the electromagnetic problem at hand. 

In Algorithm~\ref{alg:netlistEM1}, the steps to generate the netlist representing the \gls*{EM} circuit for the entire grid are listed.
A circuit stamp, as shown in Figure~\ref{fig:circuitStampEMelSource}, needs to be created for every edge $\ensuremath{L}_{m}$ in the grid which is realised by a loop in the code.
In each iteration, a resistor, inductor and capacitor with the values taken from the material matrices is added (lines~2--4).
If an impressed current source shall be placed on $\ensuremath{L}_{m}$, an independent current source with a predefined value $\ensuremath{\protect\bbow{j}}_{\text{i};\protect\tilde{m}}$ is used (lines~5--7).
Finally, an inner double loop is required to insert the controlled current sources that model the influence of the edges in the neighbourhood.
For this purpose, we choose to insert \glspl*{CCCS}\footnote{Due to the integral term in the expression for $\ensuremath{\protect\bbow{j}}_{\text{c};\protect\tilde{m}\protect\tilde{k} n}$, a direct translation into \glspl*{VCCS} is not possible. Instead, behavioural sources or \glspl*{CCCS} can be used.} being controlled by the current $M_{\nu;\protect\tilde{n} n}^{\Sigma}\int\ensuremath{\protect\bow{e}}_{k;n}\,\mathrm{d} t$ with a gain of $g_{\protect\tilde{m}\protect\tilde{k} n}^{\text{I}}:=\ensuremath{\protect\widetilde{C}}_{\protect\tilde{m}\protect\tilde{k}}M_{\nu;\protect\tilde{k} k}\ensuremath{C}_{kn}(M_{\nu;\protect\tilde{n} n}^{\Sigma})^{-1}$ (lines~8--12).
By abuse of notation, we denote the controlling device by $\ensuremath{\protect\bow{e}}_{k;n}$.

\begin{algorithm}[t]
    \caption{Electromagnetic SPICE netlist generation based on the E-H formulation.}
    \label{alg:netlistEM1}
    \begin{algorithmic}[1]
        \For{edge $\ensuremath{L}_{m}\in G$}
            \State write R$m$\quad{}n$m$\quad{}gnd\quad{}$M_{\sigma;\protect\tilde{m} m}^{-1}$
            \State write L$m$\quad{}n$m$\quad{}gnd\quad{}$(M_{\nu;\protect\tilde{m} m}^{\Sigma})^{-1}$
            \State write C$m$\quad{}n$m$\quad{}gnd\quad{}$M_{\varepsilon;\protect\tilde{m} m}$
            \If{an impressed current is placed on $\ensuremath{L}_{m}$}
                \State write I$m$\quad{}gnd\quad{}n$m$\quad{}$\ensuremath{\protect\bbow{j}}_{\text{i};\protect\tilde{m}}$ 
            \EndIf
            \For{edge $\ensuremath{\protect\widetilde{L}}_{k}\in\protect\widetilde{\mathcal{L}}_{\protect\tilde{m}}^{m}$}
                \For{edge $\ensuremath{L}_{n}\in\mathcal{L}_{k}^{m,0}$}
                    \State write F$mn$\quad{}gnd\quad{}n$m$\quad{}$\ensuremath{\protect\bow{e}}_{k;n}$\quad{}$g_{\protect\tilde{m}\protect\tilde{k} n}^{\text{I}}$
                \EndFor
            \EndFor
        \EndFor
    \end{algorithmic}
\end{algorithm}

We remark that a similar analysis on~\eqref{eq:MGEmagCharge} in which magnetic conductivities and sources are considered instead of electric ones can be done\footnote{Although magnetic carriers have not been observed in nature, there may be situations in which one can profit from the inclusion of an equivalent magnetic conductivity in \textsc{Maxwell}\xspace's equations~\cite{Wait_1992aa,Kharzeev_2009aa}.}. This approach would lead to a circuit stamp that is dual to the one in Figure~\ref{fig:circuitStampEMelSource}.
Namely, in this dual stamp, the \gls*{KVL} is guaranteed for each dual edge and $\ensuremath{\protect\bow{h}}_{\protect\tilde{k}}$ represents the electric current in the circuit.
Furthermore, the resulting lumped elements, stemming from the material matrices, are placed in series with the discrete impressed magnetic current $\ensuremath{\protect\bbow{m}}_{\text{i};k}$ that plays the role of an independent voltage source exciting the circuit.
The interaction between the dual stamps is mediated via \glspl*{CCVS}.
In the general case, when both electric and magnetic sources are present, the electric circuit will consist of both the primal and dual stamps, which interact via \glspl*{CCCS} and \glspl*{VCVS}, accordingly.

\subsubsection{Circuit Representation Based on the E-A Formulation}
\label{subsec:circuitVectorPotForm}

\begin{figure}[t]
    \centering
    \subfloat[\label{fig:facetsInterlocked}]{\includegraphics{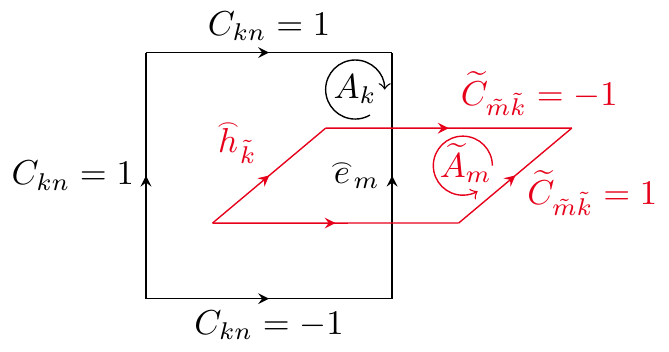}}
    \hspace{3em}
    \subfloat[\label{fig:fitCurlEdge}]{\includegraphics{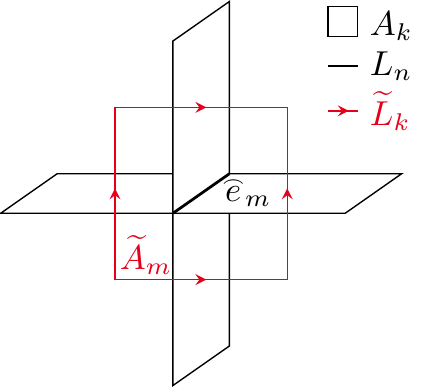}}
    \caption{(a) A pair of interlocked primal and dual facet $\ensuremath{A}_{k}$ and $\ensuremath{\protect\widetilde{A}}_{m}$ for the case of a regular hexahedral grid, respectively. The values of $\ensuremath{C}_{kn}$ and $\ensuremath{\protect\widetilde{C}}_{\protect\tilde{m}\protect\tilde{k}}$ for the edges $\ensuremath{L}_{n}$ of $\ensuremath{A}_{k}$ and the edges $\ensuremath{\protect\widetilde{L}}_{k}$ of $\ensuremath{\protect\widetilde{A}}_{m}$ are annotated. (b) The collection of edges $\ensuremath{L}_{n}$ and $\ensuremath{\protect\widetilde{L}}_{k}$ that contribute to the computation of $\ensuremath{\protect\bow{e}}_{m}$.} 
\end{figure}

\begin{figure}
    \centering
    \includegraphics[width=0.4\textwidth]{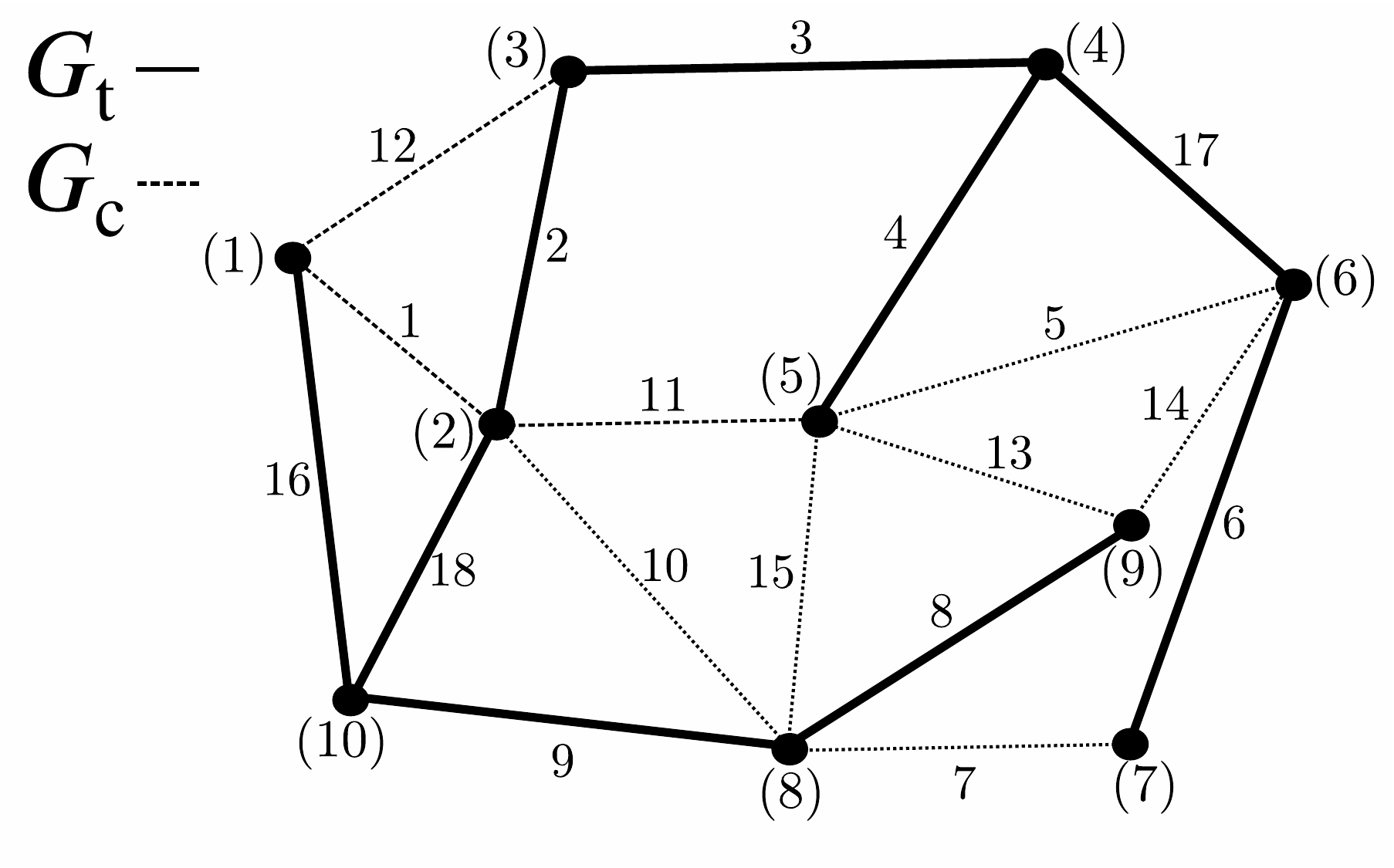}
    \caption{Example of a \emph{spanning} tree $G_{\text{t}}$ (solid) and cotree $G_{\text{c}}$ (dashed) in a \emph{generic} graph $G$. The case studied here consists of an~\emph{orthogonal} pair of primal and dual grids.}
    \label{fig:treeCotree}
\end{figure}

In this section, we describe a circuit representation based on the E-A formulation of~\eqref{eq:MGE}--\eqref{eq:gaugeAfit}.
To guarantee uniqueness of the solution, the magnetic potential $\ensuremath{\protect\bow{\mathrm{\mathbf{a}}}}\xspace$ is gauged by means of a \emph{tree}-\emph{cotree} decomposition~\cite{Munteanu_2002aa}.
Therefore, although the inferred circuit stamps are gauge dependent, the solution obtained for
$\ensuremath{\protect\bow{\mathrm{\mathbf{e}}}}\xspace$ is unique.
We start by considering one primal facet $\ensuremath{A}_{k}\in\mathcal{A}^{m}$ (cf. Figure~\ref{fig:facetsInterlocked}).
For this facet, it suffices to consider the $k$-th row of \eqref{eq:MGEfaraday} and the $\protect\tilde{m}$-th row of \eqref{eq:MGEampere} such that the E-A formulation of the \gls*{MGEs} for a generic edge $\ensuremath{L}_{m}$ reads
\begin{subequations}
    \begin{align}
        \sum_{n\in\mathcal{L}_{k}^{m}}\ensuremath{C}_{kn}\ensuremath{\protect\bow{e}}_{k;n}&=-\sum_{n\in\mathcal{L}_{k}^{m}}\ensuremath{C}_{kn}\frac{\mathrm{d}\ensuremath{\protect\bow{a}}_{k;n}}{\mathrm{d}t},\label{eq:MGElocalRevisitFaraday}\\
            \sum_{\protect\tilde{k}\in\protect\widetilde{\mathcal{L}}_{\protect\tilde{m}}^{m}}\sum_{n\in\mathcal{L}_{k}^{m}}\ensuremath{\protect\widetilde{C}}_{\protect\tilde{m}\protect\tilde{k}}M_{\nu;\protect\tilde{k} k}\ensuremath{C}_{kn}\ensuremath{\protect\bow{a}}_{k;n}
            &=M_{\varepsilon;\protect\tilde{m} m}\frac{\mathrm{d}\ensuremath{\protect\bow{e}}_{m}}{\mathrm{d}t}+M_{\sigma;\protect\tilde{m} m}\ensuremath{\protect\bow{e}}_{m}+\ensuremath{\protect\bbow{j}}_{\text{i};\protect\tilde{m}}.
        \label{eq:MGElocalRevisitAmpere}
    \end{align}
    \label{eq:MGElocalRevisit}\end{subequations}
As in Section~\ref{subsec:circuitFieldForm}, we aim at finding a unique circuit representation of edge $\ensuremath{L}_{m}$.
We start by observing that the system matrix $\ensuremath{\protect\widetilde{\mathbf{C}}}\xspace\ensuremath{\mathbf{M}_{\nu}}\xspace\ensuremath{\mathbf{C}}\xspace$ of~\eqref{eq:MGE} is singular\footnote{Note that this is also true for the corresponding continuous operator.}.
This singularity manifests itself in the non-uniqueness of $\ensuremath{\protect\bow{\mathrm{\mathbf{a}}}}\xspace$.
In fact, only the curl of $\ensuremath{\protect\bow{\mathrm{\mathbf{a}}}}\xspace$ is uniquely defined.
As a remedy, we must explicitly impose the gauging~\eqref{eq:gaugeAfit} upon~\eqref{eq:MGE}.
To this end, let us assume that we have constructed a suitable \emph{tree} $G_{\text{t}}$ and a \emph{cotree} $G_{\text{c}}$ out of the primal grid $G$, as exemplified in Figure~\ref{fig:treeCotree}.
Then, we symbolically introduce the orthogonal permutation matrix $\mathbf{P}_{\text{G}}$ to partition $\ensuremath{\protect\bow{\mathrm{\mathbf{a}}}}\xspace$ into its \emph{tree} $\ensuremath{\protect\bow{\mathrm{\mathbf{a}}}}\xspace_{\text{t}}\in G_{\text{t}}$ and \emph{cotree} $\ensuremath{\protect\bow{\mathrm{\mathbf{a}}}}\xspace_{\text{c}}\in G_{\text{c}}$ components with $N_{\text{t}}$ and $N_{\text{c}}$ entries, respectively.
Symbolically, this reads
\begin{equation}
    \mathbf{P}_{\text{G}}\ensuremath{\protect\bow{\mathrm{\mathbf{a}}}}\xspace=
    \begin{bmatrix}
        \ensuremath{\protect\bow{\mathrm{\mathbf{a}}}}\xspace_{\text{c}}\\
        \ensuremath{\protect\bow{\mathrm{\mathbf{a}}}}\xspace_{\text{t}} 
    \end{bmatrix}.
    \label{eq:treeCotreeSplit}
\end{equation}

Similarly, we also divide $\mathbf{M}_{\text{G}}$ and $\ensuremath{\protect\widetilde{\mathbf{S}}}\xspace$ into their tree and cotree components by applying the $\mathbf{P}_{\text{G}}$ matrix accordingly, viz.
\begin{equation}
    \mathbf{P}_{\text{G}}\mathbf{M}_{\text{G}}\mathbf{P}_{\text{G}}^{\top}=
    \begin{bmatrix}
        \mathbf{M}_{\text{G}_{\text{c}}} & \ensuremath{\mathbf{0}} \\
        \ensuremath{\mathbf{0}} & \mathbf{M}_{\text{G}_{\text{t}}} 
    \end{bmatrix},
    \quad\ensuremath{\protect\widetilde{\mathbf{S}}}\xspace\mathbf{P}_{\text{G}}^{\top}=
    \begin{bmatrix}
        \ensuremath{\protect\widetilde{\mathbf{S}}}\xspace_{\mathbbm{1}\text{c}} & \ensuremath{\protect\widetilde{\mathbf{S}}}\xspace_{\mathbbm{1}\text{t}} 
    \end{bmatrix}.
    \label{eq:DGsplit}
\end{equation}
The identities of~\eqref{eq:treeCotreeSplit} and~\eqref{eq:DGsplit} together with the orthogonality property of $\mathbf{P}_{\text{G}}$ enable us to rewrite the gauging of~\eqref{eq:gaugeAfit} as
\begin{equation}
    \ensuremath{\protect\widetilde{\mathbf{S}}}\xspace_{\mathbbm{1}\text{c}}\mathbf{M}_{\text{G}_{\text{c}}}\ensuremath{\protect\bow{\mathrm{\mathbf{a}}}}\xspace_{\text{c}}+
    \ensuremath{\protect\widetilde{\mathbf{S}}}\xspace_{\mathbbm{1}\text{t}}\mathbf{M}_{\text{G}_{\text{t}}}\ensuremath{\protect\bow{\mathrm{\mathbf{a}}}}\xspace_{\text{t}}=\ensuremath{ \mathbf{F} }. 
    \label{eq:gaugeAfitSplit} 
\end{equation}
Additionally, since $\ensuremath{\protect\widetilde{\mathbf{S}}}\xspace_{\mathbbm{1}\text{t}}\mathbf{M}_{\text{G}_{\text{t}}}$ is a square and invertible matrix\footnote{These two properties come from two facts: the squareness is a consequence of removing the row associated with the ground node required in circuit analysis.
The invertibility is due to removing the non-null kernel space of the matrix $\ensuremath{\protect\widetilde{\mathbf{C}}}\xspace\ensuremath{\mathbf{M}_{\nu}}\xspace\ensuremath{\mathbf{C}}\xspace$ by means of the tree-cotree decomposition.},
we may finally express the tree component $\ensuremath{\protect\bow{\mathrm{\mathbf{a}}}}\xspace_{\text{t}}$ as
\begin{equation*}
    \ensuremath{\protect\bow{\mathrm{\mathbf{a}}}}\xspace_{\text{t}}=\mathbf{M}_{\text{G}_{\text{t}}}^{-1}\ensuremath{\protect\widetilde{\mathbf{S}}}\xspace_{\mathbbm{1}\text{t}}^{-1}\ensuremath{ \mathbf{F} }-\mathbf{M}_{\text{G}_{\text{t}}}^{-1}\ensuremath{\protect\widetilde{\mathbf{S}}}\xspace_{\mathbbm{1}\text{t}}^{-1}\ensuremath{\protect\widetilde{\mathbf{S}}}\xspace_{\mathbbm{1}\text{c}}\mathbf{M}_{\text{G}_{\text{c}}}\ensuremath{\protect\bow{\mathrm{\mathbf{a}}}}\xspace_{\text{c}}.
\end{equation*}
The column vector $\ensuremath{ \mathbf{F} }$, which represents the grid counterpart of the scalar function $f$ quantifying the divergence of $\ensuremath{\mathbf{A}}$, is of free choice.
Therefore, we conveniently choose $\ensuremath{ \mathbf{F} }=\ensuremath{\mathbf{0}}$ impressing the \textsc{Coulomb}\xspace gauge~\cite{Balanis_2012aa} to straightforwardly arrive at 
\begin{equation}
    \ensuremath{\protect\bow{\mathrm{\mathbf{a}}}}\xspace_{\text{t}}=-\underbrace{\mathbf{M}_{\text{G}_{\text{t}}}^{-1}\ensuremath{\protect\widetilde{\mathbf{S}}}\xspace_{\mathbbm{1}\text{t}}^{-1}\ensuremath{\protect\widetilde{\mathbf{S}}}\xspace_{\mathbbm{1}\text{c}}\mathbf{M}_{\text{G}_{\text{c}}}}_{\mathbf{E}_{\text{tc}}}\ensuremath{\protect\bow{\mathrm{\mathbf{a}}}}\xspace_{\text{c}}.
    \label{fit_cir_eq24}
\end{equation}
The matrix $\mathbf{E}_{\text{tc}}\in\mathbb{R}^{N_{\text{t}}\times N_{\text{c}}}$ is known as the \emph{essential} incidence matrix~\cite{Munteanu_2002aa} and establishes a direct relation between the tree and cotree components $\ensuremath{\protect\bow{\mathrm{\mathbf{a}}}}\xspace_{\text{t}}$ and $\ensuremath{\protect\bow{\mathrm{\mathbf{a}}}}\xspace_{\text{c}}$, respectively\footnote{Owing to this property, we identify in the rows of $\mathbf{E}_{\text{tc}}\in\mathbb{R}^{N_{\text{t}}\times N_{\text{c}}}$ the collection of all fundamental cut-sets associated with the selected tree and cotree.
We recall that a fundamental cut-set is a set formed by the union of a single tree edge and the unique set of adjoining cotree edges. In this manner, the fundamental cut-sets are used to express \textsc{Kirchhoff}\xspace's current law in the general form of~\eqref{fit_cir_eq24}.}.
By inspection and expansion of~\eqref{fit_cir_eq24}, we express the $m$-th component of the column vector $\ensuremath{\protect\bow{\mathrm{\mathbf{a}}}}\xspace_{\text{t}}$ as
\begin{equation}
    \ensuremath{\protect\bow{a}}_{\text{t};m}=-\sum_{n\in G_{\text{c}}}E_{\text{tc};mn}\ensuremath{\protect\bow{a}}_{\text{c};n},
    \label{eq:afittm}\end{equation}
with
\begin{equation*}
    E_{\text{tc};mn}:=\sum_{i\in G}M_{G_{\text{t}};m\protect\tilde{m}}^{-1}\ensuremath{\protect\widetilde{S}}_{\mathbbm{1}\text{t};\protect\tilde{m}\protect\tilde{i}}^{-1}\ensuremath{\protect\widetilde{S}}_{\mathbbm{1}\text{c};\protect\tilde{i}\protect\tilde{n}} M_{G_{\text{c}};\protect\tilde{n} n}
\end{equation*}
and $i$ spanning over the nodes in $G$.
In this manner, \eqref{eq:afittm} removes the redundancy associated with $\ensuremath{\protect\bow{\mathrm{\mathbf{a}}}}\xspace$ and guarantees a unique solution of~\eqref{eq:MGElocalRevisit}.

Let us now get back to the generic primal edge $\ensuremath{L}_{m}$ once again.
As introduced in Section~\ref{subsec:setsEdgesFacets}, we use specific sets to refer to edges in the tree and cotree denoted by $\mathcal{L}_{k;\text{t}}^{m}$, $\mathcal{L}_{k;\text{t}}^{m,0}$ and $\mathcal{L}_{k;\text{c}}^{m}$, $\mathcal{L}_{k;\text{c}}^{m,0}$, respectively.
Isolating the voltage $\ensuremath{\protect\bow{e}}_{m}$ along edge $\ensuremath{L}_{m}$ embedded in facet $\ensuremath{A}_{k}$ from~\eqref{eq:MGElocalRevisitFaraday}, we arrive at 
\begin{equation*}
    \ensuremath{\protect\bow{e}}_{m}+\sum_{n\in\mathcal{L}_{k}^{m,0}}\left(\frac{\ensuremath{C}_{kn}}{\ensuremath{C}_{km}}\right)\ensuremath{\protect\bow{e}}_{k;n}
    =-\sum_{n\in\mathcal{L}_{k}^{m}}\left(\frac{\ensuremath{C}_{kn}}{\ensuremath{C}_{km}}\right)\frac{\mathrm{d}\ensuremath{\protect\bow{a}}_{k;n}}{\mathrm{d}t}. 
\end{equation*}
We may then take the sum over all facets $\ensuremath{A}_{k}\in\mathcal{A}^{m}$ (cf. Figure~\ref{fig:fitCurlEdge}). Then, by splitting $\lbrace\ensuremath{\protect\bow{a}}_{k;n}\rbrace$ into tree and cotree components, we arrive at 
\begin{multline}
    \ensuremath{\protect\bow{e}}_{m}+\frac{1}{N_{\text{F};m}}\sum_{k\in\mathcal{A}^{m}}\sum_{n\in\mathcal{L}_{k}^{m,0}}\left(\frac{\ensuremath{C}_{kn}}{\ensuremath{C}_{km}}\right)\ensuremath{\protect\bow{e}}_{k;n}\\
    \quad=-\frac{1}{N_{\text{F};m}}\sum_{k\in\mathcal{A}^{m}}\sum_{n\in\mathcal{L}_{k;\text{c}}^{m}}\left(\frac{\ensuremath{C}_{kn}}{\ensuremath{C}_{km}}\right)\frac{\mathrm{d}\ensuremath{\protect\bow{a}}_{k;n}}{\mathrm{d}t}
          -\frac{1}{N_{\text{F};m}}\sum_{k\in\mathcal{A}^{m}}\sum_{n\in\mathcal{L}_{k;\text{t}}^{m}}\left(\frac{\ensuremath{C}_{kn}}{\ensuremath{C}_{km}}\right)\frac{\mathrm{d}\ensuremath{\protect\bow{a}}_{k;n}}{\mathrm{d}t},
    \label{eq:MGElocalSplitAfit}
\end{multline}
where $N_{\text{F};m}$ is the number of facets in $G$ containing the edge $\ensuremath{L}_{m}$.
Note that $N_{\text{F};m}=4$ for a regular hexahedral grid.
Let us now introduce the auxiliary definitions
\begin{equation*}
    V_{e;m}:=\sum_{k\in\mathcal{A}^{m}}\sum_{n\in\mathcal{L}_{k}^{m,0}}V_{e;mkn},\quad
    V_{\text{c};m}:=\sum_{k\in\mathcal{A}^{m}}\sum_{n\in\mathcal{L}_{k;\text{c}}^{m}}V_{\text{c};mkn},\quad
    V_{\text{t};m}:=\sum_{k\in\mathcal{A}^{m}}\sum_{n\in\mathcal{L}_{k;\text{t}}^{m}}V_{\text{t};mkn},
\end{equation*}
where
\begin{align*}
    V_{e;mkn}&:=\frac{1}{N_{\text{F};m}}\left(\frac{\ensuremath{C}_{kn}}{\ensuremath{C}_{km}}\right)\ensuremath{\protect\bow{e}}_{k;n},\\
    V_{\text{c};mkn}&:=\frac{1}{N_{\text{F};m}}\left(\frac{\ensuremath{C}_{kn}}{\ensuremath{C}_{km}}\right)\frac{\mathrm{d}\ensuremath{\protect\bow{a}}_{k;n}}{\mathrm{d}t},\\
    V_{\text{t};mkn}&:=\frac{1}{N_{\text{F};m}}\left(\frac{\ensuremath{C}_{kn}}{\ensuremath{C}_{km}}\right)\frac{\mathrm{d}\ensuremath{\protect\bow{a}}_{k;n}}{\mathrm{d}t},
\end{align*}
which allow to express \eqref{eq:MGElocalSplitAfit} as
\begin{equation}
    \ensuremath{\protect\bow{e}}_{m}+V_{e;m}+V_{\text{c};m}+V_{\text{t};m}=0.
    \label{eq:MGElocalSplitAfitCompact}
\end{equation}
Similarly, we expand the left-hand side of~\eqref{eq:MGElocalRevisitAmpere} as 
\begin{multline*}
    \sum_{\protect\tilde{k}\in\protect\widetilde{\mathcal{L}}_{\protect\tilde{m}}^{m}}\sum_{n\in\mathcal{L}_{k}^{m}}\ensuremath{\protect\widetilde{C}}_{\protect\tilde{m}\protect\tilde{k}}M_{\nu;\protect\tilde{k} k}\ensuremath{C}_{kn}\ensuremath{\protect\bow{a}}_{k;n}\\
    =\left(\sum_{\protect\tilde{k}\in\protect\widetilde{\mathcal{L}}_{\protect\tilde{m}}^{m}}\ensuremath{\protect\widetilde{C}}_{\protect\tilde{m}\protect\tilde{k}}M_{\nu;\protect\tilde{k} k}\ensuremath{C}_{km}\right)\ensuremath{\protect\bow{a}}_{m}
     +\sum_{\protect\tilde{k}\in\protect\widetilde{\mathcal{L}}_{\protect\tilde{m}}^{m}}\sum_{n\in\mathcal{L}_{k}^{m,0}}\ensuremath{\protect\widetilde{C}}_{\protect\tilde{m}\protect\tilde{k}}M_{\nu;\protect\tilde{k} k}\ensuremath{C}_{kn}\ensuremath{\protect\bow{a}}_{k;n},
\end{multline*}
which, upon substitution in~\eqref{eq:MGElocalRevisitAmpere} and by splitting $\lbrace\ensuremath{\protect\bow{a}}_{k;n}\rbrace$ into tree and cotree components, yields
\begin{multline}
    \ensuremath{\protect\bow{a}}_{m}+\frac{\sum_{\protect\tilde{k}\in\protect\widetilde{\mathcal{L}}_{\protect\tilde{m}}^{m}}\sum_{n\in\mathcal{L}_{k;\text{c}}^{m,0}}\ensuremath{\protect\widetilde{C}}_{\protect\tilde{m}\protect\tilde{k}}M_{\nu;\protect\tilde{k} k}\ensuremath{C}_{kn}\ensuremath{\protect\bow{a}}_{k;n}}{M_{\nu;\protect\tilde{m} m}^{\Sigma}}
    +\frac{\sum_{\protect\tilde{k}\in\protect\widetilde{\mathcal{L}}_{\protect\tilde{m}}^{m}}\sum_{n\in\mathcal{L}_{k;\text{t}}^{m,0}}\ensuremath{\protect\widetilde{C}}_{\protect\tilde{m}\protect\tilde{k}}M_{\nu;\protect\tilde{k} k}\ensuremath{C}_{kn}\ensuremath{\protect\bow{a}}_{k;n}}{M_{\nu;\protect\tilde{m} m}^{\Sigma}}\\
    -\frac{M_{\varepsilon;\protect\tilde{m} m}}{M_{\nu;\protect\tilde{m} m}^{\Sigma}}\frac{\mathrm{d}\ensuremath{\protect\bow{e}}_{m}}{\mathrm{d}t}-\frac{M_{\sigma;\protect\tilde{m} m}}{M_{\nu;\protect\tilde{m} m}^{\Sigma}}\ensuremath{\protect\bow{e}}_{m}-\frac{\ensuremath{\protect\bbow{j}}_{\text{i};\protect\tilde{m}}}{M_{\nu;\protect\tilde{m} m}^{\Sigma}}=0,
    \label{eq:MGEampereAfit2cir} 
\end{multline}
where
\begin{equation*}
    M_{\nu;\protect\tilde{m} m}^{\Sigma}:=\sum_{\protect\tilde{k}\in\protect\widetilde{\mathcal{L}}_{\protect\tilde{m}}^{m}}\ensuremath{\protect\widetilde{C}}_{\protect\tilde{m}\protect\tilde{k}}M_{\nu;\protect\tilde{k} k}\ensuremath{C}_{km}=\sum_{\protect\tilde{k}\in\protect\widetilde{\mathcal{L}}_{\protect\tilde{m}}^{m}}M_{\nu;\protect\tilde{k} k}
\end{equation*}
since $\ensuremath{\protect\widetilde{C}}_{\protect\tilde{m}\protect\tilde{k}}=\ensuremath{C}_{km}$.
By means of the auxiliary definitions
\begin{alignat*}{3}
    I_{R;m}&:=\frac{M_{\sigma;\protect\tilde{m} m}}{M_{\nu;\protect\tilde{m} m}^{\Sigma}}\ensuremath{\protect\bow{e}}_{m},\qquad
    &&I_{\text{c};m}&&:=\sum_{\protect\tilde{k}\in\protect\widetilde{\mathcal{L}}_{\protect\tilde{m}}^{m}}\sum_{n\in\mathcal{L}_{k;\text{c}}^{m,0}}I_{\text{c};m\protect\tilde{k} n},\\
    I_{C;m}&:=\frac{M_{\varepsilon;\protect\tilde{m} m}}{M_{\nu;\protect\tilde{m} m}^{\Sigma}}\frac{\mathrm{d}\ensuremath{\protect\bow{e}}_{m}}{\mathrm{d}t},
    &&I_{\text{t};m}&&:=\sum_{\protect\tilde{k}\in\protect\widetilde{\mathcal{L}}_{\protect\tilde{m}}^{m}}\sum_{n\in\mathcal{L}_{k;\text{t}}^{m,0}}I_{\text{t};m\protect\tilde{k} n},\\
    I_{\text{i};m}&:=\frac{\ensuremath{\protect\bbow{j}}_{\text{i};\protect\tilde{m}}}{M_{\nu;\protect\tilde{m} m}^{\Sigma}},\\
\end{alignat*}
where
\begin{equation*}
    I_{\text{c};m\protect\tilde{k} n}:=\frac{\ensuremath{\protect\widetilde{C}}_{\protect\tilde{m}\protect\tilde{k}}M_{\nu;\protect\tilde{k} k}\ensuremath{C}_{kn}\ensuremath{\protect\bow{a}}_{k;n}}{M_{\nu;\protect\tilde{m} m}^{\Sigma}},\quad
    I_{\text{t};m\protect\tilde{k} n}:=\frac{\ensuremath{\protect\widetilde{C}}_{\protect\tilde{m}\protect\tilde{k}}M_{\nu;\protect\tilde{k} k}\ensuremath{C}_{kn}\ensuremath{\protect\bow{a}}_{k;n}}{M_{\nu;\protect\tilde{m} m}^{\Sigma}},
\end{equation*}
we express~\eqref{eq:MGEampereAfit2cir} as
\begin{equation}
    \ensuremath{\protect\bow{a}}_{m}+I_{\text{c};m}+I_{\text{t};m}-I_{C;m}-I_{R;m}-I_{\text{i};m}=0.
    \label{eq:MGEampereAfit2cirCompact}
\end{equation}

\begin{figure}[t]
    \centering
    \subfloat[\label{fig:circuitStampAfitCotree}]{\includegraphics[width=\columnwidth]{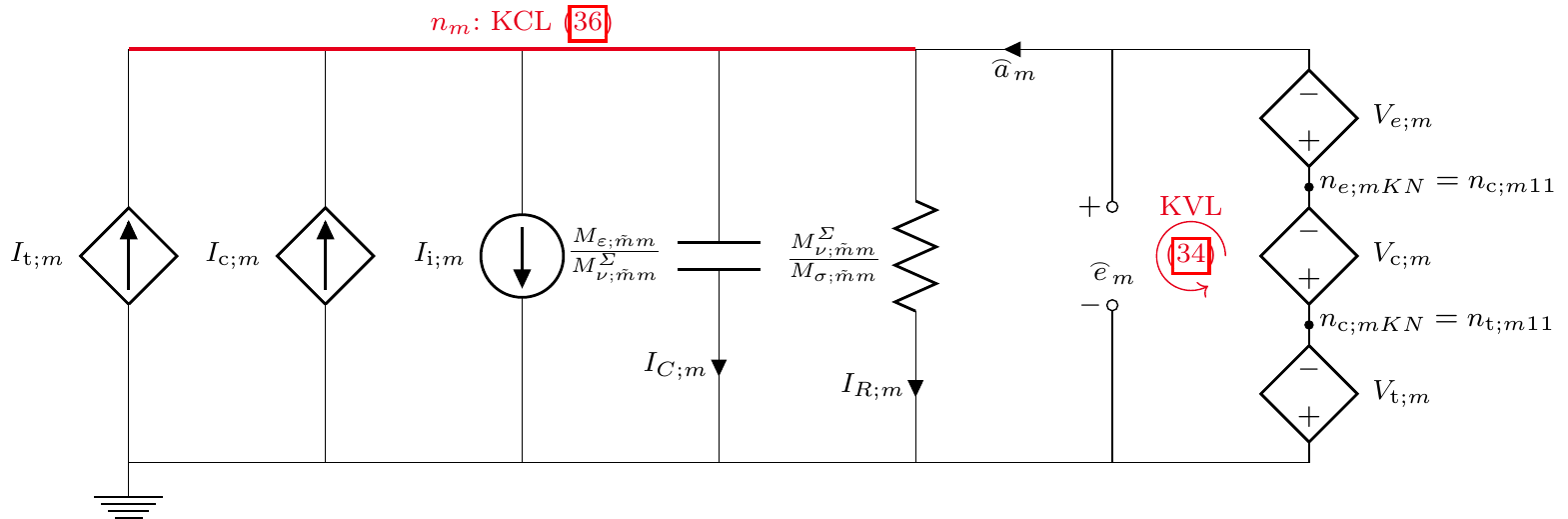}
 }\\
    \subfloat[\label{fig:circuitStampAfitTree}]{\includegraphics[width=\columnwidth]{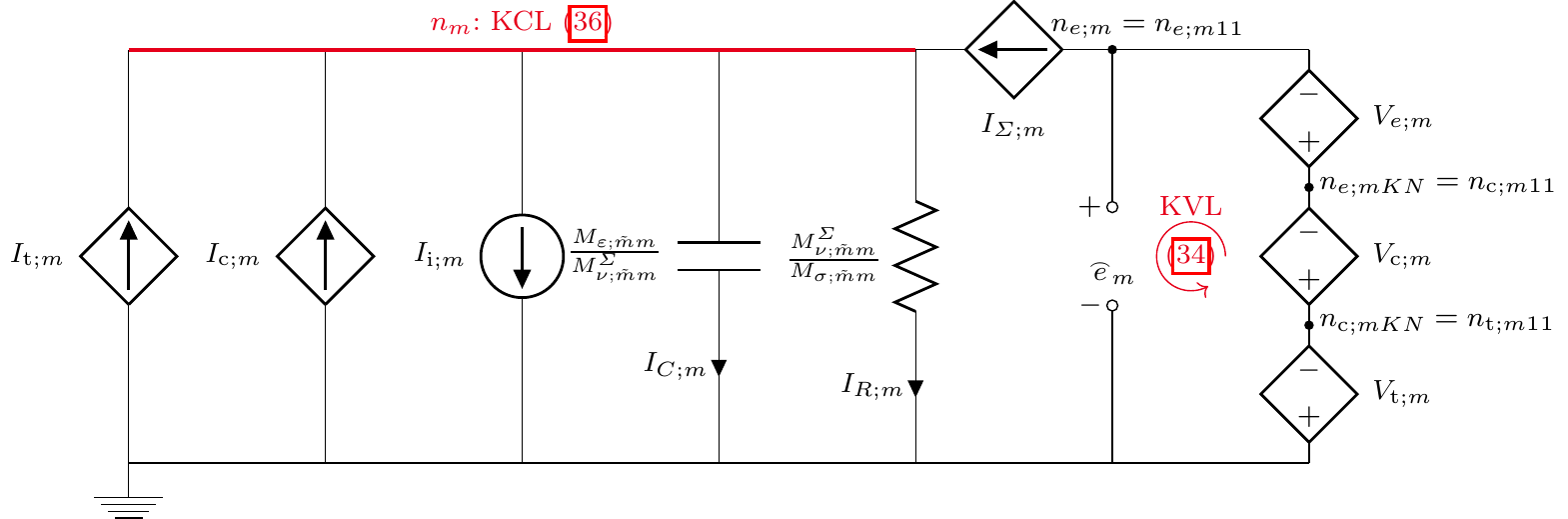}
 }
    \caption{Circuit stamp of (a) a cotree edge and (b) a tree edge.}
\end{figure}

We observe that~\eqref{eq:MGElocalSplitAfitCompact} and~\eqref{eq:MGEampereAfit2cirCompact} can be interpreted as the \gls*{KVL} and \gls*{KCL} of an arbitrary primal edge $\ensuremath{L}_{m}$ provided that $\ensuremath{\protect\bow{e}}_{m}$ and $\ensuremath{\protect\bow{a}}_{m}$ represent the sought voltage and current, respectively.
As a matter of fact, we observe that 
\begin{itemize}
    \item $\ensuremath{\protect\bow{e}}_{m}$ has unit of \si{V} and represents the voltage drop along $\ensuremath{L}_{m}$ that is either in the tree or cotree set. 
    \item $\ensuremath{\protect\bow{a}}_{m}$ has unit of \textsc{Weber}\xspace (\si{Wb}) and represents the current along $\ensuremath{L}_{m}$. If $\ensuremath{L}_{m}\in G_{\text{c}}$, this current is a degree of freedom. Otherwise, it is modelled by a \gls*{CCCS} as stated in~\eqref{eq:afittm}. For compact notation, we label this \gls*{CCCS} as ${I_{\Sigma;m}:=\sum_{n\in G_{\text{c}}}I_{\Sigma;mn}}$, where ${I_{\Sigma;mn}:=-E_{\text{tc};mn}\ensuremath{\protect\bow{a}}_{\text{c};n}}$.
    \item $M_{\nu;\protect\tilde{m} m}^{\Sigma}$ has unit of \si{H\tothe{-1}} and is expected to be positive since $M_{\nu;\protect\tilde{k} k}>0$ for all materials. This term scales the displacement, conduction and impressed current as seen in~\eqref{eq:MGEampereAfit2cir}.
\end{itemize}
With these observations, we may depict~\eqref{eq:MGElocalSplitAfit} and~\eqref{eq:MGEampereAfit2cir} by means of the circuit stamps of Figure~\ref{fig:circuitStampAfitCotree} and Figure~\ref{fig:circuitStampAfitTree} for a cotree and tree edge, respectively.
As we can see therein, $\ensuremath{\protect\bow{a}}_{m}$ is regarded as the electric current along the primal edge $\ensuremath{L}_{m}$, while the voltage drop between its terminals is established by the \gls*{VCVS} $V_{e;m}$ and the \glspl*{CCVS} $V_{\text{c};m}$ and $V_{\text{t};m}$ that mediate the interaction with neighbouring edges.
When the edge $\ensuremath{L}_{m}$ belongs to the tree $G_{\text{t}}$, then the current $\ensuremath{\protect\bow{a}}_{m}$ is modelled by the \gls*{CCCS} $I_{\Sigma;m}$ as demanded by~\eqref{eq:afittm}.
Otherwise, for cotree edges, $\ensuremath{\protect\bow{a}}_{m}$ is a degree of freedom.
The current $\ensuremath{\protect\bow{a}}_{m}$ is then split into several branches where a resistance, capacitance, the impressed current source, and \glspl*{CCCS} are connected.
The concatenation of these fundamental stamps forms the electric circuit representing the magnetic vector potential formulation of the electromagnetic problem. 

In Algorithm~\ref{alg:netlistEM2}, we show the pseudocode to generate the netlist of the circuit stamps depicted in Figures~\ref{fig:circuitStampAfitCotree} and \ref{fig:circuitStampAfitTree}. 
The iteration over all edges $\ensuremath{L}_{m}$ in the grid is done in the outermost loop.
In each iteration, one resistor and capacitor connecting the node $n_{m}$ to ground (gnd) must be added to the netlist (lines~2 and 3).
If an impressed current is present at $\ensuremath{L}_{m}$, an independent current source with value $I_{\text{i};m}$ is added between $n_{m}$ and gnd (lines~4--6).
For every tree edge, the currents $I_{\Sigma;mn}$ from the cotree edges in the neighbourhood are modelled by a parallel connection of \glspl*{CCCS} between $n_{e;m}$ and $n_{m}$ (lines~7--11).
Using a double loop, the inductive currents $I_{\text{t};m\protect\tilde{k} n}$ and $I_{\text{c};m\protect\tilde{k} n}$ from the tree and cotree branches, respectively, are added as a parallel connection of \glspl*{CCCS} between $n_{m}$ and gnd (lines~12--19).
The controlling current is given by the current through device FIc$n11$ and the gain is ${g_{m\protect\tilde{k} n}^{\text{I}}=\ensuremath{\protect\widetilde{C}}_{\protect\tilde{m}\protect\tilde{k}}M_{\nu;\protect\tilde{k} k}\ensuremath{C}_{kn}(M_{\nu;\protect\tilde{m} m}^{\Sigma})^{-1}}$.
Finally, the voltages $V_{e;mkn}$, $V_{\text{c};mkn}$ and $V_{\text{t};mkn}$ are added as a series connection between $n_{e;m}$ and gnd with intermediate nodes indexed by $k$, $n$ and $\hat{n}=n+1$ (lines~20--30).
For the voltage $V_{e;mkn}$, the voltage between node n$n$ and gnd controls a \gls*{VCVS} with a gain of $g_{mkn}^{\text{V}}=N_{\text{F};m}^{-1}\ensuremath{C}_{kn}\ensuremath{C}_{km}^{-1}$
On the other hand, the voltages $V_{\text{c};mkn}$ and $V_{\text{t};mkn}$ are added as behavioural sources using DDT as the SPICE\xspace syntax for time derivatives and a gain of $g_{mkn}^{\text{I}}=N_{\text{F};m}^{-1}\ensuremath{C}_{kn}\ensuremath{C}_{km}^{-1}$.
Note that for a cotree edge, the node $n_{e;m}$ coincides with the node $n_{m}$.

\begin{algorithm}
    \caption{Electromagnetic SPICE netlist generation based on the E-A formulation.}
    \label{alg:netlistEM2}
    \begin{algorithmic}[1]
        \For{edge $\ensuremath{L}_{m}\in G$}
            \State write R$m$\quad{}n$m$\quad{}gnd\quad{}$M_{\nu;\protect\tilde{m} m}^{\Sigma}M_{\sigma;\protect\tilde{m} m}^{-1}$
            \State write C$m$\quad{}n$m$\quad{}gnd\quad{}$M_{\varepsilon;\protect\tilde{m} m}(M_{\nu;\protect\tilde{m} m}^{\Sigma})^{-1}$
            \If{an impressed current is placed on $\ensuremath{L}_{m}$}
                \State write I$m$\quad{}n$m$\quad{}gnd\quad{}$I_{\text{i};m}$ 
            \EndIf
            \If{$\ensuremath{L}_{m}\in G_{\text{t}}$}
                \For{edge $\ensuremath{L}_{n}\in G_{\text{c}}$}
                    \State write FIsum$mn$\quad{}ne$m$\quad{}n$m$\quad{}FIc$n11$\quad{}$-E_{\text{tc};mn}$
                \EndFor
            \EndIf
            \For{edge $\ensuremath{\protect\widetilde{L}}_{k}\in\protect\widetilde{\mathcal{L}}_{\protect\tilde{m}}^{m}$}
                \For{edge $\ensuremath{L}_{n}\in\mathcal{L}_{k;\text{c}}^{m,0}$}
                    \State write FIc$mkn$\quad{}gnd\quad{}n$m$\quad{}FIc$n11$\quad{}$g_{m\protect\tilde{k} n}^{\text{I}}$
                \EndFor
                \For{edge $\ensuremath{L}_{n}\in\mathcal{L}_{k;\text{t}}^{m,0}$}
                    \State write FIt$mkn$\quad{}gnd\quad{}n$m$\quad{}FIc$n11$\quad{}$g_{m\protect\tilde{k} n}^{\text{I}}$
                \EndFor
            \EndFor
            \For{edge $\ensuremath{A}_l\in\mathcal{A}^{m}$ with $k=1,\dots,K$}
                \For{edge $\ensuremath{L}_n\in\mathcal{L}_{k}^{m,0}$ with $n=1,\dots,N-1$}
                    \State write EVe$mkn$\;\;{}ne$mk\hat{n}$\;\;{}ne$mkn$\;\;{}n$n$\;\;{}gnd\;\;{}$g_{mkn}^{\text{V}}$
                \EndFor
                \For{edge $\ensuremath{L}_n\in\mathcal{L}_{k;\text{c}}^{m}$ with $n=1,\dots,N-1$}
                    \State write BVc$mkn$\quad{}nc$mk\hat{n}$\quad{}nc$mkn$\quad{}\dots\\\hspace{17em}V=$g_{mkn}^{\text{I}}\text{DDT}(\ensuremath{\protect\bow{a}}_{k;n})$
                \EndFor
                \For{edge $\ensuremath{L}_n\in\mathcal{L}_{k;\text{t}}^{m}$ with $n=1,\dots,N-1$}
                    \State write BVt$mkn$\quad{}nt$mk\hat{n}$\quad{}nt$mkn$\quad{}\dots\\\hspace{17em}V=$g_{mkn}^{\text{I}}\text{DDT}(\ensuremath{\protect\bow{a}}_{k;n})$
                \EndFor
            \EndFor
        \EndFor
    \end{algorithmic}
\end{algorithm}

We remark that a similar analysis on~\eqref{eq:MaxwellContInt} by considering only magnetic conductivities and sources instead of electric ones is also possible.
Thus, with the auxiliary electric potential $\ensuremath{\mathbf{D}}=\nabla \times\ensuremath{\mathbf{F}}$ and $\nabla \cdot\ensuremath{\mathbf{F}}=a$, where $a$ is an arbitrary gauging function, circuit stamps which are dual to those shown in Figure~\ref{fig:circuitStampAfitCotree} and Figure~\ref{fig:circuitStampAfitTree} can be found.
In these dual stamps, $\ensuremath{\protect\bow{h}}_{m}$ represents an electric current while $\ensuremath{\protect\bow{f}}_{m}$, namely the grid counterpart of $\ensuremath{\mathbf{F}}$, would be regarded as a voltage drop.
Finally, if both electric and magnetic sources were present, then the entire circuit would consist of the aggregate of interacting primal and dual stamps.

\subsection{Absorbing Boundary Conditions}
\label{subsec:theoryABC}

In many electromagnetic field simulation set-ups, the computational domain must be bounded.
To simulate free wave propagation, one must impose according conditions for the fields at the boundaries of the domain.
These conditions are known as \glspl*{ABC}~\cite{Engquist_1977aa} and aim at minimising (ideally cancelling) unphysical incoming reflections. 
At the boundary of the domain, a distinction is made between normal and tangential components and between longitudinal and transverse derivatives\footnote{In this regard, the longitudinal (transverse) derivative coincides with the normal (tangent) derivative at the boundary.}. 

For starters, let $\partial_{r}$, $\nabla_{\text{t}}$ and $\partial_{t}$ denote the longitudinal, transversal and temporal derivative operators, respectively.
Furthermore, within the context of this analysis, we define $\nabla:=\left(\partial^{2}_{r}+\nabla_{\text{t}}^{2}\right)^{1/2}$.
We start the realisation of circuit stamps for \glspl*{ABC} by considering the time-domain wave equation for the electric field $\ensuremath{\mathbf{E}}$ in a homogeneous and isotropic medium\footnote{Although we restrict ourselves to homogeneous and isotropic media, the analysis can also be extended to more general cases.}, viz.
\begin{equation}
    \nabla^{2}\ensuremath{\mathbf{E}}=\mu\varepsilon\frac{\partial^2\ensuremath{\mathbf{E}}}{\partial t^2}.    
    \label{eq:waveEqTime}  
\end{equation}
Equation~\eqref{eq:waveEqTime} can be expanded in terms of the so-called travelling wave operators as 
\begin{equation}
    \left(\nabla+\sqrt{\mu\varepsilon}\frac{\partial}{\partial{}t}\,\right)\left(\nabla-\sqrt{\mu\varepsilon}\frac{\partial}{\partial{}t}\,\right)\ensuremath{\mathbf{E}}=\mathbf{0}. 
    \label{eq:waveEqTimeExpand} 
\end{equation}
Above, we observe that in an arbitrary point in space, the field $\ensuremath{\mathbf{E}}$ can in general be considered as the superposition of an inward and outward travelling wave $\ensuremath{\mathbf{E}}^{+}$ and $\ensuremath{\mathbf{E}}^{-}$, respectively, viz.
\begin{equation*}
    \ensuremath{\mathbf{E}}=\ensuremath{\mathbf{E}}^{+}+\ensuremath{\mathbf{E}}^{-}, 
\end{equation*}
which upon substitution in \eqref{eq:waveEqTimeExpand} straightforwardly leads to the following set of travelling wave equations,
\begin{equation*}
    \left(\nabla+\sqrt{\mu\varepsilon}\frac{\partial}{\partial{}t}\,\right)\left(\nabla-\sqrt{\mu\varepsilon}\frac{\partial}{\partial{}t}\,\right)\ensuremath{\mathbf{E}}^{+}=\mathbf{0},\quad
    \left(\nabla+\sqrt{\mu\varepsilon}\frac{\partial}{\partial{}t}\,\right)\left(\nabla-\sqrt{\mu\varepsilon}\frac{\partial}{\partial{}t}\,\right)\ensuremath{\mathbf{E}}^{-}=\mathbf{0},
\end{equation*}
inasmuch as both wave components $\ensuremath{\mathbf{E}}^{\pm}$ are linearly independent.
Owing to their definition, the wave components $\ensuremath{\mathbf{E}}^{\pm}$ satisfy independently and simultaneously the following\footnote{This property can be easily visualised if we consider, for a moment, one-dimensional wave propagation along the $x$-axis. In this circumstance, we have that $\nabla\equiv\partial_{x}$ and $e^{\phi\left(x\pm\nu t\right)}$ representing backward and forward travelling waves
at speed $\nu=1/\sqrt{\mu\varepsilon}$ with $\phi$ an arbitrary function.},
\begin{equation}
    \left(\nabla\mp\sqrt{\mu\varepsilon}\frac{\partial}{\partial{}t}\,\right)\ensuremath{\mathbf{E}}^{\pm}=\mathbf{0}, 
    \label{eq:inoutwardsim} 
\end{equation}
which tells us explicitly that the outward (inward) travelling wave operator cancels out the inward (outward) travelling wave at any point of interest.  

Now, let us get back to~\eqref{eq:waveEqTimeExpand} to further expand the operators therein to obtain
\begin{equation}
    \left(\frac{\partial}{\partial r}+\frac{\sqrt{\mu\varepsilon}\frac{\partial}{\partial{}t}\,}{\sqrt{1+\frac{\nabla_{\text{t}}^{2}}{\frac{\partial^2}{\partial r^2}}}}\right)
    \left(\frac{\partial}{\partial r}-\frac{\sqrt{\mu\varepsilon}\frac{\partial}{\partial{}t}\,}{\sqrt{1+\frac{\nabla_{\text{t}}^{2}}{\frac{\partial^2}{\partial r^2}}}}\right)\ensuremath{\mathbf{E}}=\mathbf{0}.
    \label{eq:waveEqTimeExpandMore} 
\end{equation}
As we can see, the above wave operators entail the calculation of the inverse of $\sqrt{1+\nabla_{\text{t}}^{2}/\partial_{r}^{2}}$, which generally translates into a global integral operator~\cite{Zhukovsky_2014aa}. 
In principle, the expansion of this integral operator around the observation point yields the exact explicit representation of the travelling wave operators in \eqref{eq:waveEqTimeExpandMore}.
However, this approach is contrary to the idea of realising simple and efficient \glspl*{ABC} for circuit simulations.

As a remedy, we may expand $1/\sqrt{1+\nabla_{\text{t}}^{2}/\partial_{r}^{2}}$ as in a Taylor series to yield
\begin{equation}
    \begin{split}
        &\left(\frac{\partial}{\partial r}+\sqrt{\mu\varepsilon}\frac{\partial}{\partial{}t}-\frac{1}{2}\sqrt{\mu\varepsilon}\frac{\partial}{\partial{}t}\,\frac{\nabla_{\text{t}}^{2}}{\frac{\partial^2}{\partial r^2}}+\ensuremath{{\mathcal{O}}}\left(\nabla_{\text{t}}^{4}\right)\right)\\
        &\left(\frac{\partial}{\partial r}-\sqrt{\mu\varepsilon}\frac{\partial}{\partial{}t}+\frac{1}{2}\sqrt{\mu\varepsilon}\frac{\partial}{\partial{}t}\,\frac{\nabla_{\text{t}}^{2}}{\frac{\partial^2}{\partial r^2}}+\ensuremath{{\mathcal{O}}}\left(\nabla_{\text{t}}^{4}\right)\right)\ensuremath{\mathbf{E}}=\mathbf{0}.
    \end{split}
    \label{eq:waveEqTimeExpandMoreSimple}\end{equation}
The above equation holds for any component of $\ensuremath{\mathbf{E}}$.
Furthermore, if the field $\ensuremath{\mathbf{E}}$ propagates in~\emph{free} space, we may assume that the operation $\nabla^{2}_{\text{t}}$ is negligible in the neighbourhood of an observation point within the spherical wavefront.
Hence, we may rewrite~\eqref{eq:waveEqTimeExpandMoreSimple} in terms of simplified inward and outward travelling wave propagators as 
\begin{equation}
    \left(\frac{\partial}{\partial r}+\sqrt{\mu\varepsilon}\frac{\partial}{\partial{}t}\,\right)
    \left(\frac{\partial}{\partial r}-\sqrt{\mu\varepsilon}\frac{\partial}{\partial{}t}\,\right)\ensuremath{\mathbf{E}}=\mathbf{0},
    \label{eq:waveEqTimeExpandMoreSimpleMore} 
\end{equation}
since the general expression of $\ensuremath{\mathbf{E}}$ in~\eqref{eq:waveEqTimeExpandMoreSimpleMore} admits the superposition of inward and outward travelling waves $\ensuremath{\mathbf{E}}^{+}$ and $\ensuremath{\mathbf{E}}^{-}$.
Then, to minimise unwanted incoming reflections at a certain boundary given by $r=r_{\text{B}}$, we may impose on $\ensuremath{\mathbf{E}}$ the condition
\begin{equation}
    \left.\left(\frac{\partial}{\partial r}-\sqrt{\mu\varepsilon}\frac{\partial}{\partial{}t}\,\right)\ensuremath{\mathbf{E}}\right|_{r=r_{\text{B}}}=\mathbf{0},
    \label{eq:EngquistMadja} 
\end{equation}
in agreement with~\eqref{eq:inoutwardsim}.
The condition given by~\eqref{eq:EngquistMadja} is a first-order \gls*{ABC} known as Engquist-Madja condition~\cite{Engquist_1977aa}.
It is a local condition because it is evaluated pointwise taking only into account the wave component propagating perpendicularly to the boundary.
Thereby, the condition in~\eqref{eq:EngquistMadja} states that a practical transparent boundary at $r=r_{\text{B}}$ can be realised if it is guaranteed that the phase-amplitude of the field on the boundary at a certain time $t_{\text{B}}$ is equal to that one the field had at some previous instant $t_{\text{B}}-\Delta t$ at a point $r=r_{\text{B}}-\Delta t/\sqrt{\mu\varepsilon}$.  
The practical relevance of~\eqref{eq:EngquistMadja} comes from its simplicity.

Let us now interpret~\eqref{eq:EngquistMadja} within the context of the \textsc{Maxwell}\xspace grid equations in order to realise circuit stamps associated with the \glspl*{ABC} of~\eqref{eq:EngquistMadja}.
To this end, let us consider the grid wave equation for the electric grid voltage $\ensuremath{\protect\bow{\mathrm{\mathbf{e}}}}\xspace$, which for time-independent constitutive parameters and non-conducting source-free regions can be obtained from the grid curl equations of~\eqref{eq:MGEmagCharge}, viz.
\begin{equation*}
    \ensuremath{\protect\widetilde{\mathbf{C}}}\xspace\ensuremath{\mathbf{M}_{\nu}}\xspace\ensuremath{\mathbf{C}}\xspace\ensuremath{\protect\bow{\mathrm{\mathbf{e}}}}\xspace=-\ensuremath{\mathbf{M}_{\varepsilon}}\xspace\frac{\mathrm{d}^2}{\mathrm{d}t^2}\ensuremath{\protect\bow{\mathrm{\mathbf{e}}}}\xspace.
\end{equation*}
With the structure of the matrices \ensuremath{\mathbf{C}}\xspace, \ensuremath{\mathbf{M}_{\varepsilon}}\xspace and \ensuremath{\mathbf{M}_{\nu}}\xspace given by \eqref{eq:fitMatBlockStruct} and the subdivision $\ensuremath{\protect\bow{\mathrm{\mathbf{e}}}}\xspace=\left(\ensuremath{\protect\bow{\mathrm{\mathbf{e}}}}\xspace_{x},\ensuremath{\protect\bow{\mathrm{\mathbf{e}}}}\xspace_{y},\ensuremath{\protect\bow{\mathrm{\mathbf{e}}}}\xspace_{z}\right)^{\top}$, the expression
\begin{equation}
    \left(\ensuremath{\mathbf{P}}\xspace_{z}^{\top}\ensuremath{ \mathbf{M} }_{\nu;y}\ensuremath{\mathbf{P}}\xspace_{z}
    +\ensuremath{\mathbf{P}}\xspace_{y}^{\top}\ensuremath{ \mathbf{M} }_{\nu;z}\ensuremath{\mathbf{P}}\xspace_{y}\right)\ensuremath{\protect\bow{\mathrm{\mathbf{e}}}}\xspace_{x}
    -\ensuremath{\mathbf{P}}\xspace_{y}^{\top}\ensuremath{ \mathbf{M} }_{\nu;z}\ensuremath{\mathbf{P}}\xspace_{x}\ensuremath{\protect\bow{\mathrm{\mathbf{e}}}}\xspace_{y}
    -\ensuremath{\mathbf{P}}\xspace_{z}^{\top}\ensuremath{ \mathbf{M} }_{\nu;y}\ensuremath{\mathbf{P}}\xspace_{x}\ensuremath{\protect\bow{\mathrm{\mathbf{e}}}}\xspace_{z}
    =-\ensuremath{ \mathbf{M} }_{\varepsilon;x}\frac{\mathrm{d}^2}{\mathrm{d}t^2}\ensuremath{\protect\bow{\mathrm{\mathbf{e}}}}\xspace_{x}
    \label{eq:waveEqMGEoneComp}
\end{equation}
for the component $\ensuremath{\protect\bow{\mathrm{\mathbf{e}}}}\xspace_{x}$ is obtained.
Similar expressions can be also obtained for the other two components $\ensuremath{\protect\bow{\mathrm{\mathbf{e}}}}\xspace_{y}$ and $\ensuremath{\protect\bow{\mathrm{\mathbf{e}}}}\xspace_{z}$.

The grid counterpart of the \gls*{ABC} in~\eqref{eq:EngquistMadja} is obtained from~\eqref{eq:waveEqMGEoneComp} by extracting the grid wave equation associated with a generic primal edge $\ensuremath{L}_{x;m}\in G$ oriented along the $x$-direction.
This yields
\begin{multline}
         \sum_{\protect\tilde{k}}\sum_{n}\ensuremath{P}^{\top}_{z;\protect\tilde{m}\protect\tilde{k}}M_{\nu;y;\protect\tilde{k} k}\ensuremath{P}_{z;kn}\ensuremath{\protect\bow{e}}_{x;n}
        +\sum_{\protect\tilde{k}}\sum_{n}\ensuremath{P}^{\top}_{y;\protect\tilde{m}\protect\tilde{k}}M_{\nu;z;\protect\tilde{k} k}\ensuremath{P}_{y;kn}\ensuremath{\protect\bow{e}}_{x;n}\\
        -\sum_{\protect\tilde{k}}\sum_{n}\ensuremath{P}^{\top}_{y;\protect\tilde{m}\protect\tilde{k}}M_{\nu;z;\protect\tilde{k} k}\ensuremath{P}_{x;kn}\ensuremath{\protect\bow{e}}_{y;n}
        -\sum_{\protect\tilde{k}}\sum_{n}\ensuremath{P}^{\top}_{z;\protect\tilde{m}\protect\tilde{k}}M_{\nu;y;\protect\tilde{k} k}\ensuremath{P}_{x;kn}\ensuremath{\protect\bow{e}}_{z;n}
        =-M_{\varepsilon;x;\protect\tilde{m} m}\frac{\mathrm{d}^2}{\mathrm{d}t^2}\ensuremath{\protect\bow{e}}_{x;m},   
    \label{eq:gridwaveEq}
\end{multline}
with $n$ and $\protect\tilde{k}$ spanning over the corresponding primal and dual edges oriented along the indicated directions.
We make the following observations upon the above grid wave equation.
\begin{itemize}
\item Two main grid wave components contribute to the time variation of $\ensuremath{\protect\bow{e}}_{x;m}$ along $\ensuremath{L}_{x;m}$. The first of these grid wave components propagates along the $z$-direction and stems from the spatial variation of $\ensuremath{\protect\bow{\mathrm{\mathbf{e}}}}\xspace_{x}$ along this direction as stated by the grid derivatives in $\sum_{\protect\tilde{k}}\sum_{n}\ensuremath{P}^{\top}_{z;\protect\tilde{m}\protect\tilde{k}}M_{\nu;y;\protect\tilde{k} k} \ensuremath{P}_{z;kn}\ensuremath{\protect\bow{e}}_{x;n}$. The second one propagates along the $y$-direction and stems from the spatial variation of $\ensuremath{\protect\bow{\mathrm{\mathbf{e}}}}\xspace_{x}$ along this direction as stated by $\sum_{\protect\tilde{k}}\sum_{n}\ensuremath{P}^{\top}_{y;\protect\tilde{m}\protect\tilde{k}}M_{\nu;z;\protect\tilde{k} k}\ensuremath{P}_{y;kn}\ensuremath{\protect\bow{e}}_{x;n}$.
\item Two secondary grid wave components contribute to the time variation of $\ensuremath{\protect\bow{e}}_{x;m}$ along $\ensuremath{L}_{x;m}$. The first one propagates along the $y$-direction and stems from the spatial variation of $\ensuremath{\protect\bow{\mathrm{\mathbf{e}}}}\xspace_{y}$ along the $x$-direction as stated by $\sum_{\protect\tilde{k}}\sum_{n}\ensuremath{P}^{\top}_{y;\protect\tilde{m}\protect\tilde{k}}M_{\nu;z;\protect\tilde{k} k} \ensuremath{P}_{x;kn}\ensuremath{\protect\bow{e}}_{y;n}$. The second one propagates along the $z$-direction and stems from the spatial variation of $\ensuremath{\protect\bow{\mathrm{\mathbf{e}}}}\xspace_{z}$ along the $x$-direction as stated by $\sum_{\protect\tilde{k}}\sum_{n}\ensuremath{P}^{\top}_{z;\protect\tilde{m}\protect\tilde{k}}M_{\nu;y;\protect\tilde{k} k} \ensuremath{P}_{x;kn}\ensuremath{\protect\bow{e}}_{z;n}$.
\item Owing to both the structure of~\eqref{eq:gridwaveEq} and the grid-like embedding where the propagation takes place, we may apply superposition to treat each propagation direction separately. Therefore, each grid wave component satisfies its own one-dimensional grid wave equation.
\end{itemize}
\begin{figure}
    \centering
    \includegraphics{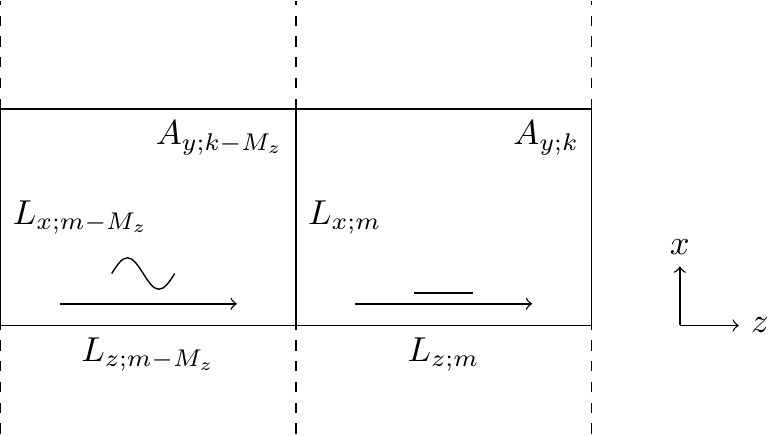}
    \caption{The \gls*{ABC} is imposed at edge $\ensuremath{L}_{x;m}$ in the grid $G$. The impinging propagating grid wave coming from edge $\ensuremath{L}_{x;m-M_{z}}$ is then absorbed. The two facets $\ensuremath{A}_{y;k-M_{z}}$ and $\ensuremath{A}_{y;k}$ that contribute to the averaging of $\overline{M}_{\nu;y;\protect\tilde{k} k}$ are shown.}
    \label{fig:waveBndEdges}
\end{figure}
We construct the grid counterpart of~\eqref{eq:EngquistMadja} by adhering to the same principle that led to it. Namely, we take only into account grid wave propagation perpendicular to the boundary of interest.
Thereby, let us assume that our boundary is located for example at $z=z_{\text{B}}$.
Thus, by invoking the superposition principle, we may write the grid wave equation associated with    the component $\ensuremath{\protect\bow{e}}^{(z)}_{x;m}$ that propagates along the $z$-direction as
\begin{equation}
         \sum_{\protect\tilde{k}}\sum_{n}\ensuremath{P}^{\top}_{z;\protect\tilde{m}\protect\tilde{k}}M_{\nu;y;\protect\tilde{k} k}\ensuremath{P}_{z;kn}\ensuremath{\protect\bow{e}}^{(z)}_{x;n}
        -\sum_{\protect\tilde{k}}\sum_{n}\ensuremath{P}^{\top}_{z;\protect\tilde{m}\protect\tilde{k}}M_{\nu;y;\protect\tilde{k} k}\ensuremath{P}_{x;kn}\ensuremath{\protect\bow{e}}^{(x)}_{z;n}
        =-M_{\varepsilon;x;\protect\tilde{m} m}\frac{\mathrm{d}^2}{\mathrm{d}t^2}\ensuremath{\protect\bow{e}}^{(z)}_{x;m},   
    \label{eq:simpgridwaveEq1}
\end{equation}
where we have used the labels $^{(z)}$ and $^{(x)}$ to explicitly indicate directions of propagation.
By considering that for sufficiently well-refined grids $G$ and $\protect\widetilde{G}$ we may expect $M_{\nu;y;\protect\tilde{k} k}$ to not vary significantly for the two relevant facets $\ensuremath{A}_{y;k}$ (cf. Figure~\ref{fig:waveBndEdges}), we may define an average value $\overline{M}_{\nu;y;k\protect\tilde{k}}$ that enables us to write the leftmost term on the left-hand side of \eqref{eq:simpgridwaveEq1} as
\begin{equation*}
    \sum_{\protect\tilde{k}}\sum_{n}\ensuremath{P}^{\top}_{z;\protect\tilde{m}\protect\tilde{k}}M_{\nu;y;\protect\tilde{k} k}\ensuremath{P}_{z;kn}\ensuremath{\protect\bow{e}}^{(z)}_{x;n}
    =\overline{M}_{\nu;y;\protect\tilde{k} k}\sum_{\protect\tilde{k}}\sum_{n}\ensuremath{P}^{\top}_{z;\protect\tilde{m}\protect\tilde{k}}\ensuremath{P}_{z;kn}\ensuremath{\protect\bow{e}}^{(z)}_{x;n},
\end{equation*}
with a similar result for the rightmost term on the left-hand side of \eqref{eq:simpgridwaveEq1}.
Thus, together with the property $\ensuremath{P}^{\top}_{\xi;nk}=-\ensuremath{\widetilde{P}}_{\xi;nk}$, we may proceed to write \eqref{eq:simpgridwaveEq1} as
\begin{equation*}
     \sum_{\protect\tilde{k}}\sum_{n}\ensuremath{\widetilde{P}}_{z;\protect\tilde{m}\protect\tilde{k}}\ensuremath{P}_{z;kn}\ensuremath{\protect\bow{e}}^{(z)}_{x;n}
    -\sum_{\protect\tilde{k}}\sum_{n}\ensuremath{\widetilde{P}}_{z;\protect\tilde{m}\protect\tilde{k}}\ensuremath{P}_{x;kn}\ensuremath{\protect\bow{e}}^{(x)}_{z;n}
    =\overline{M}_{\nu;y;\protect\tilde{k} k}^{-1}M_{\varepsilon;x;\protect\tilde{m} m}\frac{\mathrm{d}^2}{\mathrm{d}t^2}\ensuremath{\protect\bow{e}}^{(z)}_{x;m}.   
\end{equation*}
Above, the term $\sum_{\protect\tilde{k}}\sum_{n}\ensuremath{\widetilde{P}}_{z;\protect\tilde{m}\protect\tilde{k}}\ensuremath{P}_{x;kn}\ensuremath{\protect\bow{e}}^{(x)}_{z;n}$ is the grid counterpart of the continuous operator in the wave equation of \eqref{eq:waveEqTime}.
Analogously, we may assume that in the vicinity of $\ensuremath{L}_{x;m}$ at the boundary $z=z_{\text{B}}$, the wavefront of the impinging grid wave is plane.
Therefore, we may neglect transverse variations and we arrive at 
\begin{equation*}
     \sum_{\protect\tilde{k}}\sum_{n}\ensuremath{\widetilde{P}}_{z;\protect\tilde{m}\protect\tilde{k}}\ensuremath{P}_{z;kn}\ensuremath{\protect\bow{e}}^{(z)}_{x;n}
    =\overline{M}_{\nu;y;\protect\tilde{k} k}^{-1}M_{\varepsilon;x;\protect\tilde{m} m}\frac{\mathrm{d}^2}{\mathrm{d}t^2}\ensuremath{\protect\bow{e}}^{(z)}_{x;m}.   
\end{equation*}
\begin{figure}
    \centering
    \includegraphics{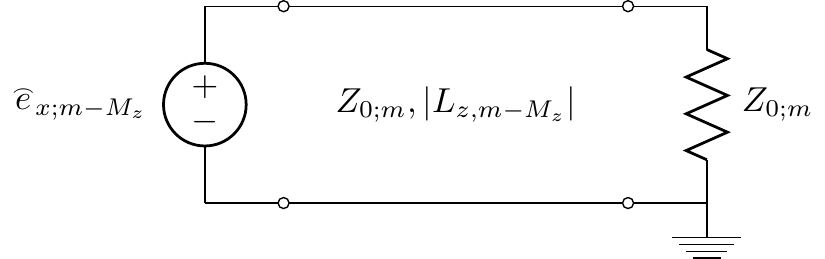}
    \caption{The space separating $\ensuremath{L}_{x;m-M_{z}}$ and $\ensuremath{L}_{x;m}$ can be regarded as a homogeneous transmission line with characteristic impedance $Z_{0;m}$. Implementing \glspl*{ABC} then entails stamping an impedance of value $Z_{0;m}$ for edge $\ensuremath{L}_{x;m}$ in the netlist.}
    \label{fig:circuitABC}
\end{figure}
We observe that $\sum_{\protect\tilde{k}}\sum_{n}\ensuremath{\widetilde{P}}_{z;\protect\tilde{m}\protect\tilde{k}}\ensuremath{P}_{z;kn}\ensuremath{\protect\bow{e}}^{(z)}_{x;n}\equiv\mathrm{d}^{2}_{z}\ensuremath{\protect\bow{e}}^{(z)}_{x;m}$ is formally the second-order grid derivative of the scalar field $\ensuremath{\protect\bow{e}}^{(z)}_{x;m}$ along the $z$-direction.
This directly leads us to the grid counterpart of \eqref{eq:waveEqTime} in the vicinity of $\ensuremath{L}_{x;m}$, viz.
\begin{equation*}
     \frac{\mathrm{d}^2}{\mathrm{d}z^2}\ensuremath{\protect\bow{e}}^{(z)}_{x;m}
    =\overline{M}_{\nu;y;\protect\tilde{k} k}^{-1}M_{\varepsilon;x;\protect\tilde{m} m}\frac{\mathrm{d}^2}{\mathrm{d}t^2}\ensuremath{\protect\bow{e}}^{(z)}_{x;m}.   
\end{equation*}
Again, this equation can be factorised in terms of grid wave propagators similar to~\eqref{eq:waveEqTimeExpandMoreSimpleMore} to yield the grid version of \eqref{eq:EngquistMadja}, viz.
\begin{equation}
    \left.\left(\frac{\mathrm{d}}{\mathrm{d}z}-\sqrt{\overline{M}_{\nu;y;\protect\tilde{k} k}^{-1}M_{\varepsilon;x;\protect\tilde{m} m}}\frac{\mathrm{d}}{\mathrm{d}t}\right)\ensuremath{\protect\bow{e}}^{(z)}_{x;m}\right|_{z=z_{\text{B}}}=0,  
    \label{eq:EngquistMadjaGrid}
\end{equation}
which can be easily extended to other field components and to other boundary orientations.
It also states that the grid wave impinging perpendicularly to the boundary $z=z_{\text{B}}$ at the edge $\ensuremath{L}_{x;m}$ propagates at a speed $\nu=(\overline{M}_{\nu;y;\protect\tilde{k} k}^{-1}M_{\varepsilon;x;\protect\tilde{m} m})^{-\frac{1}{2}}$.
A similar result can be obtained if we had used the grid wave equation of the magnetic field $\ensuremath{\protect\bow{\mathrm{\mathbf{h}}}}\xspace$.
Having said this, we can think of the space between edge $\ensuremath{L}_{x;m}$ and $\ensuremath{L}_{x;m-M_{z}}$, namely the preceding $x$-edge in $z$-direction, as a homogeneous transmission line with characteristic impedance
\begin{equation*}
    Z_{0;m}:=\left(\overline{M}_{\nu;y;\protect\tilde{k} k}M_{\varepsilon;x;\protect\tilde{m} m}\right)^{-1/2},
\end{equation*}
 and length $|\ensuremath{L}_{z;m-M_{z}}|$, see Figure~\ref{fig:circuitABC}.
The voltage that excites the line is given by the voltage on edge $\ensuremath{L}_{x;m-M_z}$, namely $\ensuremath{\protect\bow{e}}_{x;m-M_{z}}$.
Therefore, if one wants to implement the \gls*{ABC} at edge $\ensuremath{L}_{m}$, an impedance $Z_{0;m}$ must be assigned to $\ensuremath{L}_{m}$.
A subsequent circuit extraction can be carried out by implementing either Algorithm~\ref{alg:netlistEM1} or \ref{alg:netlistEM2}.
   
\section{Numerical Examples}
\label{sec:numerics}

In this section, the presented methodology to generate electric circuit stamps representing 3D field problems is applied to several representative numerical examples.
In Section~\ref{subsec:resultsET}, we use our netlist extraction method as described in Section~\ref{sec:circuitsET} on an \gls*{ET} problem.
The considered \gls*{ET} problem is a 3D field problem corresponding to the series connection of a capacitor and a resistor.
While applying an external voltage, the transient heating due to the resulting current is simulated using SPICE\xspace and then compared to a field solver reference solution.
Additionally, a circuit representation for the \gls*{ET} field problem of a microelectronic chip package is obtained and used for circuit simulation.
In Section~\ref{subsec:resultsEM}, we apply our method of circuit extraction for \gls*{EM} field problems as described in Section~\ref{sec:circuitsEM} to compute the resonant frequencies of a rectangular cavity with \gls*{PEC} boundaries.
This example is quite illustrative and easy to implement because of the required \glspl*{BC} on the cavity walls.
It simply suffices not to print the circuit stamp associated with those edges on the wall, meaning that the associated stamps are short-circuited.
Furthermore, the availability of an analytic formula for the resonant frequencies permits a direct error assessment.
Finally, in Section~\ref{subsec:resultsABC}, the implementation of \glspl*{ABC} as discussed in Section~\ref{subsec:theoryABC} is carried out to investigate reflections at the end of a rectangular coaxial waveguide.
For all presented examples, the Matlab\textsuperscript{\textregistered}\xspace code to generate the corresponding netlists from the discretised 3D field problem is openly available~\cite{Casper_2018ab}.

\subsection{Electrothermal Circuit Validation}
\label{subsec:resultsET}

\begin{figure}
    \centering
    \includegraphics{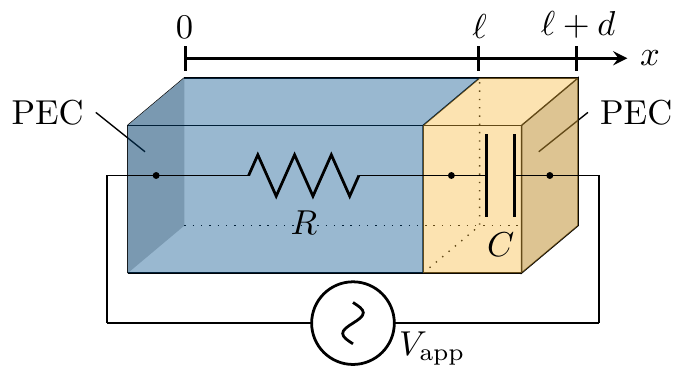}
    \caption{Geometry of the \gls*{ET} validation example. The series connection of a resistive part and a capacitive part is excited with a voltage source $V_{\text{app}}$ imposed as a \textsc{Dirichlet}\xspace condition.}
    \label{fig:benchmarkStructure}
\end{figure}

To validate our netlist extraction method on an \gls*{ET} problem, we consider the \textsc{Joule}\xspace heating in a 3D field problem represented by a series connection of an electric resistor and a capacitor.
The temperature dependence of the electric conductivity is manifested via the temperature coefficient $\alpha=\SI{3.9e-3}{1\per K}$.
The relevant configuration is realised by a brick of two different materials as shown in Figure~\ref{fig:benchmarkStructure}.
The brick is of dimension $0.4\times 0.1\times\SI{0.1}{\micro\metre\tothe{3}}$, with the resistive part having a length of $\ell=\SI{0.3}{\micro\metre}$ and the capacitive part having a length of $d=\SI{0.1}{\micro\metre}$.
At $x=0$ and $x=\ell+d$, \gls*{PEC} electrodes are used.
In Table~\ref{tab:materialPropET}, all material properties are summarised for a reference temperature of $T_{0}=\SI{293}{K}$.

\begin{figure}[t]
    \centering
    \subfloat[\label{fig:testcaseETresultsPotential}]{\includegraphics[width=0.48\columnwidth]{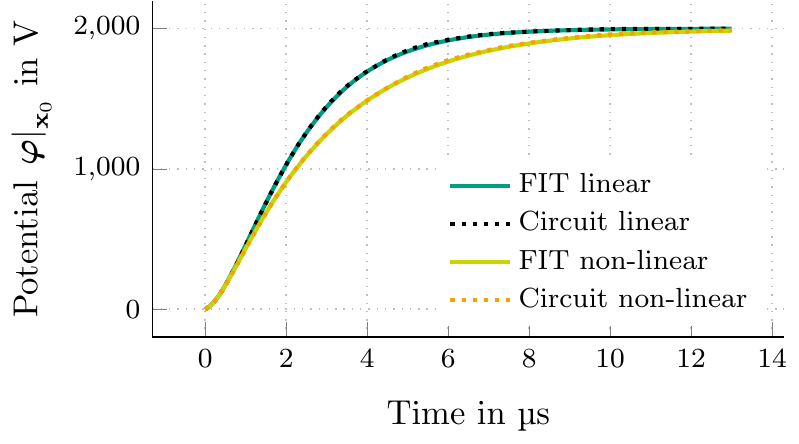}}
    \hspace{0.02\columnwidth}
    \subfloat[\label{fig:testcaseETresultsTemperature}]{\includegraphics[width=0.48\columnwidth]{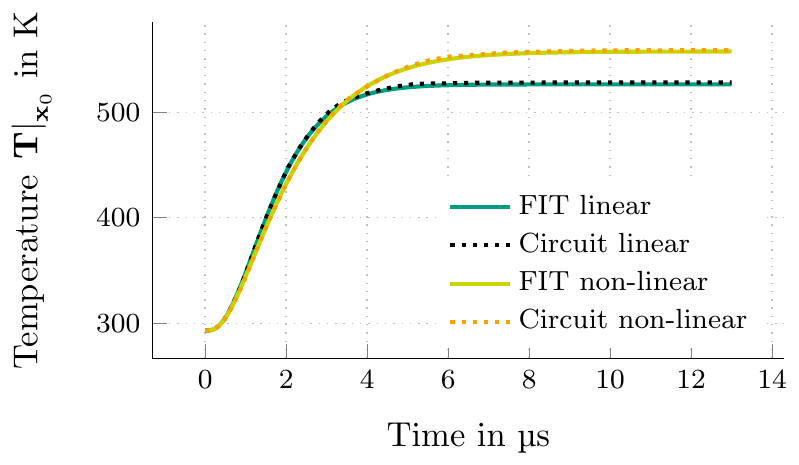}}
    \caption{\gls*{ET} validation example results at $\mathbf{x}=\mathbf{x}_{0}$ for (a) the electric potential $\ensuremath{\boldsymbol{\mathrm{\varphi}}}$ and (b) the temperature $\ensuremath{\mathbf{T}}$. A comparison between the \gls*{FIT} and circuit simulation results is shown for the linear and non-linear cases.}
    \label{fig:resultsET}
\end{figure}

A spatial grid with $9\times 9\times 9$ cells is employed and the field problem is solved by using an in-house implementation of the \gls*{FIT} method with a first-order implicit \textsc{Euler}\xspace scheme as time integrator.
The simulation time amounts to $t_{\text{end}}=\SI{13}{\micro\second}$.
For the simulation of the extracted electric circuit, we use the freely available LTspice\xspace software\footnote{All circuit simulations in this paper have been done using LTspice\xspace in its version 4.22x with default settings.}.
LTspice\xspace uses adaptive refinement in time for which an initial time step of $\Delta t_{\text{init}}=\SI{0.13}{\mu s}$ is used.
The resulting non-equidistant time axis is refined by a factor of three and then used for the \gls*{FIT} solver.
A voltage $V_{\text{app}}=\SI{1}{kV}(1-\exp(-t/\tau))$ with $\tau=0.1 t_{\text{end}}$ is applied at the electrodes as shown in Figure~\ref{fig:benchmarkStructure}.
Using this setting, two simulations are run.
The first neglects the temperature dependence of the conductivities and thus a linear setting ensues.
The second neglects only the temperature dependence of the thermal conductivity but accounts for that of the electric conductivity via the temperature coefficient $\alpha$ entailing a non-linear setting.
To observe the transient behaviour, we select the resistor-capacitor interface point $\mathbf{x}_{0}=(\ell,0,0)$ as observation point and plot the results in Figure~\ref{fig:resultsET}.

\begin{table}[t]
    \centering
    \begin{tabular}{cccc}\toprule
    Symbol                                       & Description             & $0<x<\ell$   & $\ell<x<\ell+d$ \\\bottomrule\toprule
    $\sigma$ $\left(\si{S\per m}\right)$         & electric conductivity   & \num{1e-4}   & \num{0}         \\
    $\varepsilon_\text{r}$                       & relative permittivity   & \num{1}      & \num{3.9}       \\
    $\lambda$ $\left(\si{W/K/m}\right)$          & thermal conductivity    & \num{401}    & \num{1400}      \\
    $\rho c$ $\left(\si{J\per K\per cm^3}\right)$& volumetric heat density & \num{3.48}   & \num{2.10}      \\
    $\alpha (\si{1\per K})$                      & temperature coefficient & \num{3.9e-3} & --              \\\bottomrule
\end{tabular}
     \caption{Material properties at reference temperature $T_{0}=\SI{293}{K}$ for the \gls*{ET} test case.}
    \label{tab:materialPropET}
\end{table}

For a quantitative error assessment of the solution, we define the measures 
    \begin{equation}
        \Delta_{\ensuremath{\boldsymbol{\mathrm{\varphi}}}}=\frac{\max_{i}\lVert\ensuremath{\boldsymbol{\mathrm{\varphi}}}^{\text{cir}}(t_{i})-\ensuremath{\boldsymbol{\mathrm{\varphi}}}^\text{FIT}(t_{i})\rVert_{2}}{\max_{i}\lVert\ensuremath{\boldsymbol{\mathrm{\varphi}}}^{\text{FIT}}(t_{i})\rVert_{2}},\quad
        \Delta_{\ensuremath{\mathbf{T}}}=\frac{\max_{i}\lVert\ensuremath{\mathbf{T}}^{\text{cir}}(t_{i})-\ensuremath{\mathbf{T}}^\text{FIT}(t_{i})\rVert_{2}}{\max_{i}\lVert\ensuremath{\mathbf{T}}^{\text{FIT}}(t_{i})\rVert_{2}},
        \label{eq:solDiffs}
    \end{equation}
where $\ensuremath{\boldsymbol{\mathrm{\varphi}}}^{\text{cir}}$, $\ensuremath{\boldsymbol{\mathrm{\varphi}}}^{\text{FIT}}$, $\ensuremath{\mathbf{T}}^{\text{cir}}$ and $\ensuremath{\mathbf{T}}^{\text{FIT}}$ are the potential and temperature solution vectors obtained via circuit and \gls*{FIT} simulation, respectively.
To calculate these errors appropriately, the circuit solution is interpolated to the time axis employed by the \gls*{FIT} solution using cubic spline interpolation.
We want to remark that the quantities in \eqref{eq:solDiffs} are not errors in the classical sense since none of the solutions is exact.
The computed differences amount to $\Delta_{\ensuremath{\boldsymbol{\mathrm{\varphi}}}}^{\text{lin}}\approx\SI{0.36}{\percent}$ and $\Delta_{\ensuremath{\mathbf{T}}}^{\text{lin}}\approx\SI{0.48}{\percent}$ for the linear case and $\Delta_{\ensuremath{\boldsymbol{\mathrm{\varphi}}}}^{\text{nlin}}\approx\SI{0.42}{\percent}$ and $\Delta_{\ensuremath{\mathbf{T}}}^{\text{nlin}}\approx\SI{0.44}{\percent}$ for the non-linear case.
The remaining error is attributed mainly to the different time integrators.

\begin{figure}
	\centering
    \subfloat[\label{fig:chip}]{\includegraphics[width=0.28\textwidth]{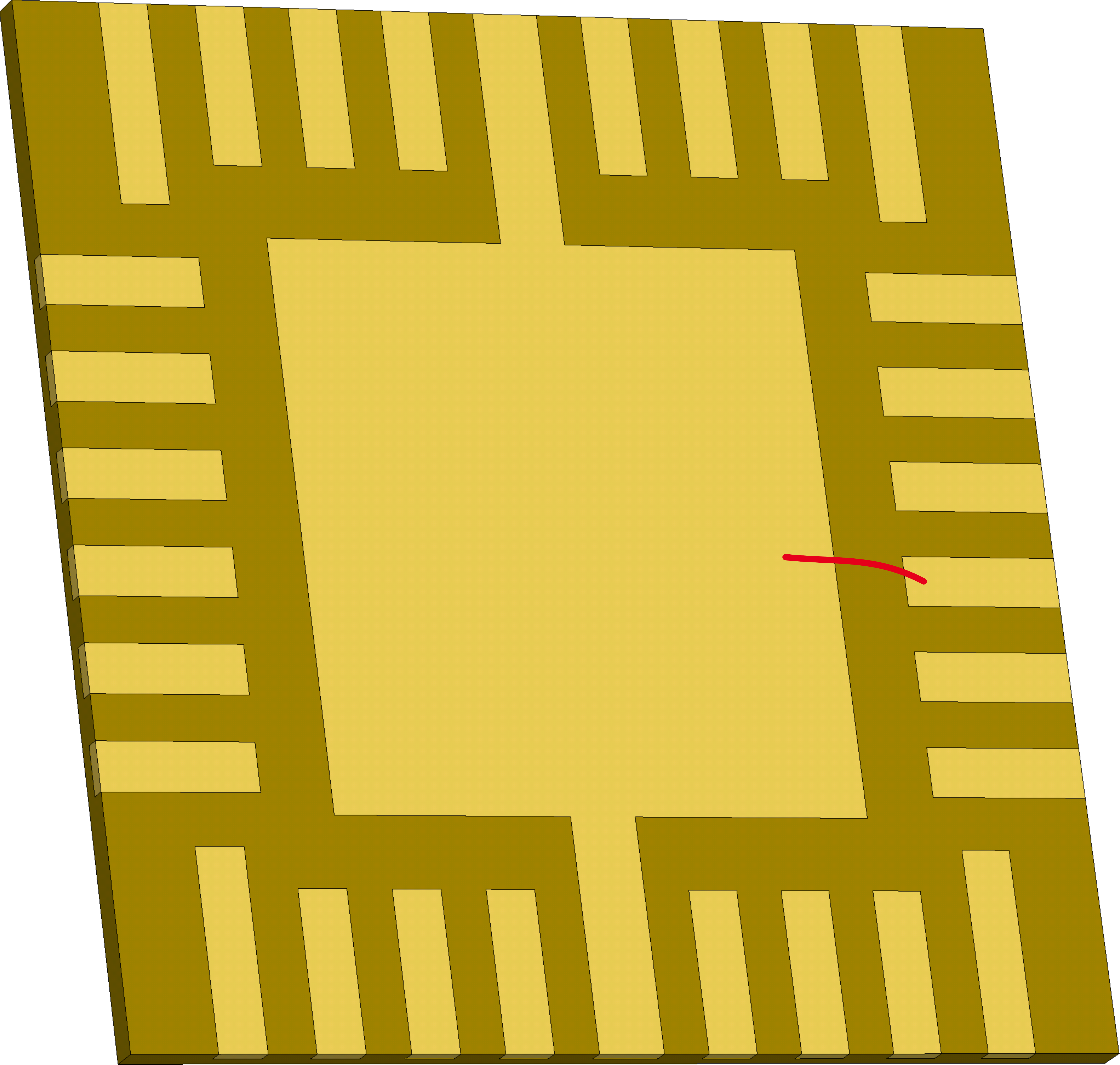}}
    \hspace{0.02\textwidth}
    \subfloat[\label{fig:testcaseChip}]{\includegraphics[width=0.33\textwidth]{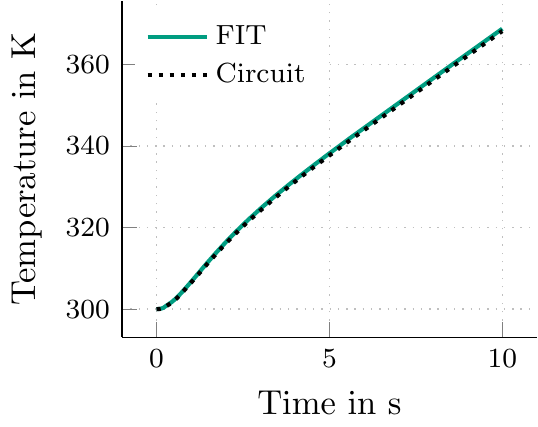}}
    \hspace{0.02\textwidth}
    \subfloat[\label{fig:chipResults3D}]{\includegraphics[width=0.3\textwidth]{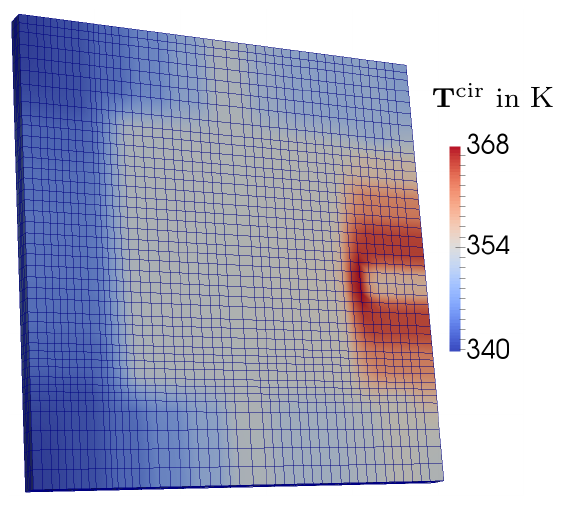}}
	\caption{(a) 3D microelectronic chip package with an attached bond wire. (b) Temperature at the hottest point of the chip package obtained by \gls*{FIT} and circuit simulation. (c) Temperature distribution in the chip package obtained by circuit simulation.}
	\label{fig:resultsChip}
\end{figure}

For an industry-relevant example, the proposed method is applied to the 3D microelectronic chip package~\cite{Casper_2016ab} as shown in Figure~\ref{fig:chip}.
The field problem is discretised as described in Section~\ref{sec:FIT} and Algorithm~\ref{alg:netlistET} is used to generate the corresponding netlist.
This netlist uses \num{101147} circuit elements to describe a field problem that has been discretised using a grid with \num{9660} nodes.
Running a transient analysis on this netlist, an error of $\Delta_{\ensuremath{\boldsymbol{\mathrm{\varphi}}}}\approx\SI{0.23}{\percent}$ and $\Delta_{\ensuremath{\mathbf{T}}}\approx\SI{0.17}{\percent}$ compared to the field simulation is achieved.
Figure~\ref{fig:testcaseChip} shows the temperature of the hottest point in the chip package obtained by \gls*{FIT} and circuit simulation.
The temperature distribution in the chip package resulting from circuit simulation is shown in Figure~\ref{fig:chipResults3D}.
Thus, a good agreement of circuit simulation results for a 3D \gls{ET} problem is achieved when compared to the corresponding field solver results.

\subsection{Electromagnetic Circuit Validation}
\label{subsec:resultsEM}

In this section, we validate the method presented in Section~\ref{sec:circuitsEM} for the circuit representation of \gls*{EM} field problems.
To this end, a lossless rectangular resonant cavity with \gls*{PEC} boundaries and outer dimensions of $a\times b\times d=0.1\times 0.2\times \SI{0.2}{\cubic\metre}$ is simulated and its resonant frequencies are computed.
The homogeneous material within the cavity is specified by the relative permittivity $\ensuremath{\varepsilon}_{\mathrm{r}}=\num{2}$ and the relative permeability $\ensuremath{\mu}_{\mathrm{r}}=\num{1}$ for which the resonant frequencies can also be calculated by means of the formula~\cite{Griffiths_1999aa}
\begin{equation*}
    f_{\text{r}}^{mnp}=\frac{c_{0}}{2\sqrt{\ensuremath{\mu}_{\mathrm{r}}\ensuremath{\varepsilon}_{\mathrm{r}}}}\sqrt{\left(\frac{m}{a}\right)^2+\left(\frac{n}{b}\right)^2+\left(\frac{p}{d}\right)^2},
\end{equation*}
where $c_{0}$ is the speed of light and $\left\{m,n,p\right\}$ are the indices of the resonant modes and are given by natural numbers including zero.
For these resonant frequencies, the longitudinal \gls*{TE} and \gls*{TM} field components are given by
\begin{subequations}
    \begin{align}
        H_{z}^{mnp}=H_{0}^{mnp}\cos\left(\frac{m\pi}{a}x\right)\cos\left(\frac{n\pi}{b}y\right)\sin\left(\frac{p\pi}{d}z\right),\\
        E_{z}^{mnp}=E_{0}^{mnp}\sin\left(\frac{m\pi}{a}x\right)\sin\left(\frac{n\pi}{b}y\right)\cos\left(\frac{p\pi}{d}z\right),
    \end{align}
    \label{eq:fieldLongitud}\end{subequations}
respectively, where $H_{0}^{mnp}$ and $E_{0}^{mnp}$ are the corresponding field amplitudes.
For a \gls*{TE} (\gls*{TM}) mode $mnp$ to exist, $H_{z}$ ($E_{z}$) must not become zero.

\begin{figure}[t]
    \centering
    \subfloat[\label{fig:testcaseEMresultsTE}]{\includegraphics[width=0.49\columnwidth]{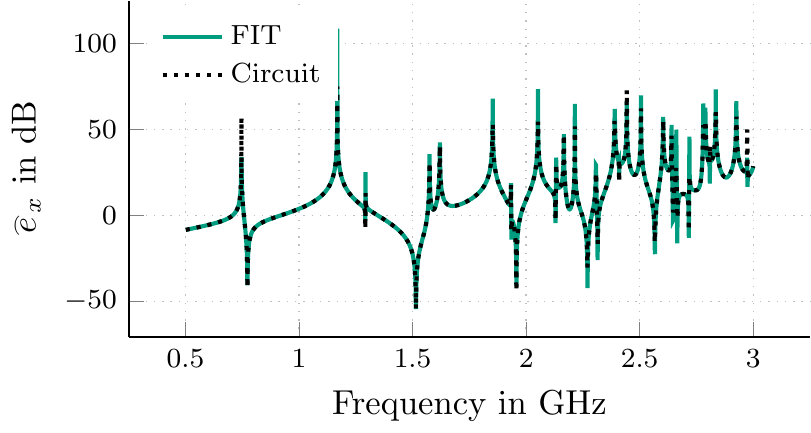}}
    \subfloat[\label{fig:testcaseEMresultsTM}]{\includegraphics[width=0.49\columnwidth]{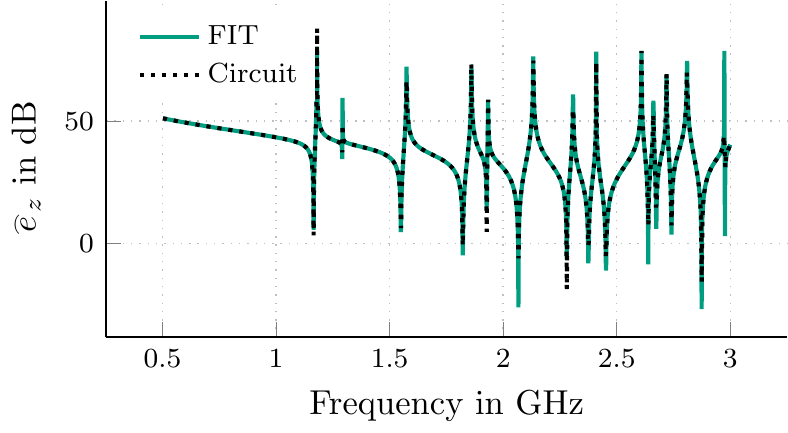}}
    \caption{(a) \gls*{TE} and (b) \gls*{TM} computed resonant frequencies by using \gls*{FIT} and by using circuit simulation. The plotted fields (voltages) $\ensuremath{\protect\bow{e}}_{x}$ and $\ensuremath{\protect\bow{e}}_{z}$ are evaluated at the edges connected to the central grid point.}
    \label{fig:resultsEM}
\end{figure}

To simulate the excitation of modes within the cavity, we discretise the interior of the cavity using a regular grid of $10$ cells in each direction and apply \glspl*{PEC} upon all cavity walls.
Then, we obtain the resonant frequencies by solving the generalised eigenvalue problem given by
\begin{equation*}
    \ensuremath{\protect\widetilde{\mathbf{C}}}\xspace\ensuremath{\mathbf{M}_{\nu}}\xspace\ensuremath{\mathbf{C}}\xspace\ensuremath{\protect\bow{\mathrm{\mathbf{e}}}}\xspace = \left(2\pi f_{\text{r};\text{E}}^{\text{FIT}}\right)^{2}\ensuremath{\mathbf{M}_{\varepsilon}}\xspace\ensuremath{\protect\bow{\mathrm{\mathbf{e}}}}\xspace,
\end{equation*}
where a resonant frequency is denoted by $f_{\text{r};\text{E}}^{\text{FIT}}$.
Alternatively, we can use appropriate excitations to analyse the resulting field at a set of given frequencies.
For example, we can use an electric current source oriented along the positive $z$-direction and attached to the central grid point to excite \gls*{TM} modes.
In a similar manner, \gls*{TE} modes are excited by means of a looping electric current source located in the cavity centre.
We then solve the discretised problem given by
\begin{equation*}
    (\ensuremath{\protect\widetilde{\mathbf{C}}}\xspace\ensuremath{\mathbf{M}_{\nu}}\xspace\ensuremath{\mathbf{C}}\xspace-\omega^{2}\ensuremath{\mathbf{M}_{\varepsilon}}\xspace)\ensuremath{\protect\bow{\mathrm{\mathbf{e}}}}\xspace = -j\omega\ensuremath{\protect\bbow{\mathrm{\mathbf{j}}}}\xspace_{\text{i}}
\end{equation*}
for a set of angular frequencies $\omega$, where $\ensuremath{\protect\bbow{\mathrm{\mathbf{j}}}}\xspace_{\text{i}}$ is the current source vector whose entries are all zero except at the corresponding source edges.
The frequency axis from \num{0.5} to \SI{3}{GHz} is discretised using \num{2000} points for \gls*{TM} excitation and \num{3000} points for \gls*{TE} excitation.
The results are evaluated on one edge for each excitation type.
For the \gls*{TM} case, an edge in positive $z$-direction connected to the central grid point is used while for the \gls*{TE} case, an edge in positive $x$-direction connected to the point (5,6,10)\SI{}{cm} is used.
In Figure~\ref{fig:resultsEM}, the voltages $\ensuremath{\protect\bow{e}}_{x}(\omega)$ and $\ensuremath{\protect\bow{e}}_{z}(\omega)$ along these edges are plotted.
From the peaks in the plots, the corresponding resonant frequencies $f_{\text{r;TE}}^{\text{FIT}}$ and $f_{\text{r;TM}}^{\text{FIT}}$ are identified\footnote{We have used the function \texttt{findpeaks} of Matlab\textsuperscript{\textregistered}\xspace R2017a to identify the peaks in the plot.}.

To validate our circuit extraction method for \gls*{EM} problems, we generate the netlist of the resonant cavity according to Algorithm~\ref{alg:netlistEM1} and simulate the resulting circuit in LTspice\xspace by performing an AC analysis in the same frequency range as before.
We then identify the circuit voltages corresponding to $\ensuremath{\protect\bow{e}}_{x}(\omega)$ and $\ensuremath{\protect\bow{e}}_{z}(\omega)$ and plot them also directly in Figure~\ref{fig:resultsEM} for a comparison.
The circuit resonant frequencies $f_{\text{r;TE}}^{\text{Cir}}$ and $f_{\text{r;TM}}^{\text{Cir}}$ are again identified by means of the peaks and we collect the computed resonant frequencies for the first few modes in Table~\ref{tab:resultsEMerrors}.
Note that, according to~\eqref{eq:fieldLongitud}, the \gls*{TM}-mode does not exist for $m=1$.
Additionally, the errors 
\begin{equation*}
    \epsilon_{\text{r;TE}} := \frac{\left|f_{\text{r;TE}}^{\text{FIT}}-f_{\text{r;TE}}^{\text{Cir}}\right|}{f_{\text{r;TE}}^{\text{FIT}}},\quad \epsilon_{\text{r;TM}} := \frac{\left|f_{\text{r;TM}}^{\text{FIT}}-f_{\text{r;TM}}^{\text{Cir}}\right|}{f_{\text{r;TM}}^{\text{FIT}}}
\end{equation*}
are presented.
For all modes, these errors are much smaller than $\SI{1}{\%}$.

\begin{table}[t]
    \centering
    \begin{tabular}{ccccccccc}\toprule
    Mode $mnp$ & $f_{\text{r};\text{E}}^{\text{FIT}}$ & $f_{\text{r;TE}}^{\text{Cir}}$ & $f_{\text{r;TE}}^{\text{FIT}}$ & $f_{\text{r;TM}}^{\text{Cir}}$ & $f_{\text{r;TM}}^{\text{FIT}}$ & $f_{\text{r}}^{mnp}$ & $\epsilon_{\text{r;TE}}$ (\%) & $\epsilon_{\text{r;TM}}$ (\%)\\\bottomrule\toprule
    011               &      \num{0.746}       &      \num{0.746}        &         \num{0.747}     &        ---              &        ---              &      \num{0.749}   &   \num{0.0471}   &        ---    \\
    110/101/012/021   &      \num{1.180}       &      \num{1.169}        &         \num{1.169}     &      \num{1.180}        &      \num{1.180}        &      \num{1.185}   &   \num{0.0030}   & \num{0.0086}  \\
    111               &      \num{1.293}       &      \num{1.293}        &         \num{1.293}     &      \num{1.293}        &      \num{1.293}        &      \num{1.298}   &   \num{0.0270}   & \num{0.0216}  \\
    121               &      \num{1.575}       &      \num{1.575}        &         \num{1.575}     &      \num{1.574}        &      \num{1.574}        &      \num{1.590}   &   \num{0.0354}   & \num{0.0262}  \\
    013               &      \num{1.620}       &      \num{1.621}        &         \num{1.620}     &        ---              &        ---              &      \num{1.676}   &   \num{0.0050}   &    ---        \\
    122               &      \num{1.853}       &      \num{1.853}        &         \num{1.853}     &      \num{1.860}        &      \num{1.861}        &      \num{1.836}   &   \num{0.0244}   & \num{0.0465}  \\\bottomrule
\end{tabular}
     \caption{Analytic and computed resonant frequencies in \si{GHz} for several resonant modes and the corresponding relative errors. As mode degeneracy in the cavity is relevant, an exact identification of the mode indices from the plots in Figure~\ref{fig:resultsEM} is not possible.}
    \label{tab:resultsEMerrors}
\end{table}

\subsection{Signal Transmission Using Absorbing Boundary Conditions}
\label{subsec:resultsABC}

\begin{figure}
    \centering
    \includegraphics{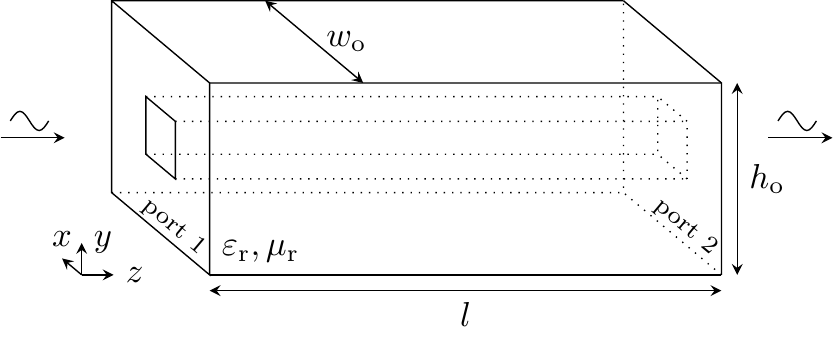}
    \caption{Geometry of a coaxial transmission line oriented along the $z$-axis and excited at port 1 by a Gaussian current pulse. A propagating wave is thus generated arriving at port~2 that is connected to an external impedance. For reasons of visibility, the annotations of the inner conductor's dimensions are not shown.}
    \label{fig:coaxialLine}
\end{figure}

In this section, based on the method described in Section~\ref{subsec:theoryABC}, we present a simple validation example for \glspl*{ABC} in the context of circuit simulation.
To this end, let us consider a coaxial transmission line of rectangular cross section oriented along the $z$-direction as depicted in Figure~\ref{fig:coaxialLine}.
For the simulation of an infinitely long line using a finite computational domain, the implementation of \glspl*{ABC} is required to counteract unwanted incoming reflections.
We use an excitation signal at port 1 and simulate its propagation in time until it has reached port~2.
Thus, we generate two simulation results in time domain.
The first one corresponds to the case when port~2 is terminated with a perfect magnetic wall, that is an open port ($Z_{2}\to\infty$).
The second one corresponds to the case when port~2 is terminated with the characteristic line impedance ($Z_{2}=Z_{0}$).
For both cases, a \gls*{PMC} ($Z_{1}\to\infty$) at port 1 is applied\footnote{According to image theory, the perfect magnetic wall at port 1 serves as a mirror which reflects uprightly the otherwise backward travelling wave.}.

The length of the coaxial line is $l=\SI{150}{cm}$, the width and height of the outer conductor are $w_{\text{o}}=h_{\text{o}}=\SI{3}{cm}$ and of the inner conductor $w_{\text{i}}=h_{\text{i}}=\SI{1}{cm}$.
While the conductors are modelled as \gls*{PEC}, the material between them is vacuum.
Due to the expected propagation in $z$-direction, the longitudinal direction requires a finer discretisation compared to the transversal direction.
Thus, we choose a grid of $3\times 3\times 150$ cells.
According to \eqref{eq:EngquistMadjaGrid} for such a discretisation grid, the characteristic impedance for the edges connecting the inner and outer conductor along the plane of port~2 should amount to\footnote{Note that in the calculation of $Z_{0;m}$, the value employed for $M_{\ensuremath{\varepsilon};x;\protect\tilde{m} m}$ is taken directly from the parallel edge just in front of the boundary edge in accordance with the impinging grid wave front speed.} $Z_{0;m}=(\overline{M}_{\nu;y;k\protect\tilde{k}}M_{\varepsilon;x;\protect\tilde{m} m})^{-1/2}\approx\SI{376.7}{\Omega}$.
For the given grid, there are eight such edges giving eight parallel conductances such that the total resistance at port~2 equals $Z_{2}=Z_{0}=8Z_{0;m}\approx\SI{47.09}{\Omega}$, which is also the characteristic impedance of the line.
To excite the signal at port 1, the edges connecting the inner and outer conductors along the plane of port 1 are impressed with a current such that the total current from inner to outer conductor is
\begin{equation*}
    I_{\text{i}}(t)=\hat{I}\exp\left(\frac{(t-t_{0})^2}{2\sigma_{\text{G}}^2}\right),
\end{equation*}
with $\hat{I}=\SI{1}{A}$, which is a \textsc{Gauss}\xspace pulse with a maximal frequency\footnote{Confining the excitation to this maximal frequency component, we assure that the TEM mode is the only propagating mode on the line} component of $f_{\text{max}}=\SI{1}{GHz}$.
The used constants are given by ${\sigma_{\text{G}}=\sqrt{\ln(10)}/(\pi f_{\text{max}})}$ and by ${t_{0}=\sqrt{6\sigma_{\text{G}}^2\ln(10)}}$.
The simulation time $t_{\text{end}}=\SI{10}{ns}$ is chosen such that the excited pulse can reach port~2.
Having defined the geometry, the excitation and the simulation time, we also generate the corresponding netlist using Algorithm~\ref{alg:netlistEM1}.

\begin{figure}[t]
    \centering
    \subfloat[\label{fig:testcaseABCresultsWavePropReflect}]{\includegraphics[width=0.499\columnwidth]{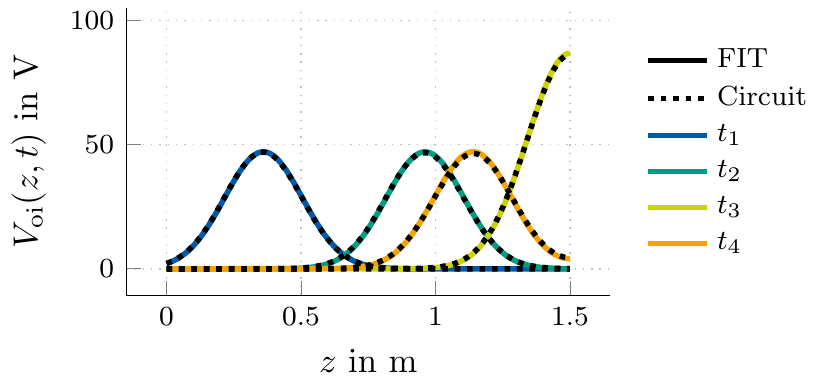}}
    \subfloat[\label{fig:testcaseABCresultsWavePropNoReflect}]{\includegraphics[width=0.499\columnwidth]{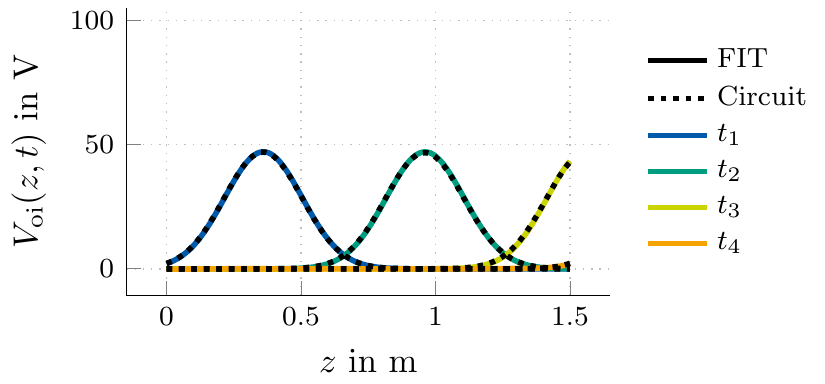}}
    \caption{Snapshots of the wave propagating along the $z$-axis at times $t_{1}=\SI{3}{ns}$, $t_{2}=\SI{5}{ns}$, $t_{3}=\SI{7}{ns}$ and $t_{4}=\SI{8}{ns}$ computed by \gls*{FIT} and circuit simulation. (a) shows the case $Z_{2}\to\infty$ to realise total reflection. (b) shows the case $Z_{2}=Z_{0}$ to realise \glspl*{ABC}.}
    \label{fig:testcaseABCresultsWaveProp}
\end{figure}
\begin{figure}[t]
    \centering
    \subfloat[\label{fig:testcaseABCresultsV2vsTimeReflect}]{\includegraphics[width=0.49\columnwidth]{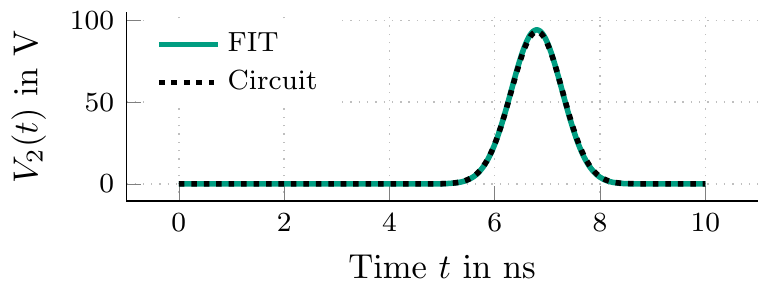}}
    \hspace{0.01\columnwidth}
    \subfloat[\label{fig:testcaseABCresultsV2vsTimeNoReflect}]{\includegraphics[width=0.49\columnwidth]{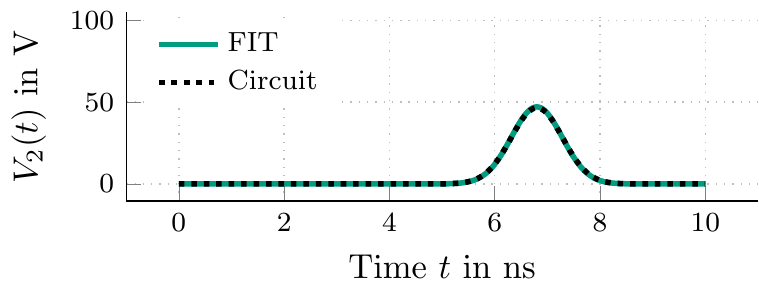}}
    \caption{Comparison of the output voltage $V_{2}$ with respect to time computed by \gls*{FIT} and circuit simulation for the case of (a) $Z_{2}\to\infty$ to realise total reflection and (b) $Z_{2}=Z_{0}$ to realise \glspl*{ABC}.}
    \label{fig:testcaseABCresultsV2vsTime}
\end{figure}

The generated netlist is simulated by means of a transient analysis in LTspice\xspace.
On the other hand, the Leapfrog scheme is used as a time integrator within the \gls*{FIT} framework to carry out the simulation directly on the 3D grid. 
In both cases, we use the time axis generated by the adaptive time stepping algorithm provided by LTspice\xspace, which satisfies the \gls*{CFL}-condition being a stability requirement for the explicit Leapfrog scheme~\cite{Yee_1966aa}.
In the following, we compare the voltage $V_{\text{oi}}(z,t)$ between outer and inner conductor and the voltage $V_{2}(t)=V_{\text{oi}}(l,t)$ at port~2.
For $Z_{2}\to\infty$ and $Z_{2}=Z_{0}$, Figure~\ref{fig:testcaseABCresultsWaveProp} shows $V_{\text{oi}}(z,t)$ at different times computed by means of \gls*{FIT} and circuit simulation.
Figure~\ref{fig:testcaseABCresultsWavePropReflect} shows the case in which port~2 is terminated by a perfect magnetic boundary (standard homogeneous \textsc{Neumann}\xspace) condition while Figure~\ref{fig:testcaseABCresultsWavePropNoReflect} shows the case when a matching impedance $Z_{2}=Z_{0}$ according to \eqref{eq:EngquistMadjaGrid} is applied at port~2.
As predicted by the theory in Section~\ref{subsec:theoryABC}, we observe that the matching impedance at port~2 counteracts incoming reflections effectively. 
In Figure~\ref{fig:testcaseABCresultsV2vsTime}, we show $V_{2}(t)$ computed by means of \gls*{FIT} and circuit simulation for the two already considered cases.
We observe therein that incoming reflections at port~2 result in an undesired overshooting of the voltage.
For a quantitative comparison, we define the relative error of $V_{2}(t)$ between \gls*{FIT} and circuit results as 
\begin{equation*}
    \Delta_{V_{2}}^{Z_{2}}=\frac{\max_{i}\lVert V_{2}^{\text{cir}}(t_{i})-V_{2}^{\text{FIT}}(t_{i})\rVert_{2}}{\max_{i}\lVert V_{2}^{\text{FIT}}(t_{i})\rVert_{2}},
\end{equation*}
and obtain $\Delta_{V_{2}}^{Z_{0}}\approx\SI{1.046}{\%}$ and $\Delta_{V_{2}}^{\infty}\approx\SI{1.136}{\%}$.

\section{Conclusion and Future Work}
\label{sec:conclusion}

A method for the automatic netlist generation of general 3D \gls*{ET} and \gls*{EM} problems has been presented.
The topology of each circuit stamp associated with edges in the regular primal grid has been derived by using \gls*{FIT} for spatial discretisation.
Using the \gls*{MNA}, the \gls*{FIT}-discretised \gls*{ET} formulation has been mapped into a circuit that can be solved by any SPICE\xspace-like program.
It has been shown that initial conditions can be easily prescribed as initial potentials for the lumped capacitances in the SPICE\xspace language.
Furthermore, the implementation of mixed boundary conditions of \textsc{Dirichlet}\xspace, homogeneous \textsc{Neumann}\xspace and \textsc{Robin}\xspace type has been discussed.
We have also shown that temperature dependent material models result in non-linearities in the lumped resistances requiring the implementation of behavioural \glspl*{VCCS} in SPICE\xspace.

From the standard E-H formulation and the E-A formulation, we have derived circuit stamps representing general \gls*{EM} problems.
In both circuit representations, the integrated electric field models the voltage between the stamp terminals while the integrated magnetic vector potential models the electric current in the E-A formulation.
To guarantee uniqueness of the solution in the latter, we have employed \textsc{Coulomb}\xspace's gauge on the magnetic vector potential, that has been implemented by means of a tree-cotree decomposition of the primal discretisation grid.
Thereby, the electric current along edges in the cotree are degrees of freedom whereas those along edges in the tree are modelled by \glspl*{CCCS} being controlled by currents in the cotree.
For both representations, a dual circuit formulation exists if magnetic sources instead of electrical sources are considered.
In the dual case, an auxiliary electric potential would be used instead of the magnetic vector potential.
To demonstrate the correctness of our formulations, several numerical examples have been shown for the primal circuits involving electric sources only.

The formulation of inhomogeneous \textsc{Neumann}\xspace \glspl*{BC} could be a further extension to the presented approach.
Furthermore, the method can also be applied to extract circuits from \gls*{FEM} models.
To account for thermal effects in \gls*{EM} problems, the methods for the extraction of \gls*{ET} and \gls*{EM} circuit stamps can be combined to generate a thermo-\gls*{EM} circuit stamp.
Methods to account for non-linear material characteristics in the \gls*{EM} case are still to be developed.
However, in principle one can follow similar ideas to those presented in the \gls*{ET} case.
For large field models, the resulting circuit can become very large.
Therefore, to efficiently simulate such circuits, dedicated \gls*{MOR} techniques for circuits can be applied.
The first of these techniques is known as the \gls*{AWE} proposed by Pillage and Rohrer~\cite{Pillage_1990aa} and extensions developed afterwards.
The most prominent ones are the \gls*{MPVL} by Feldmann and Freund~\cite{Feldmann_1995aa} and the \gls*{PRIMA}~\cite{Odabasioglu_1998aa}.
More recent approaches are based on the proper orthogonal decomposition~\cite{Hinze_2012ad} and other well-known general \gls*{MOR} techniques.

\section*{Acknowledgements}
This is a pre-print of an article published in the Journal of Computational Electronics.
The final authenticated version is available online at: https://doi.org/10.1007/s10825-019-01368-6.
The authors thank Abdul Moiz and Victoria Heinz for their passionate work on implementing the automated electrothermal netlist generation. The work is supported by the European Union within FP7-ICT-2013 in the context of the \emph{Nano-electronic COupled Problems Solutions} (nanoCOPS) project (grant no. 619166), by the \emph{Excellence Initiative} of the German Federal and State Governments and the Graduate School of Computational Engineering at Technische Universität Darmstadt.

\end{document}